\documentclass[useAMS,usenatbib,usegraphicx]{mn2e}

\usepackage{amsmath}

\usepackage{epsfig}
\title{Signature of Inverse Compton emission from  blazars}

\author[Gaur et al.]
{Haritma Gaur$^{1,2}$\thanks{E-mail: haritma@shao.ac.cn}, P. Mohan$^{1}$\thanks{E-mail: pmohan@shao.ac.cn}, Alicja Wierzcholska$^{3}$, 
Minfeng Gu$^{1,2}$  \\
\\
$^{1}$ Shanghai Astronomical Observatory, Chinese Academy of Sciences, 80 Nandan Road, Shanghai 200030, China \\
$^{2}$ Key Laboratory for Research in Galaxies and Cosmology, Shanghai Astronomical Observatory, Chinese Academy of Sciences, \\
 80 nandan road, Shanghai 200030, China   \\
$^{3}$ Institute of Nuclear Physics, Polish Academy of Sciences, ul. Radzikowskiego 152, PL-31-342 Krakow, Poland  \\
}

\begin{document}

\date{Accepted ....... Received  ......; in original form ......}

\pagerange{\pageref{firstpage}--\pageref{lastpage}} \pubyear{2010}

\maketitle

\label{firstpage}
\begin{abstract}

Blazars are classified into high, intermediate and low energy peaked sources based on the location of their synchrotron peak. 
This lies in infra-red/optical to ultra-violet bands for low and intermediate peaked blazars. The transition from synchrotron 
to inverse Compton emission falls in the X-ray bands for such sources. We present the spectral and timing analysis of 14 
low and intermediate energy peaked blazars observed with XMM–Newton spanning 31 epochs. Parametric fits to X-ray spectra 
helps constrain the possible location of transition from the high energy end of the synchrotron to the low energy end of 
the inverse Compton emission. In seven sources in our sample, we infer such a transition and constrain the break energy 
in the range 0.6 − 10 keV. The Lomb-Scargle periodogram is used to estimate the power spectral density (PSD) shape. It is 
well described by a power law in a majority of light curves, the index being flatter compared to general expectation from 
AGN, ranging here between 0.01 and 1.12, possibly due to short observation durations resulting in an absence of long term trends. 
A toy model involving synchrotron self-Compton (SSC) and external Compton (EC; disk, broad line region, torus) mechanisms 
are used to estimate magnetic field strength $\leq$ 0.03 - 0.88 G in sources displaying the energy break and infer a prominent 
EC contribution. The timescale for variability being shorter than synchrotron cooling implies steeper PSD 
slopes which are inferred in these sources.

\end{abstract}

\begin{keywords}
galaxies: active $-$ BL Lacertae objects: general $-$ BL Lacertae objects: radiation mechanisms: non-thermal; relativistic processes

individual)

\end{keywords}

\section{Introduction}

Blazars constitute a class of Active Galactic Nuclei (AGN) characterized by their extreme properties including strong variability (flux and 
polarization) and weak spectral lines (dominated by continuum), believed to be due to observer line of sight orientation based effects which 
from AGN unification models (e.g., Urry \& Padovani 1995) imply that the relativistic jet is directed towards the observer at small angles. 
Radiation from blazars span the whole electromagnetic spectrum from radio wavelengths upto $\gamma$-ray range. Their broad 
band spectral energy distribution (SED) is characterized by a double peaked structure. The low energy peak is mainly due to the synchrotron 
radiation from relativistic non-thermal electrons. The high energy peak can be due to the inverse Compton scattering of lower energy synchrotron 
photons from the same electron population (synchrotron self Compton scenario; e.g. Kirk, Rieger \& Mastichiadis 1998) or of external photons from 
accretion disc, broad line region or dusty torus in the leptonic scenarios (external Compton scenario; e.g. Sikora, Begelman \& Rees 1994); 
while it can be due to synchrotron emission from protons or from secondary decay products of charged pions in the hadronic scenarios (e.g. 
Atoyan \& Dermer 2003; Bottcher et al. 2013).

Based on the location of the low energy or synchrotron peak, these sources are classified into high, intermediate and low energy peaked blazars 
(HBL, IBL and LBL, respectively; Padovani \& Giommi 1995). Abdo et al. (2010) classified blazars based on the location of the synchrotron 
peak frequency, $\nu_{s}$. If $\nu_{s}$ $\le$ $10^{14}$ Hz (in the infrared), it is classified as a low spectral peak (LSP) source; if it is in 
optical--ultra-violet range ($10^{14}$ $\le$ $\nu_{s}$ $\le$ $10^{15}$ Hz), it is classified as an intermediate spectral peak (ISP) source; and 
if it lies in the X-ray regime ($\nu_{s}$ $\ge$ $10^{15}$ Hz), it is classified as a high spectral peak (HSP) source. BL Lacertae (BL Lac) and 
Flat Spectrum Radio Quasars (FSRQs) together constitute the blazar class of sources. Low luminosity BL Lacs which are HSPs exhibit the synchrotron 
peak in the UV-soft X-ray band and the inverse Compton peak between the GeV and the TeV gamma-ray band (Padovani \& Giommi 1995). In mid luminosity 
sources which are LSPs and ISPs, the synchrotron peak is in the near infrared band and the X-ray emission is due to either or both the synchrotron 
and inverse Compton components. In high luminosity FSRQ sources, the synchrotron peak is in the far infrared band and X-ray emission is ascribed 
to the inverse Compton component (e.g. Giommi et al. 1995; Fossati et al. 1998). The peak frequency of the synchrotron component is found to 
inversely correlate with the luminosity of the blazar and different kind of blazars can be classified based on their peak energy to form a 
``blazar sequence" (e.g. Fossati et al. 1998; Ghisellini et al. 1998; Giommi et al. 2012).

The HSP blazars are brightest in X-ray bands and show strong flux variability over diverse time-scales ranging between minutes to years(e.g. Sembay 
et al. 1993; Brinkmann et al. 2005; Zhang et al. 2005, 2008; Gaur et al. 2010; Kapanadze et al. 2014 and references therein). The variability 
time-scales can be used to constrain the emission region size (e.g. Tramacere et al. 2009; Mohan et al. 2016). The variability amplitude is found 
to be correlated with energy as the hardest synchrotron radiation is produced by the most energetic electrons with the smallest cooling time-scales
 (e.g. Zhang et al. 2005; Gliozzi et al. 2006). Their X-ray spectra are generally characterized by a soft ($\gamma$ $>$ 2) convex shape or by continuosly 
downward curved shape (e.g. Perlman et al. 2005; Zhang et al. 2008), and can originate from an energy dependent particle acceleration with suitable 
cooling timescales, and is accompanied by strong and rapid variability (e.g. Massaro et al. 2004). 

The X-ray emission from LSPs is believed to originate mainly from the inverse Compton scattering of seed photons by the low energy tail of the electron 
population. In addition, it can include a contribution from the synchrotron emission from high energy particles. In ISP blazars, there is a clear turning 
point in the SED where synchrotron and inverse Compton components intersect e.g. S5 0716$+$714 ( Giommi et al. 1999; Tagliaferri et al. 
2003; Donato et al. 2005; Ferrero et al. 2006; Wierzcholska \& Siejkowski 2015); ON 231 (Tagliaferri et al. 2000); BL Lacertae (Tanihata et al. 2000; 
Ravasio et al. 2002);  AO 0235$+$164 (Raiteri et al. 2006); OQ 530 (Tagliaferri et al. 2003); 3C 66A (Donato et al. 2005; Wierzcholska \& Wagner 2016); 
4C $+$21.35 (Wierzcholska \& Wagner 2016), amongst others. 

The X-ray variability is also observed for ISP and LSP blazars on both long timescales (e.g. Donato et al. 2005; 
Ferrero et al. 2006; Raiteri et al. 2006; Wierzcholska \& Siejkowski 2015). Recently, Gupta et al. (2016) studied a sample of LSP blazars in X-ray bands 
and found the intra-day variability to be less pronounced as compared to that in HSP blazars, expected due to longer cooling time scales for the lowest 
energy electrons responsible for the inverse Compton emission. In LSPs, short time scale ($<$hours) variability is prevalent only in the synchrotron 
component, while inverse Comption emission appears to dominate over longer time scales (∼days). Unlike HSPs, their variability amplitudes are found to 
be anti-correlated with the emission energy and no significant time lags are found between the hard and soft energy X-rays (Giommi et al. 1999; Ravasio 
et al. 2002; Ferrero et al. 2006). In addition, the X-ray band mostly lies at the transition from synchrotron to the inverse Compton component for LSP 
and ISP blazars and hence is key to disentangle the contribution of the two components to the broad band continuum. One expects that FSRQs would have 
a flatter spectra (index $\Gamma$ $<$ 2), while the LSPs and ISPs would have intermediate slopes (with index $\Gamma$=1.5--1.8), or even concave X-ray 
continua (e.g. Donato et al. 2005; Massaro et al. 2008; Wierzcholska \& Wagner 2016). In previous studies, it was inferred that 
the X-ray spectra of blazars is well described by a single power law or a broken power law (e.g. Perlmann et al. 1996; Urry et al. 1996). The log parabola 
model has also proven to describe the spectrum well with the power law index varying as $\log E$ (e.g. Massaro et al. 2004, 2008; Donato et al. 
2005; Tramacere et al. 2010) 
is not a constant but varies slowly with energy i.e. $\propto$ log E and hence the name log parabola (Massaro et al. 2004, 2008; Donato et al. 
2005; Tramacere et al. 2009). 
The model has often been invoked to fit the entire SED of blazars (e.g. Landau et al. 1986; Massaro et al. 2004; Chen et al. 2014) and such curved 
spectra of blazars are known to arise due to log parabolic electron distributions (e.g. Tramacere et al. 2007,2009; Paggi et al. 2009).

In the current study, we compile a sample of 14 LSP and ISP blazars observed with XMM$-$Newton spanning 31 observation epochs and study their spectra 
and timing information in the 0.6--10 keV energy range. Owing to a large effective area and small cadence, XMM$-$Newton offers good spectral resolution 
and timing information for our study. We fit each spectra with parametric models to identify possible transitions from low to high energy component. 
We then study the timing properties of our sample including a measure of variability and the power spectral density shape to study any possible evolving 
features. 
Our motivations include the disentangling of low and high energy contribution from synchrotron and inverse Compton components respectively, 
the comparison of the inferred break energy across sources, and a comparison of our inferences with previous studies (e.g. Wierzcholska 
\& Wagner 2016) which were similarly motivated through studies of mostly high peaked blazars and ISPs.

The paper is structured as follows: in Section 2, we give a brief description of the sample selection and data reduction method; 
in Section 3, we present the results of the analysis which are then discussed and interpreted in Section 4.

\begin{table*}
\caption{ Observation log of XMM-Newton X-ray data for low and intermediate energy peaked blazars.}

\begin{tabular}{lccclccr} \hline
Blazar Name    & $\alpha_{2000.0}$& $\delta_{2000.0}$             & redshift &Blazar & Date of Obs. & Obs. ID     & $N_{H}^{b}$(s) \\
               &                  &                               &   $z$    &Class & yyyy.mm.dd   &              & $10^{20}$  $cm^{-2}$   \\\hline

TXS 0106$+$612$^{**}$ & 01h09m46.3s       & $+$61$^{0}$33$^{'}$30$^{''}$ & 0.783    & ~LSP$^{a}$ & 2010.02.09   & 0652410201    &69.0 \\
3C 66A$^{*}$         & 02h22m39.6s      & $+$43$^{0}$02$^{'}$08$^{''}$  & 0.444    & ~ISP  & 2002.02.05   & 0002970201   &9.40   \\
PKS 0235$+$164$^{**}$ & 02h38m38.9s      & $+$16$^{0}$36$^{'}$59$^{''}$  & 0.94     & ~LSP & 2002.02.10   & 0110990101  & 10.30  \\
               &                  &                               &          & & 2004.01.18        & 0206740101    &  \\
               &                  &                               &          & & 2004.08.02        & 0206740501    &  \\
               &                  &                               &         &  & 2005.01.28        & 0206740701     &   \\
PKS 0426$-$380$^{*}$ & 04h28m40.4s& $-$37$^{0}$56$^{'}$20$^{''}$  & 1.11    &LSP$^{a}$  & 2012.02.11   & 0674330201   &2.26   \\
PKS 0521$-$365$^{**}$ & 05h22m58s        & $-$37$^{0}$27$^{'}$31$^{''}$  & 0.055    & LSP$^{a}$  & 2005.08.14   &0302580901   &15.70    \\
PKS 0537$-$441$^{**}$ & 05h38m50.3s      & $-$44$^{0}$05$^{'}$08.9$^{''}$&0.894    & ~LSP  &2010.02.27    & 0551503101    &3.54  \\  
               &                  &                               &         &       &2010.03.02    &0551503201     &   \\
               &                  &                               &         &       &2010.03.04    &0551503301  &  \\
S5 0716$+$714$^{*}$  & 07h21m53.4s      & $+$71$^{0}$20$^{'}$36$^{''}$  & 0.31    & ~ISP  & 2007.09.24   & 0502271401   & 3.60  \\
OJ 287$^{*}$   & 08h54m48.9s      & $+$20$^{0}$06$^{'}$31$^{''}$  & 0.3056  & ~ISP  & 2005.04.12   & 0300480201   & 2.71   \\
               &                  &                               &         &  & 2005.11.03        & 0300480301     &   \\
               &                  &                               &         &  & 2006.11.17        & 0401060201     &   \\
               &                  &                               &         &  & 2008.04.22        & 0502630201     &   \\
               &                  &                               &         &  & 2011.10.15        & 0679380701     &   \\
S4 0954$+$65$^{*}$   & 09h58m47.0s      &$+$65$^{0}$33$^{'}$55$^{''}$   &0.368    &~LSP$^{a}$  & 2007.09.30    & 0502430201     &6.30  \\
               &                  &                               &         &     &2007.10.27      &0502430701   &  \\
ON 231$^{*}$   & 12h21m31.7s      &$+$28$^{0}$13$^{'}$59$^{''}$  & 0.102   & ~ISP  &2002.06.26   &0104860501   &2.18  \\   
               &                  &                              &         &       &2008.06.14   &0502211301    &      \\            
3C 279$^{*}$   & 12h56m11.1s      &$-$05$^{0}$47$^{'}$22$^{''}$  & 0.5362  & ~LSP$^{a}$  & 2009.01.21     & 0556030101   &2.22   \\
               &                  &                               &         &  & 2011.01.18        & 0651610101     &  \\
PKS 1334$-$127 & 13h37m39.8s      &$-$12$^{0}$57$^{'}$25$^{''}$  &  0.539  & ~LSP$^{a}$  &2003.01.31    &0147670201       &6.53 \\
BL Lac$^{*}$   & 22h02m43.3s      &$+$42$^{0}$16$^{'}$40$^{''}$  & 0.0686  & ~ISP  & 2007.07.10    & 0501660201    & 30.30  \\
               &                  &                               &         &  & 2007.12.05        & 0501660301     &  \\
               &                  &                               &         &  & 2008.01.08        & 0501660401     &  \\
3C 454.3$^{**}$ & 22h53m57.7s      &$+$16$^{0}$08$^{'}$54$^{''}$  & 0.859   & ~LSP$^{a}$  & 2006.07.02     & 0401700201   &6.86  \\
               &                  &                               &         &  & 2007.05.23        & 0401700401     &   \\
               &                  &                               &         &  & 2006.12.18        & 0401700501     &  \\
               &                  &                               &         &  & 2007.05.31        & 0401700601     &  \\  \hline
\end{tabular}     \\
$^{a}$ LSPs are FSRQs \\
$^{b}$ $N_{H}$: Galactic absorption and values are taken from Willingale et al. (2013) \\
$^{*}$: http://tevcat.uchicago.edu/ \\
$^{**}$: Tsujimoto et al. (2015) \\
\end{table*}

\begin{figure*}
\centering
\includegraphics[width=0.48\textwidth]{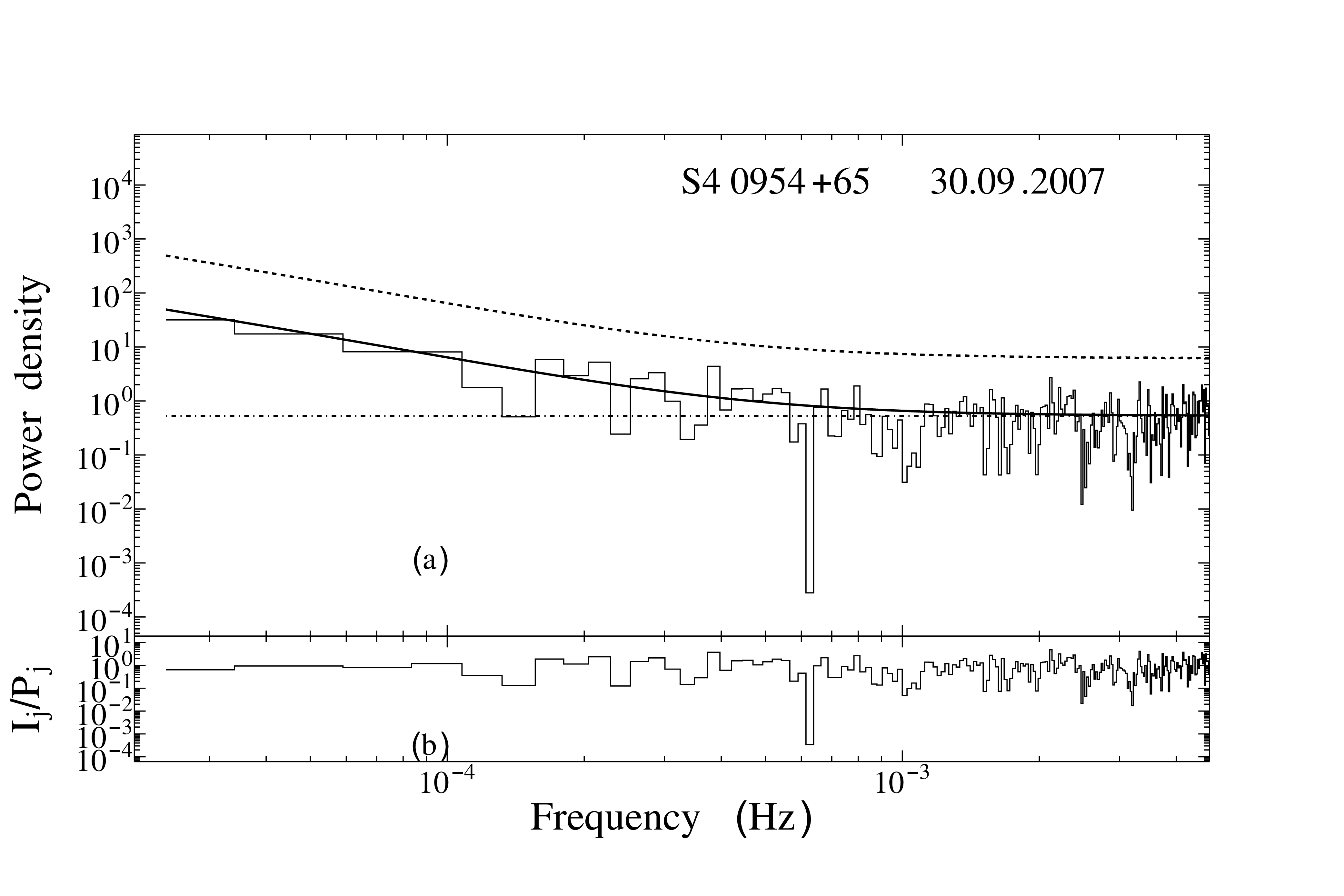}
\includegraphics[width=0.48\textwidth]{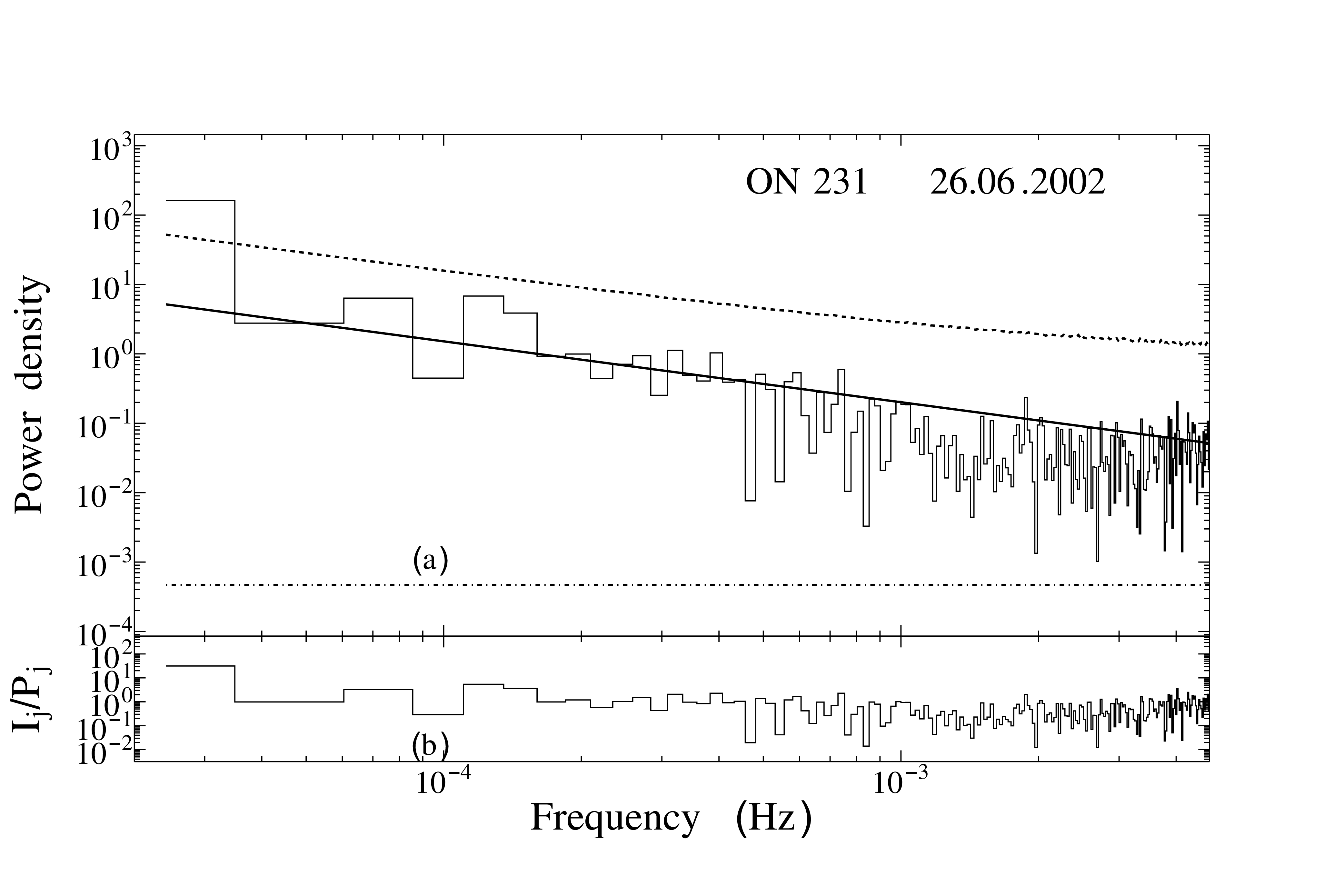}
\includegraphics[width=0.48\textwidth]{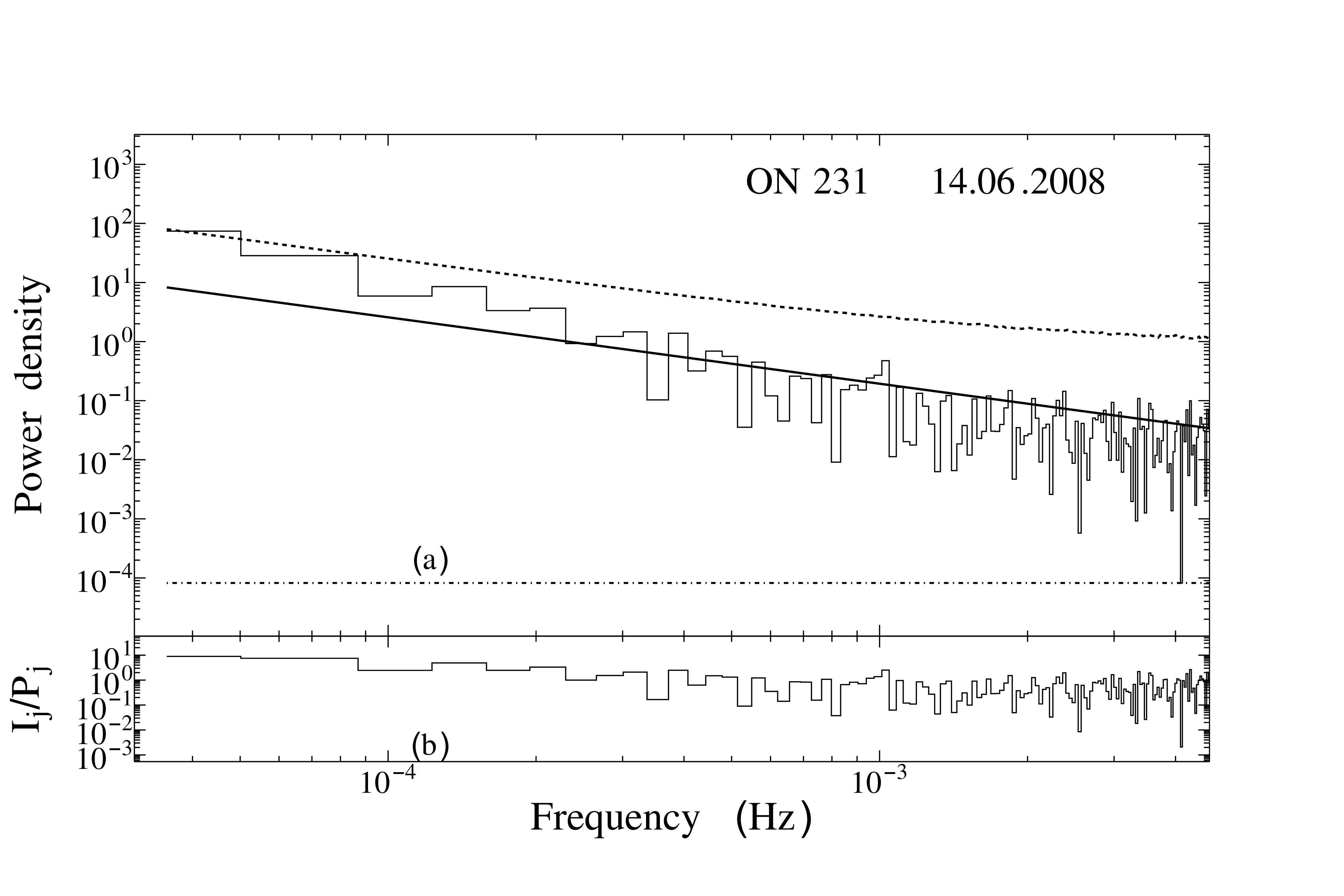}
\includegraphics[width=0.48\textwidth]{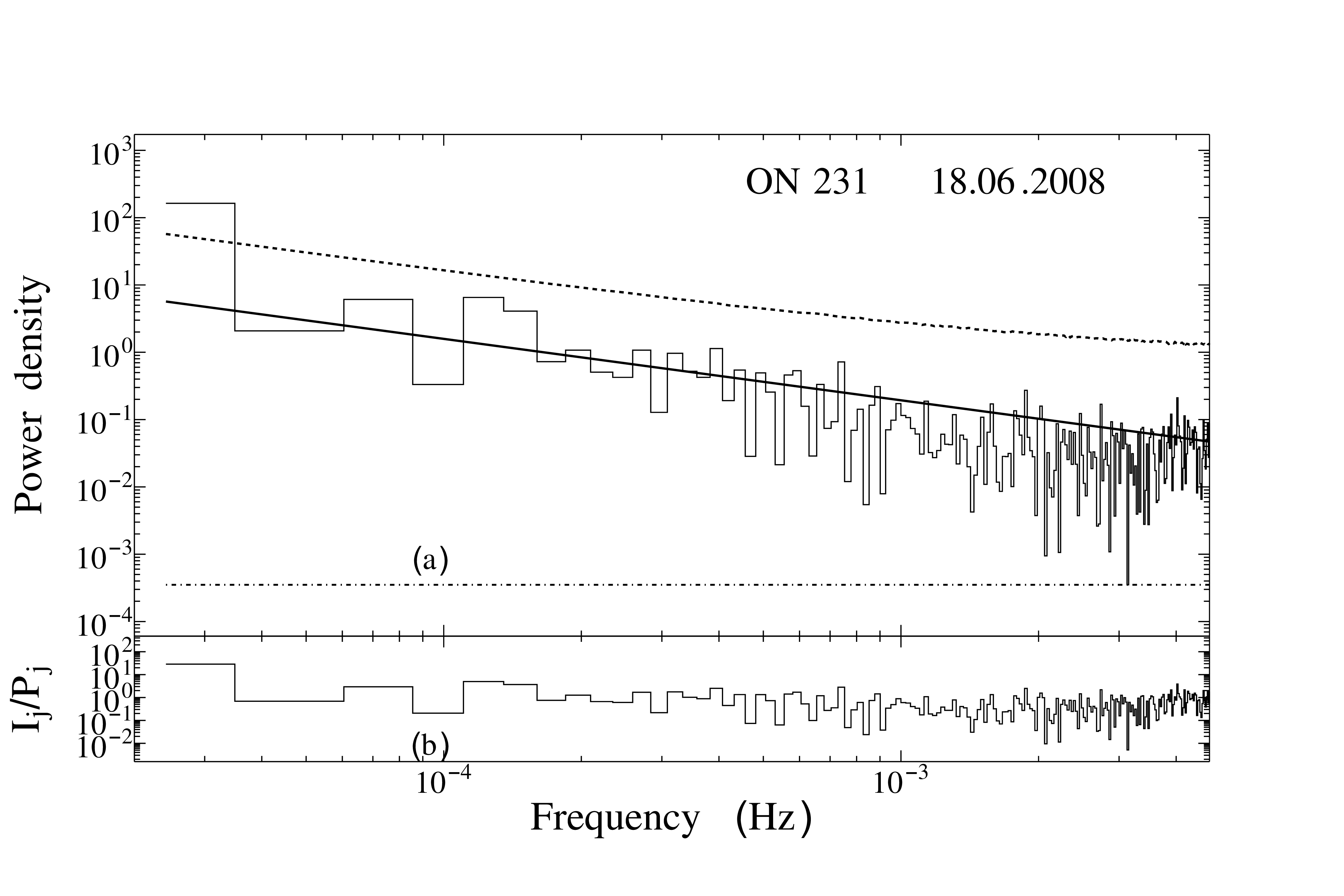}
\includegraphics[width=0.48\textwidth]{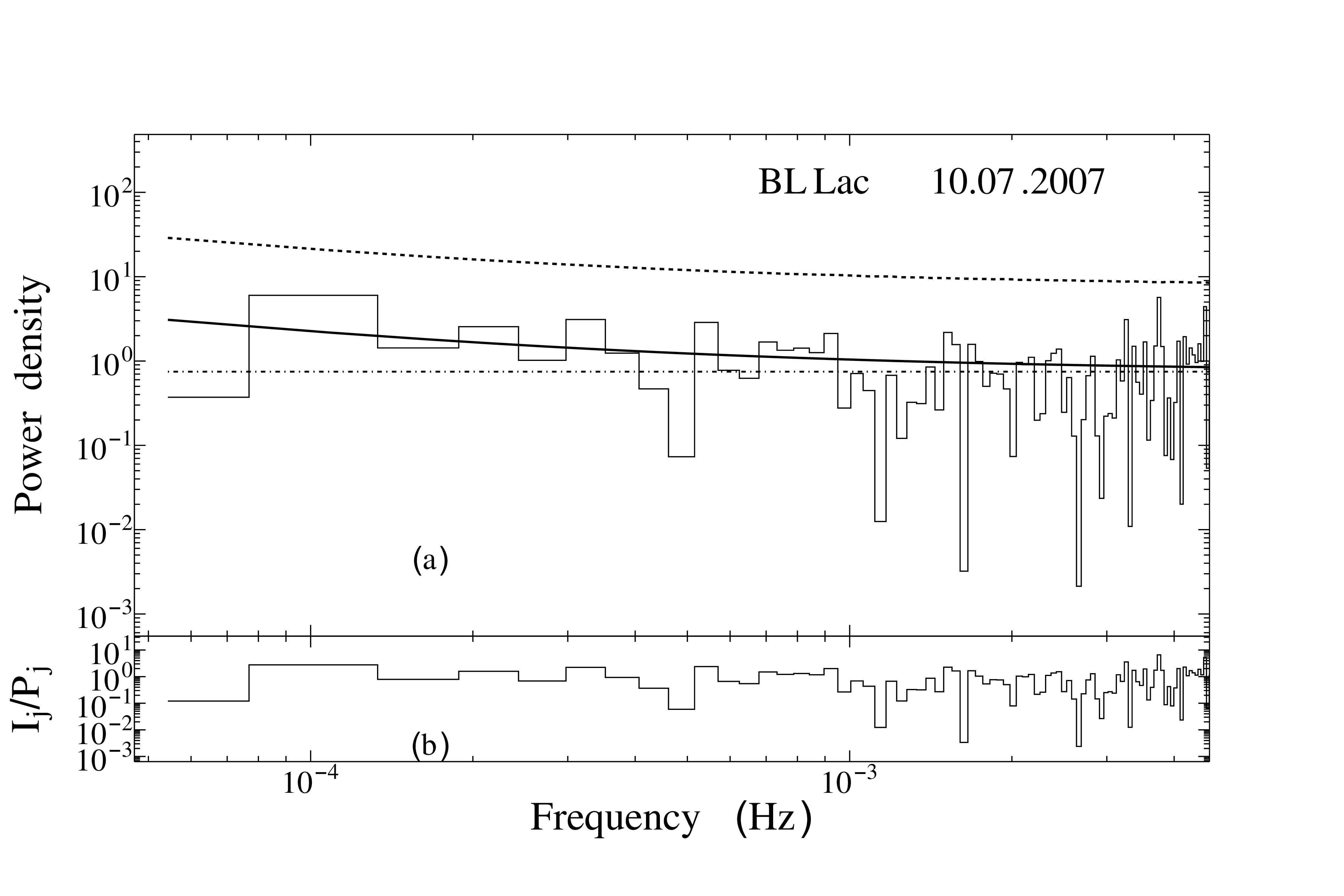}
\includegraphics[width=0.48\textwidth]{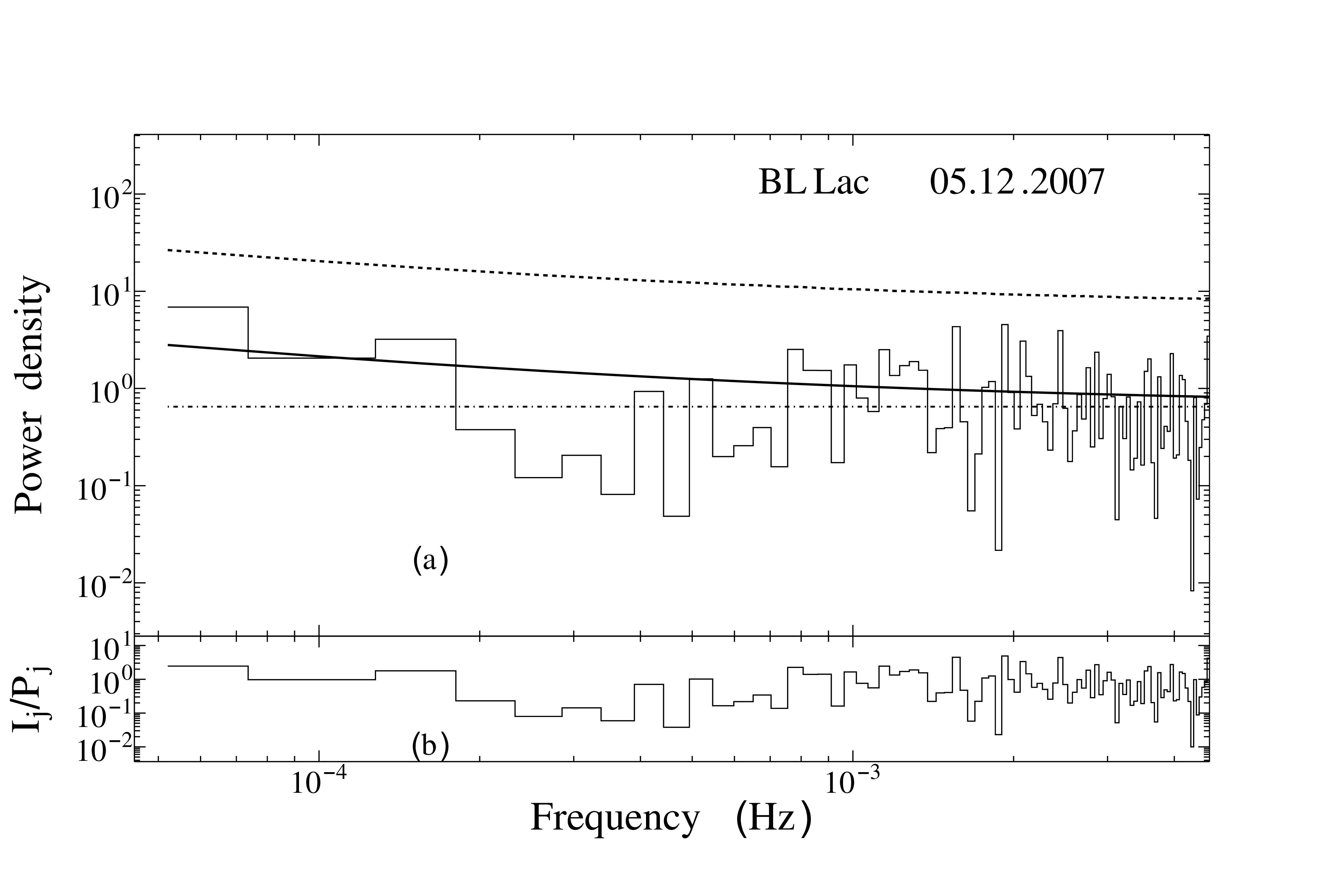}
\caption{Lomb-Scargle periodograms of the low peaked blazars. The best fit model is the solid curve, the dashed curve above it is the 99 \% significance 
contour (accounting for model errors) which can identify statistically significant QPOs, the dot-dashed horizontal line is the white noise level and 
the plot below each periodogram panel shows the fit residuals.}
\label{LSPp1}
\end{figure*}

\begin{figure*}
\centering
\includegraphics[width=0.44\textwidth]{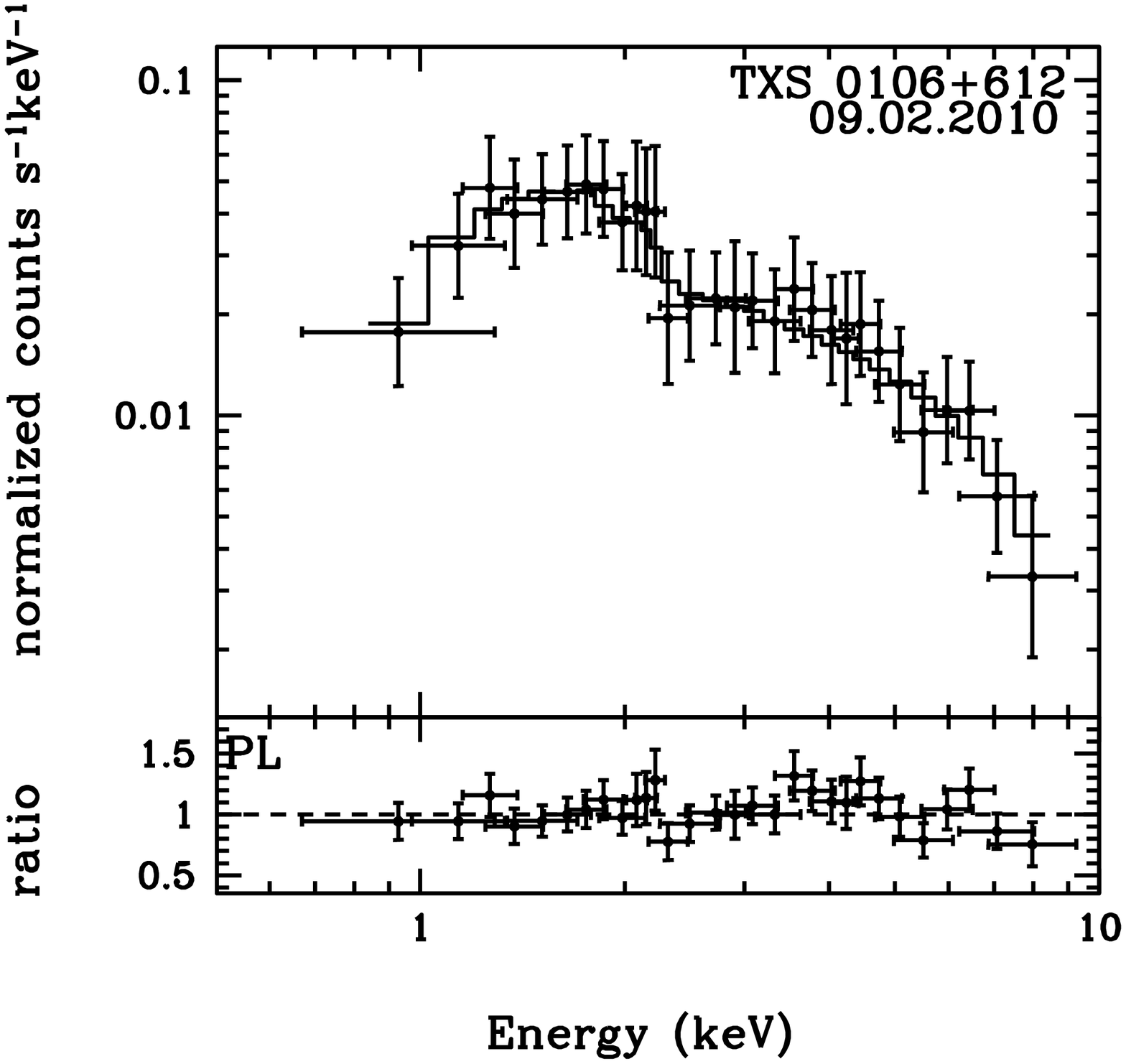}
\includegraphics[width=0.44\textwidth]{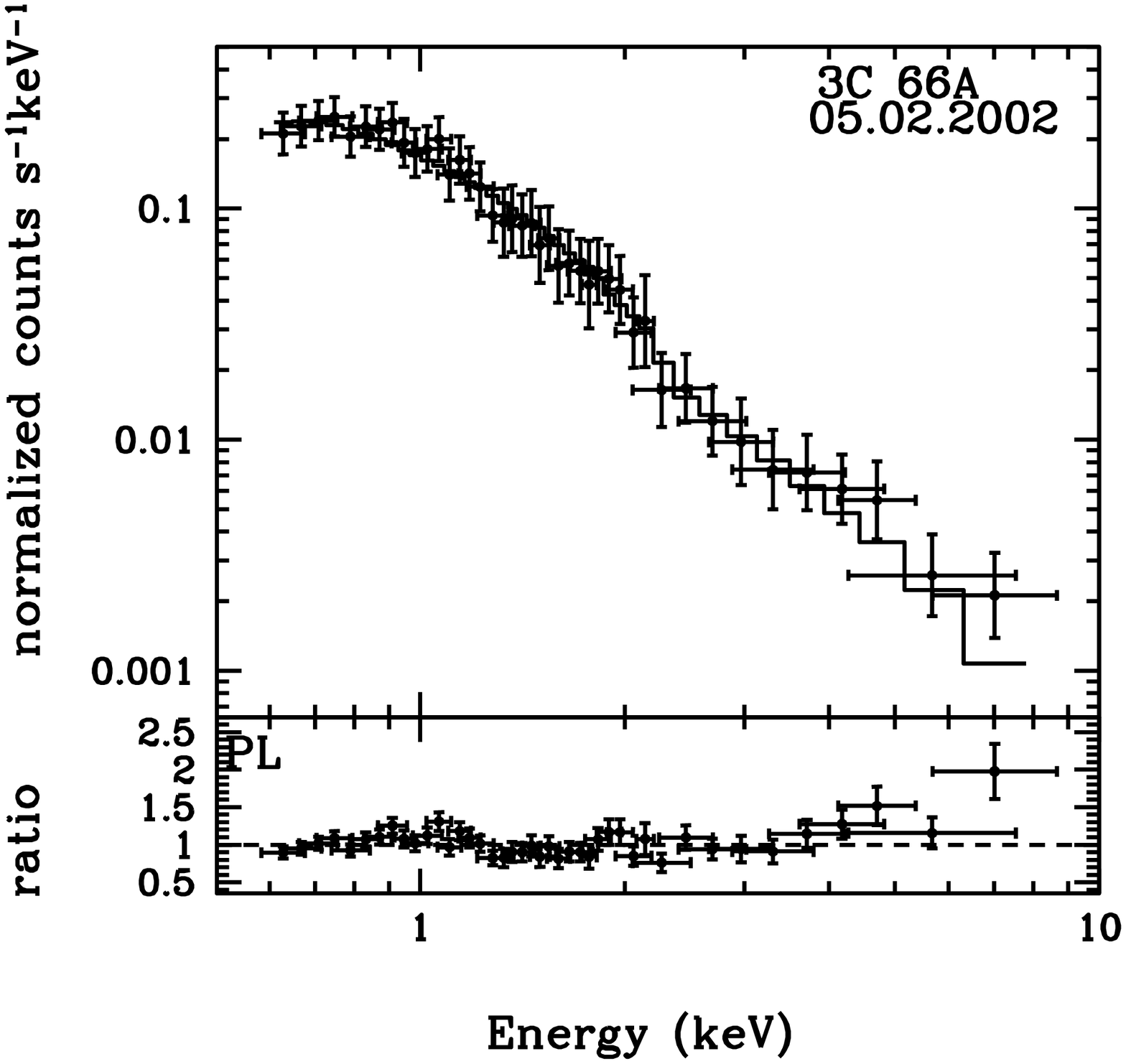}
\includegraphics[width=0.44\textwidth]{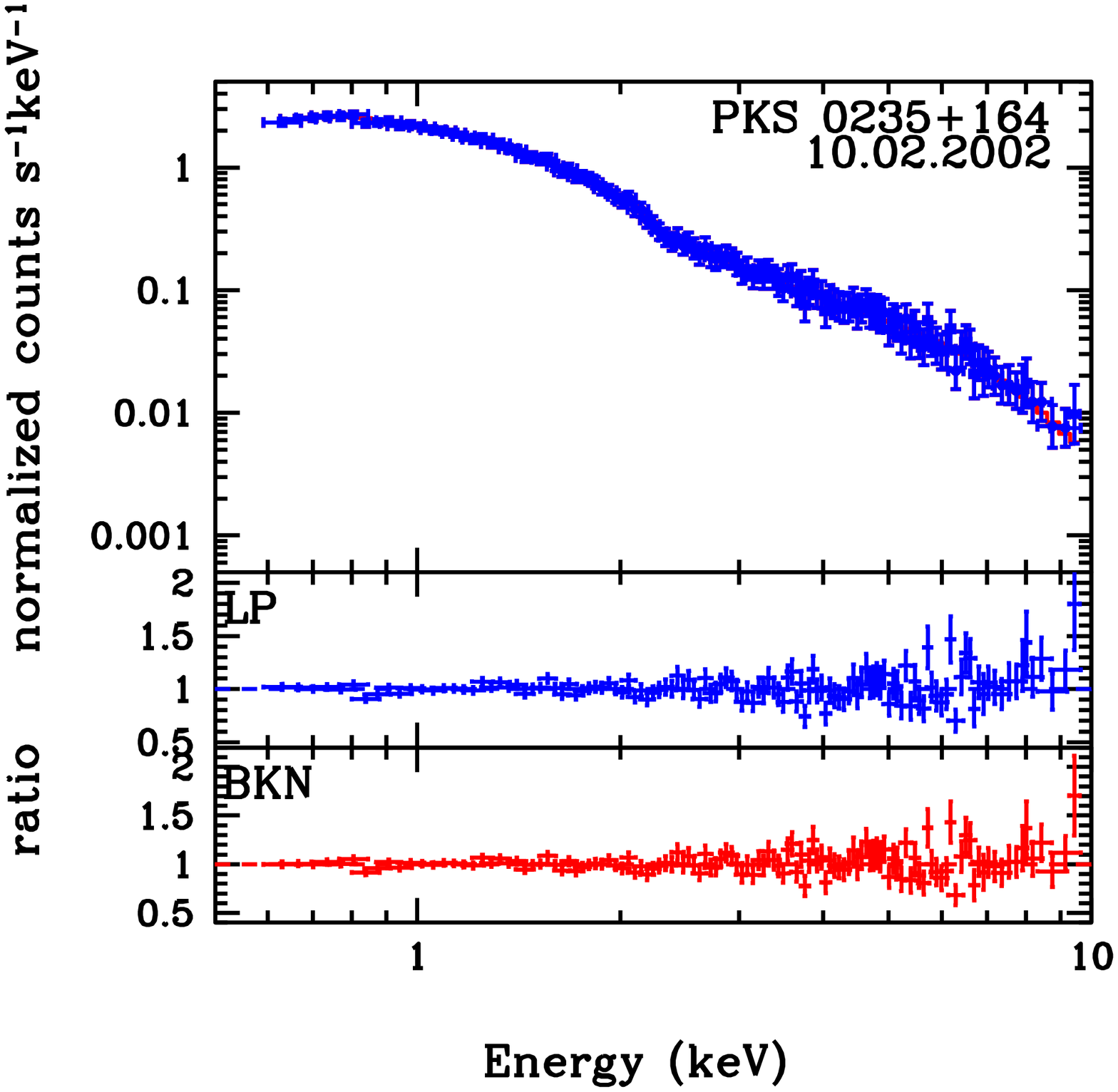}
\includegraphics[width=0.44\textwidth]{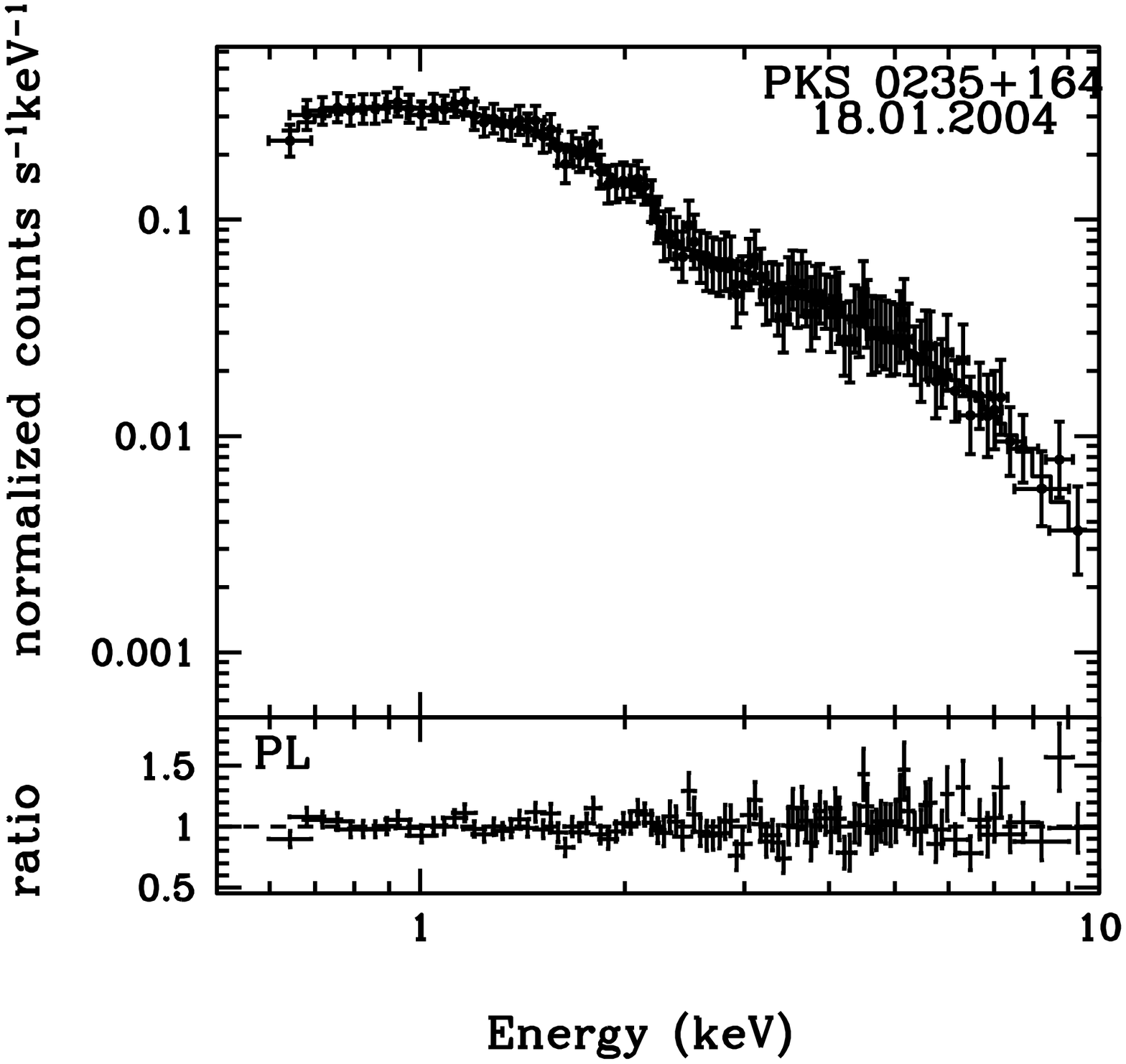}
\caption{X-ray spectra fitted with their respective best models are shown in each figure.  Power law, log parabolic model and broken power law
with their respective ratios are presented in each figure with black, blue and red colors, respectively. The name and date of observation 
of each object is also provided in upper right corner of each figure. }
\end{figure*}

\begin{figure*}
\centering
\includegraphics[width=0.44\textwidth]{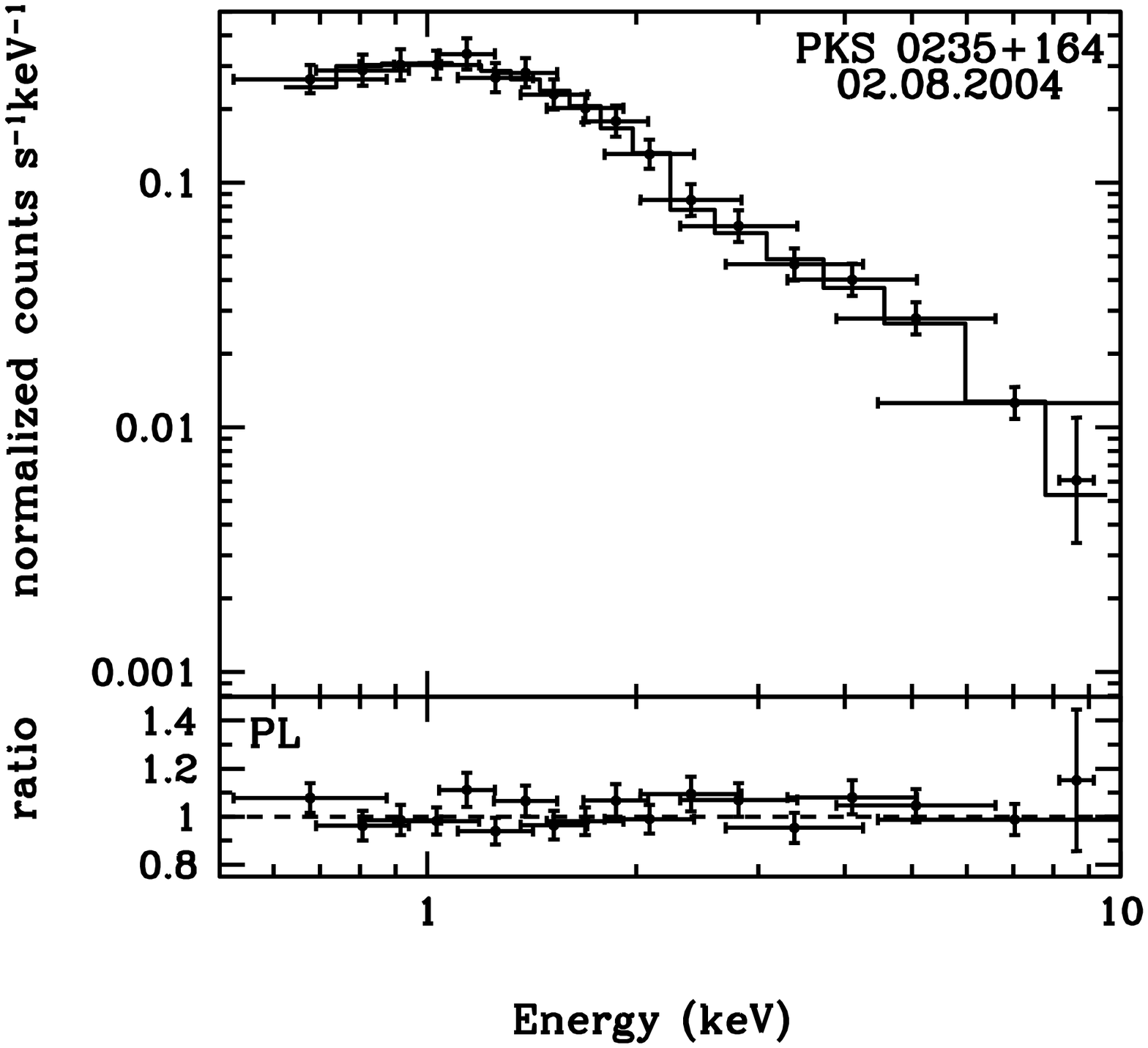}
\includegraphics[width=0.44\textwidth]{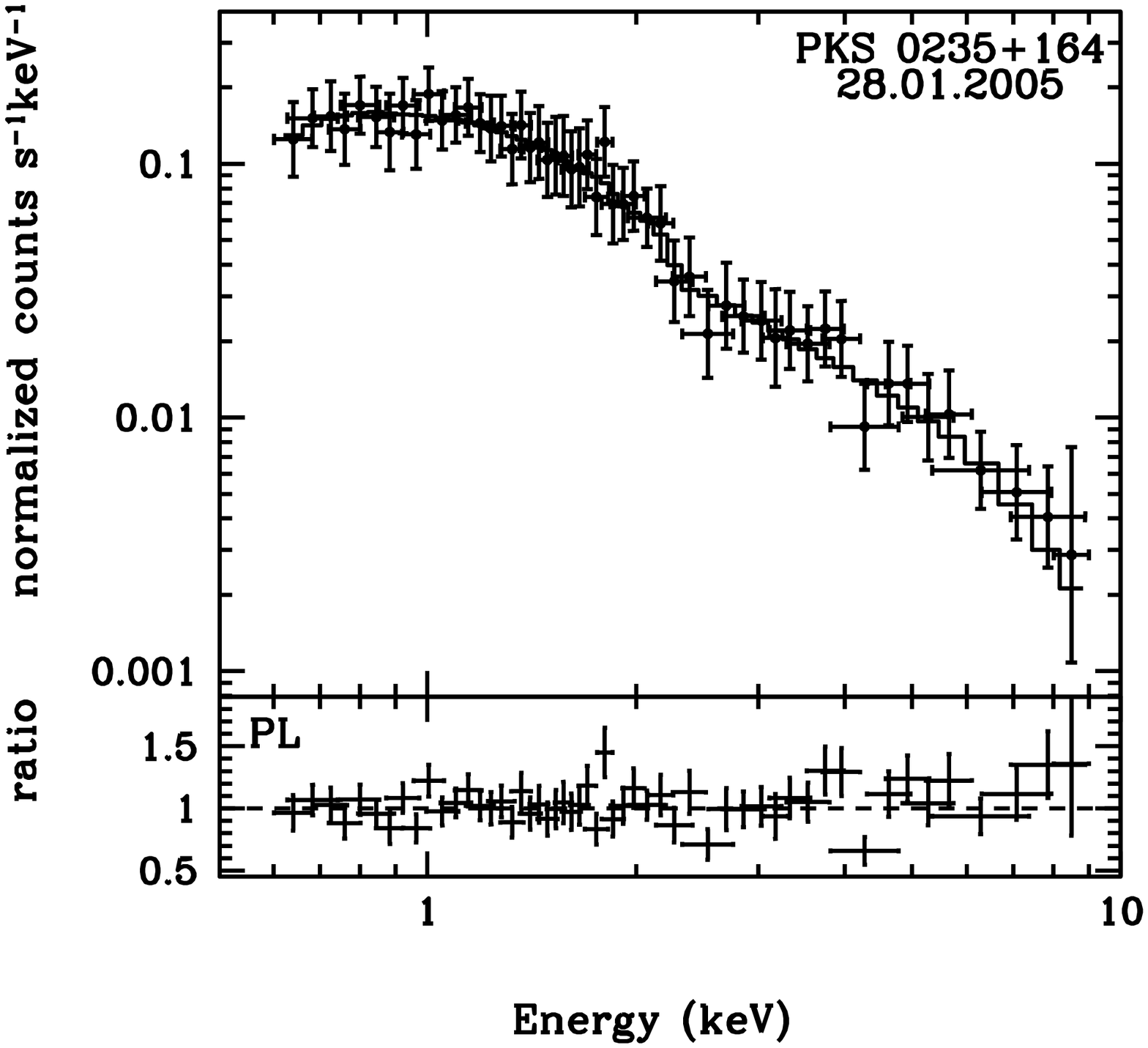}
\includegraphics[width=0.44\textwidth]{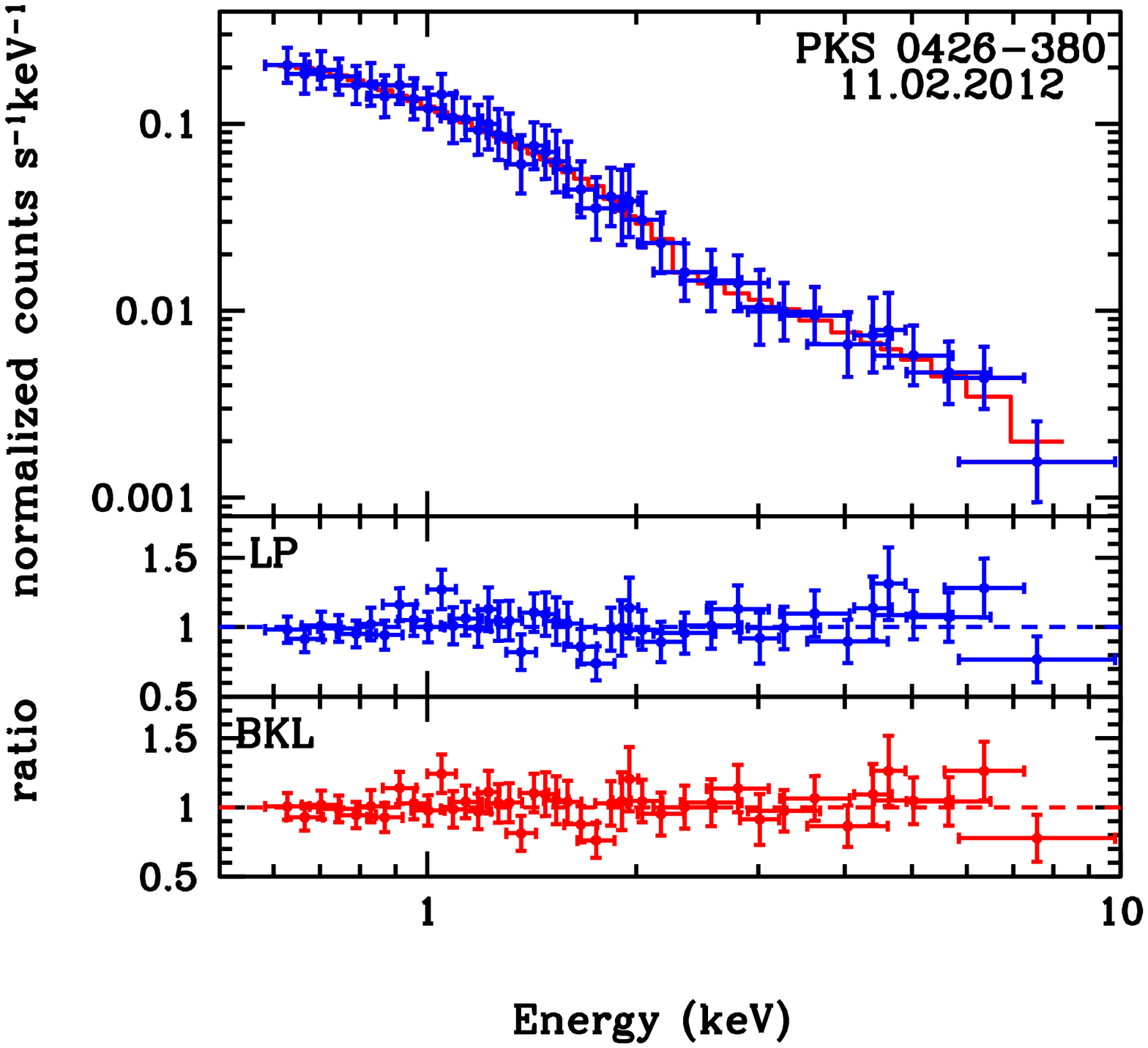}
\includegraphics[width=0.44\textwidth]{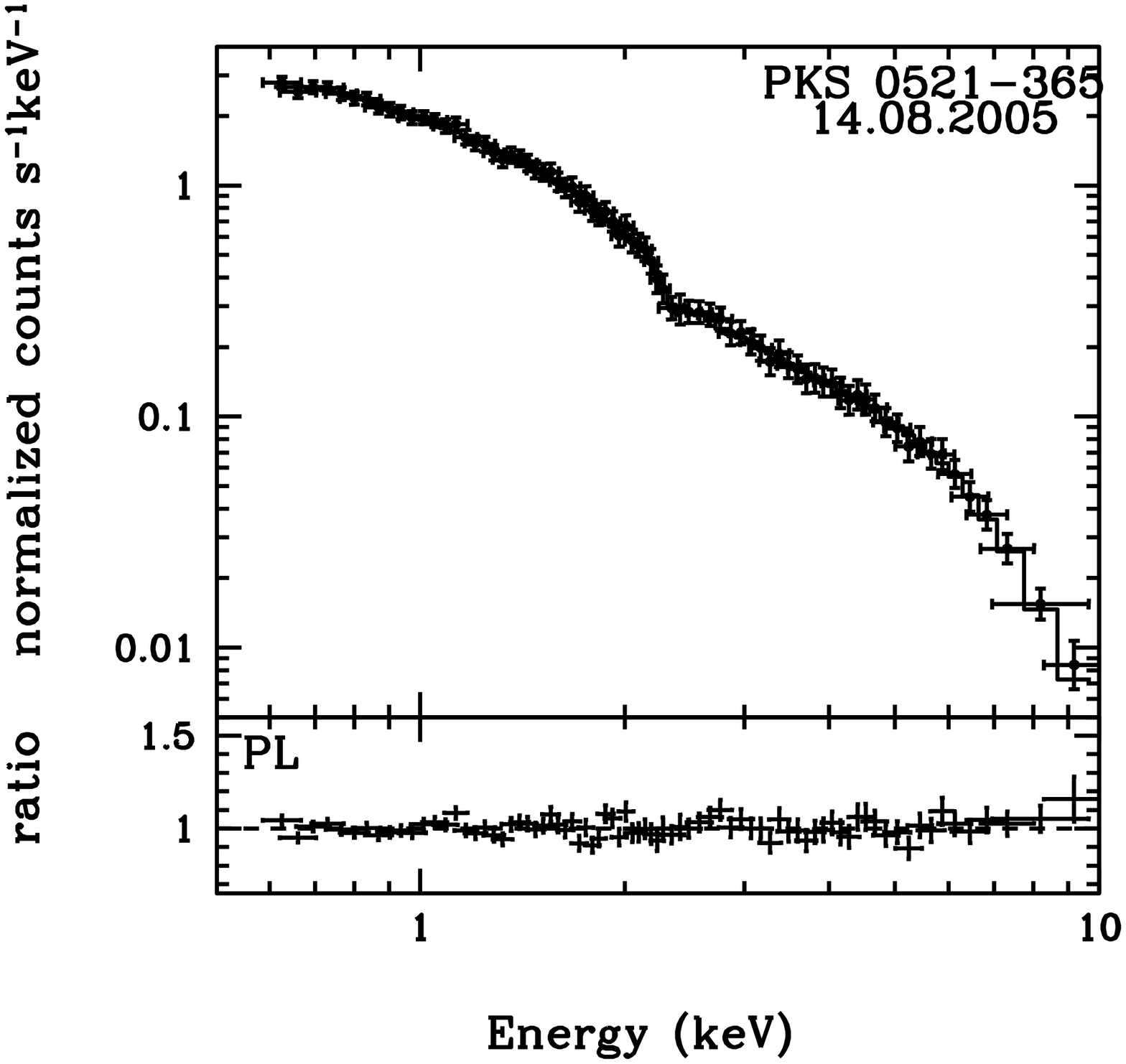}
\caption{Same as in figure 2. }
\end{figure*}

\begin{figure*}
\centering
\includegraphics[width=0.44\textwidth]{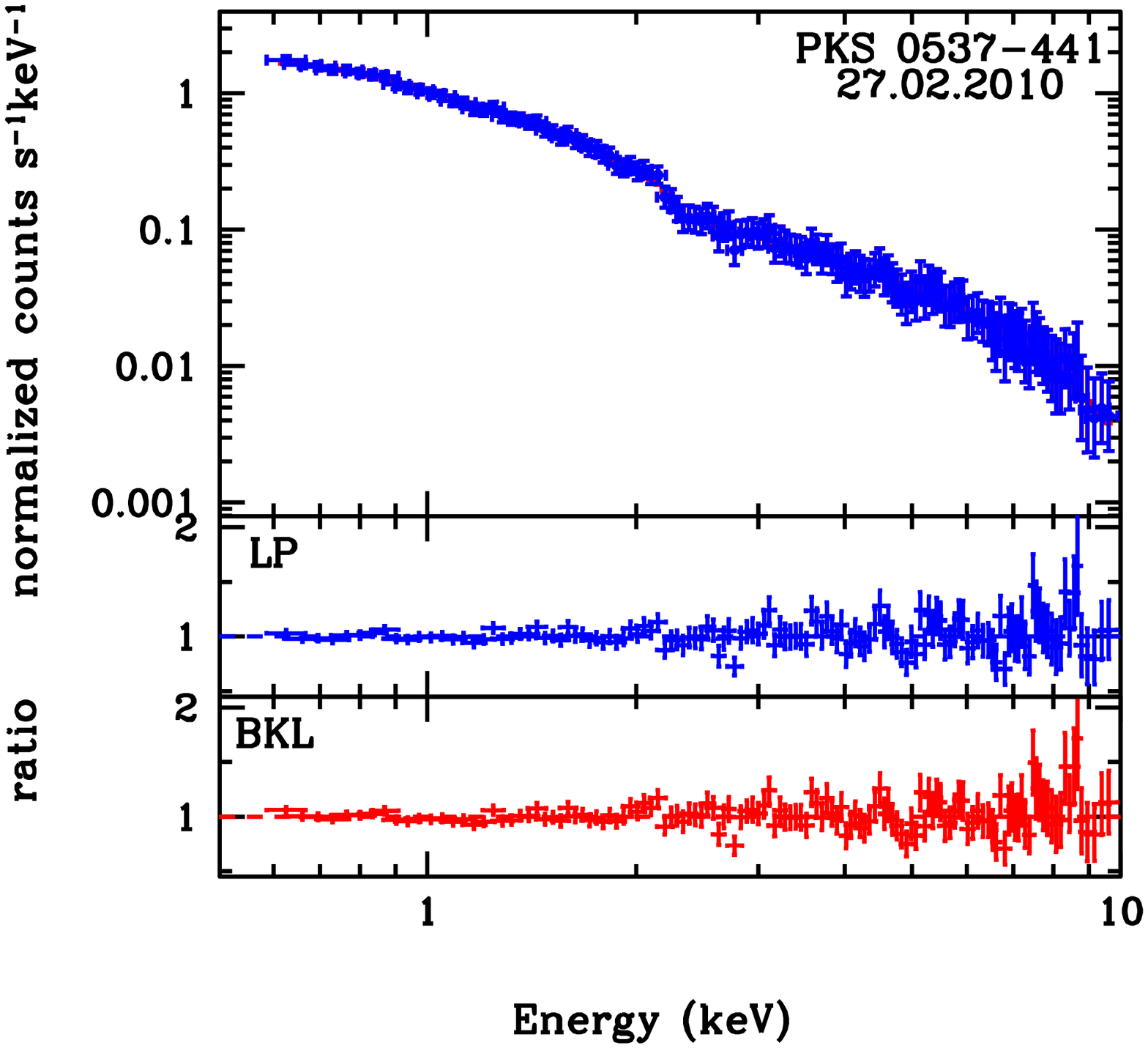}
\includegraphics[width=0.44\textwidth]{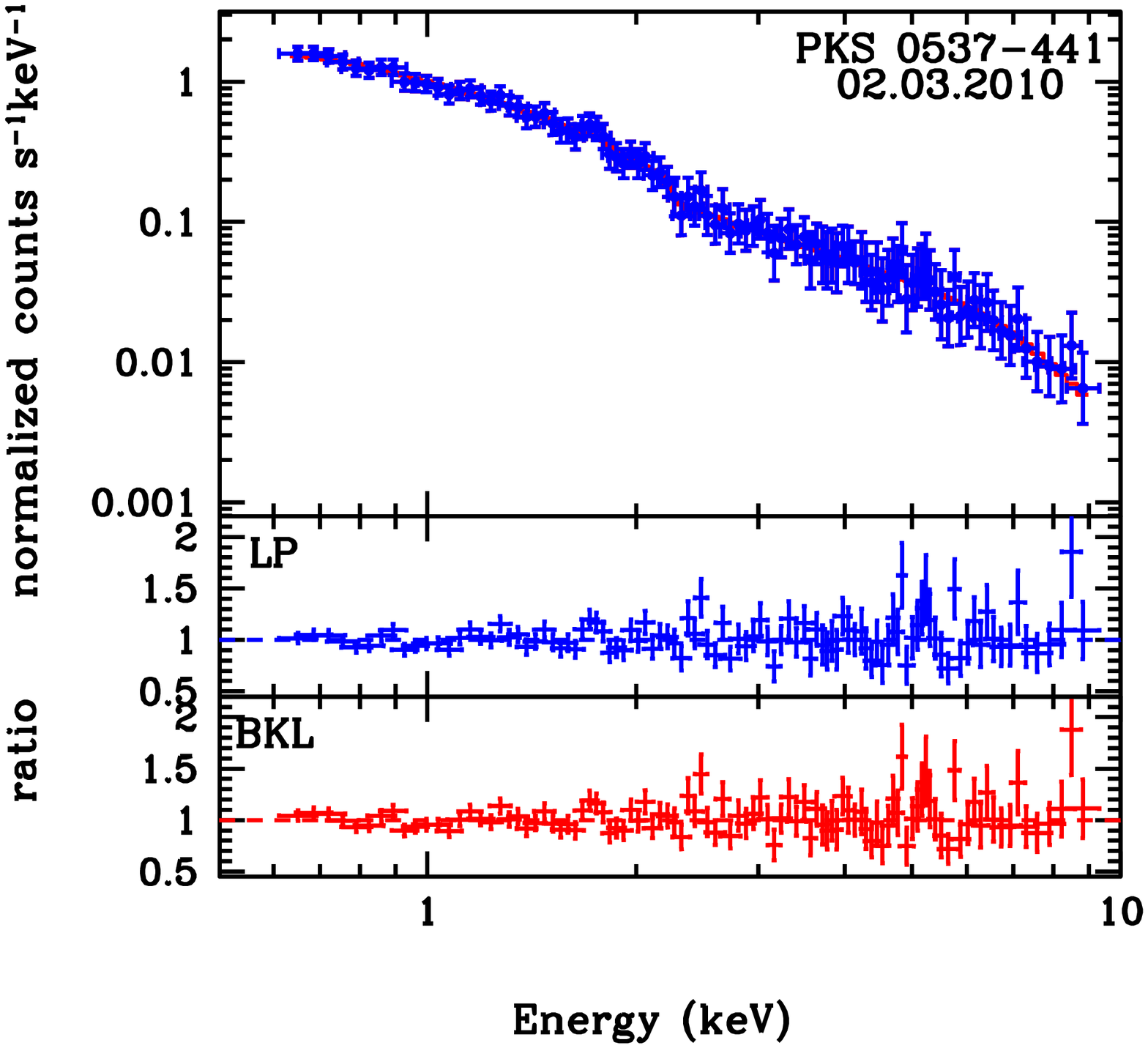}
\includegraphics[width=0.44\textwidth]{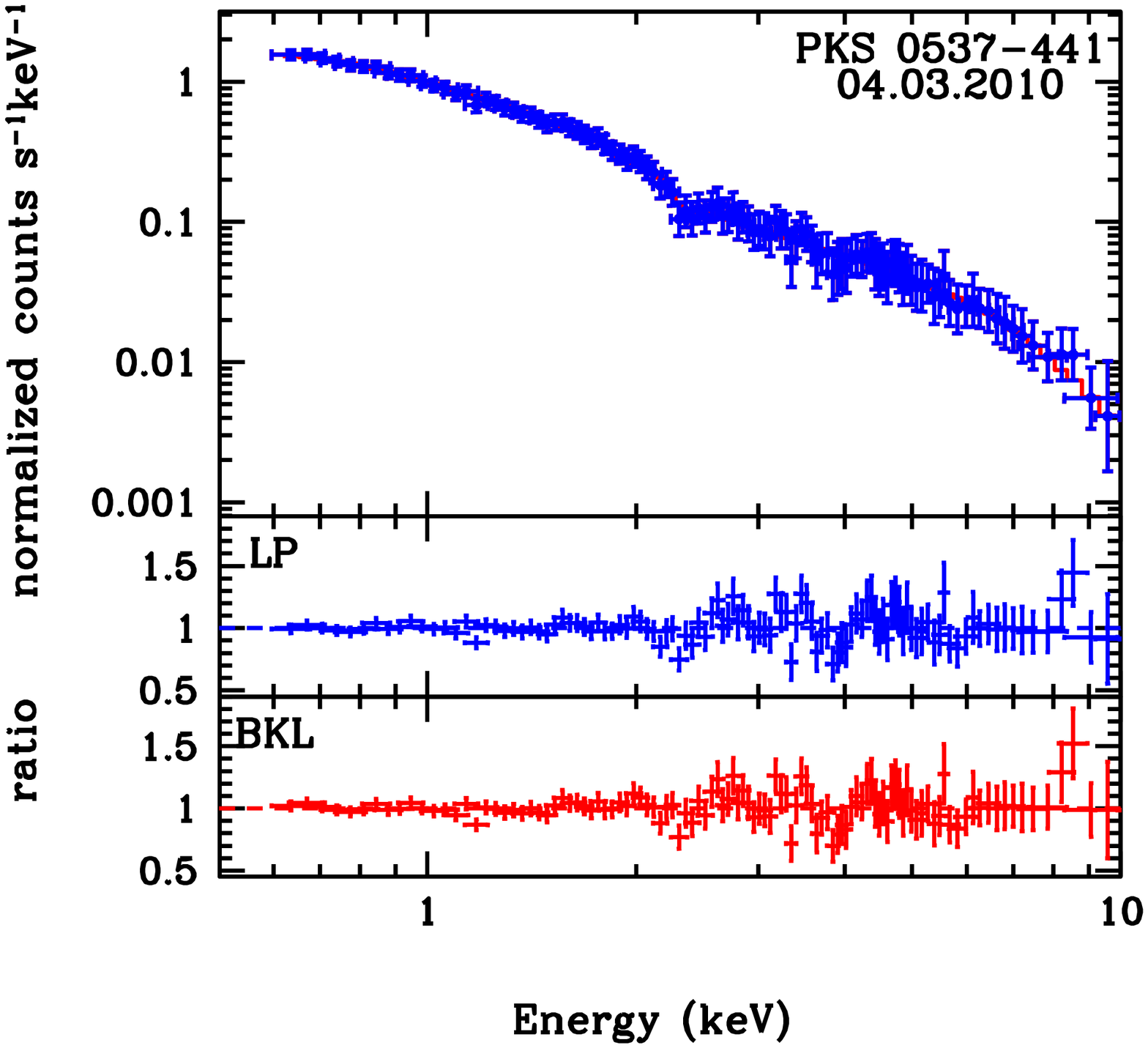}
\includegraphics[width=0.44\textwidth]{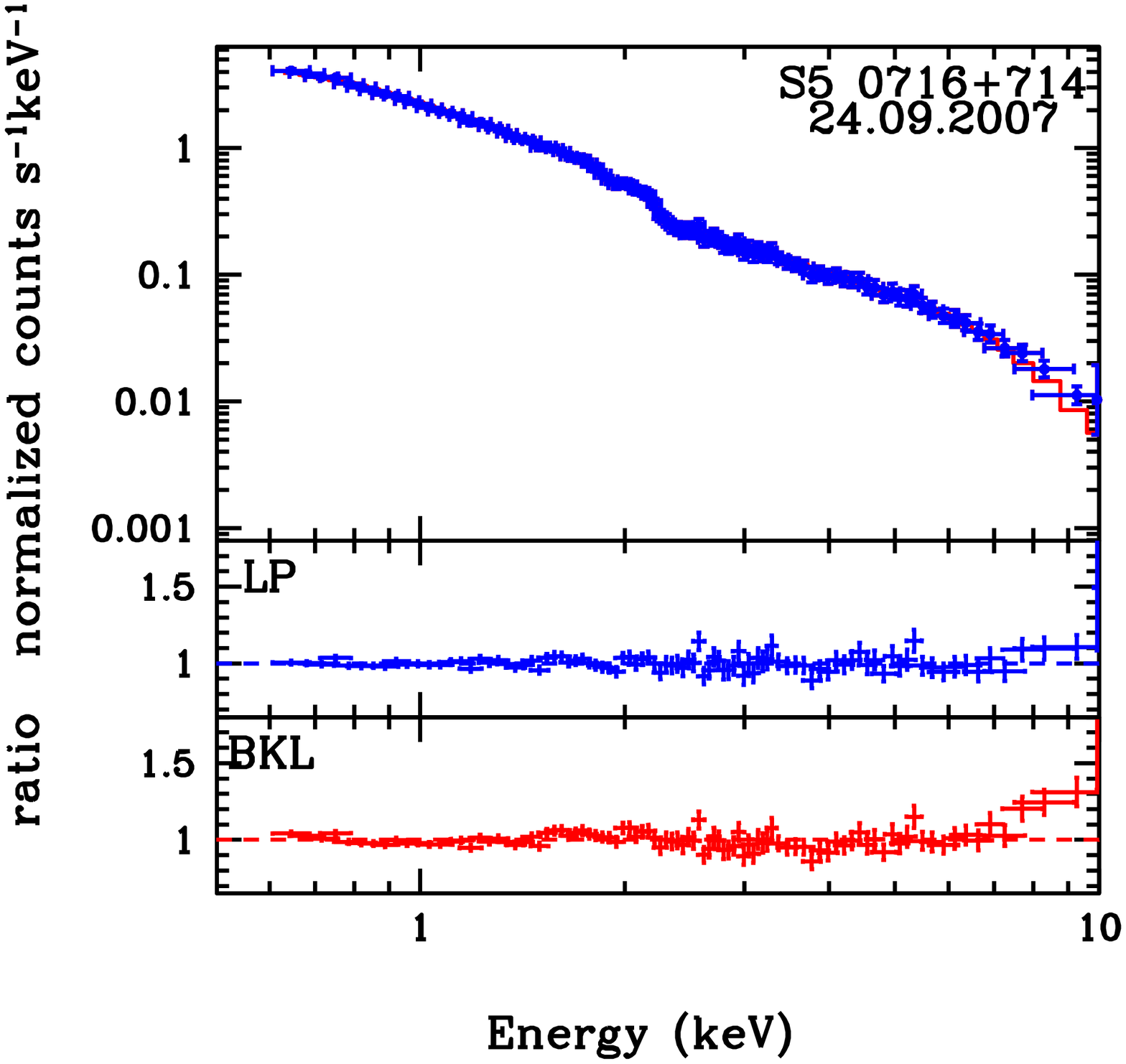}

\caption{Same as in figure 2. }
\end{figure*}

\begin{figure*}
\centering
\includegraphics[width=0.44\textwidth]{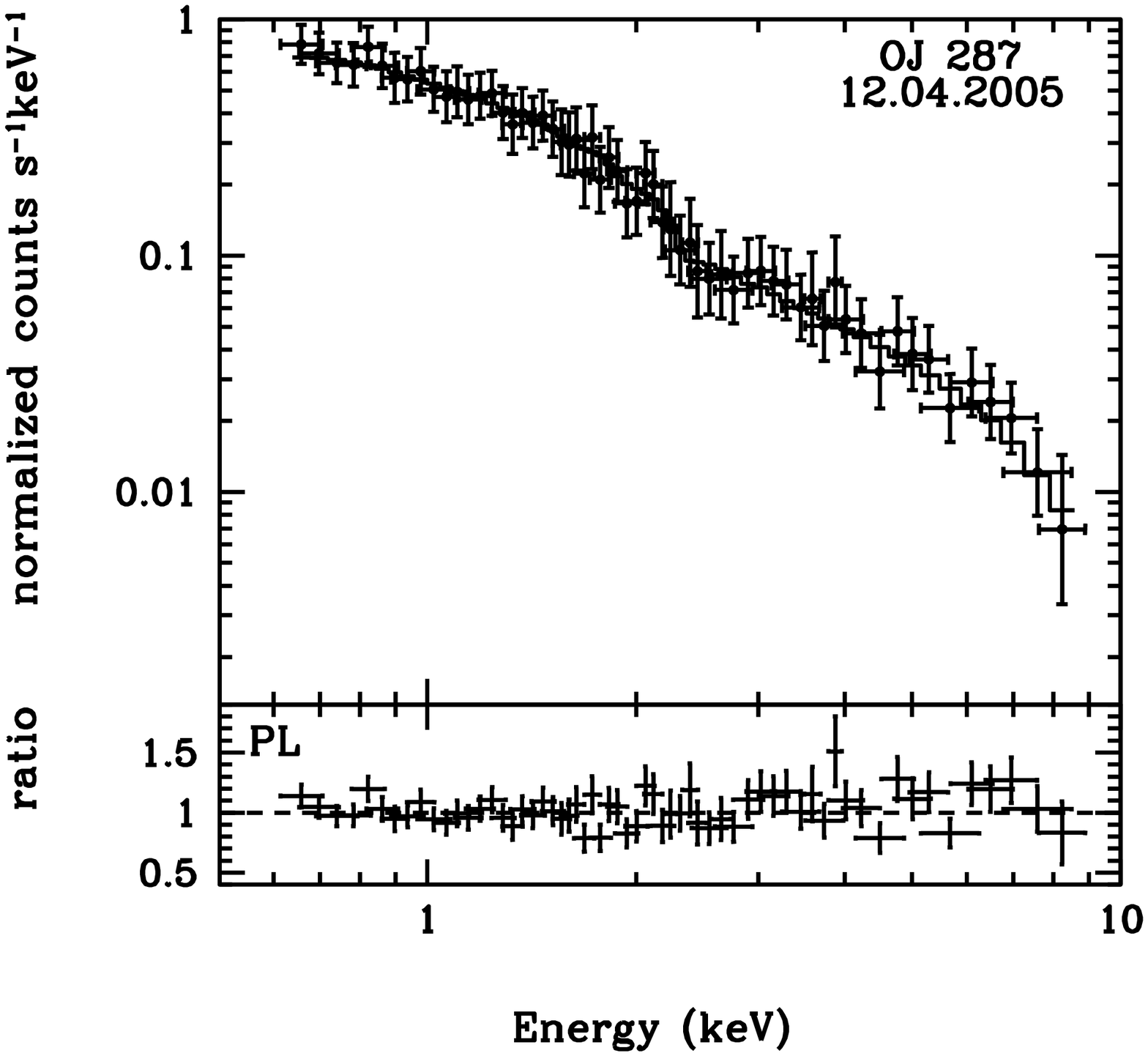}
\includegraphics[width=0.44\textwidth]{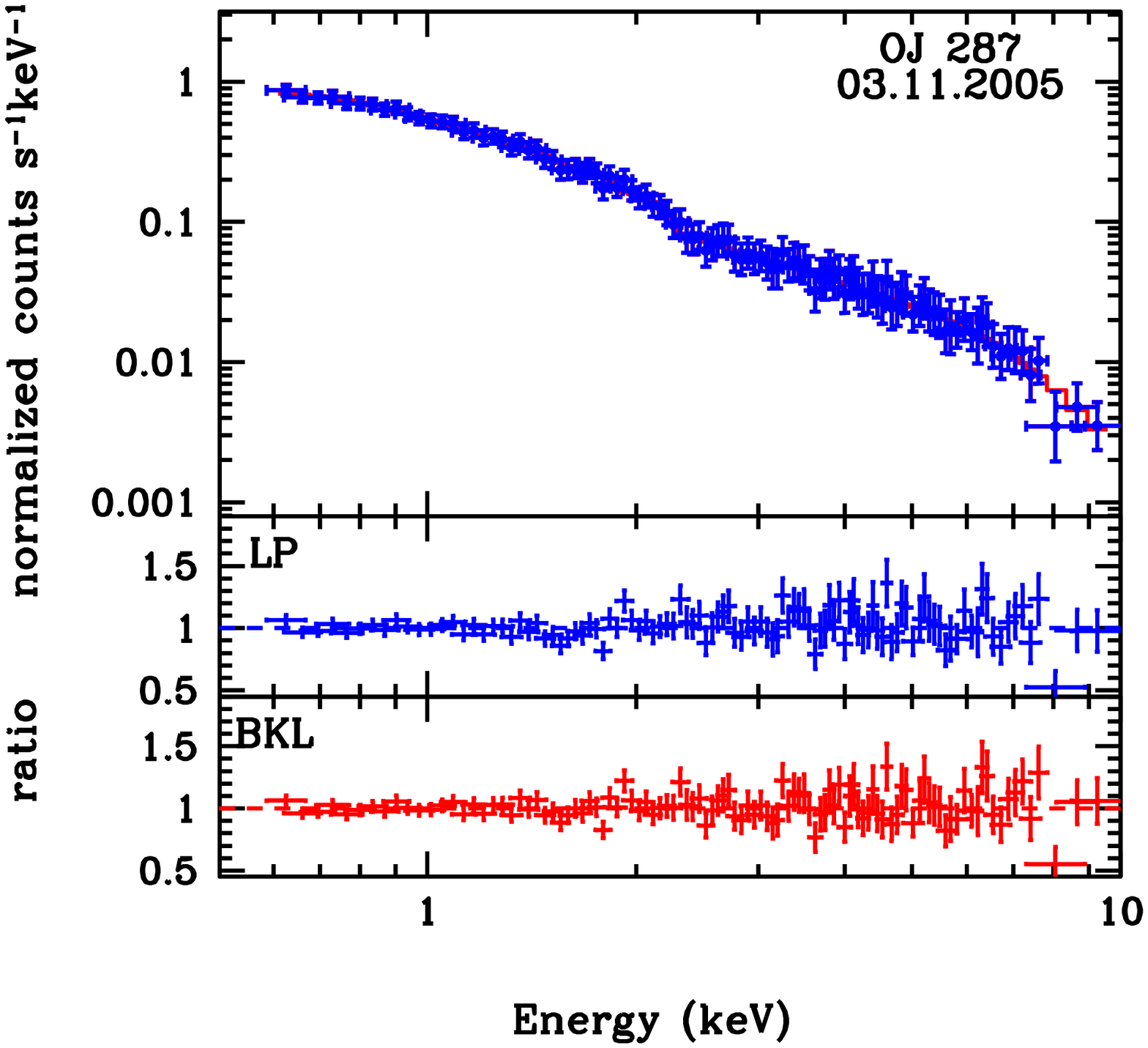}
\includegraphics[width=0.44\textwidth]{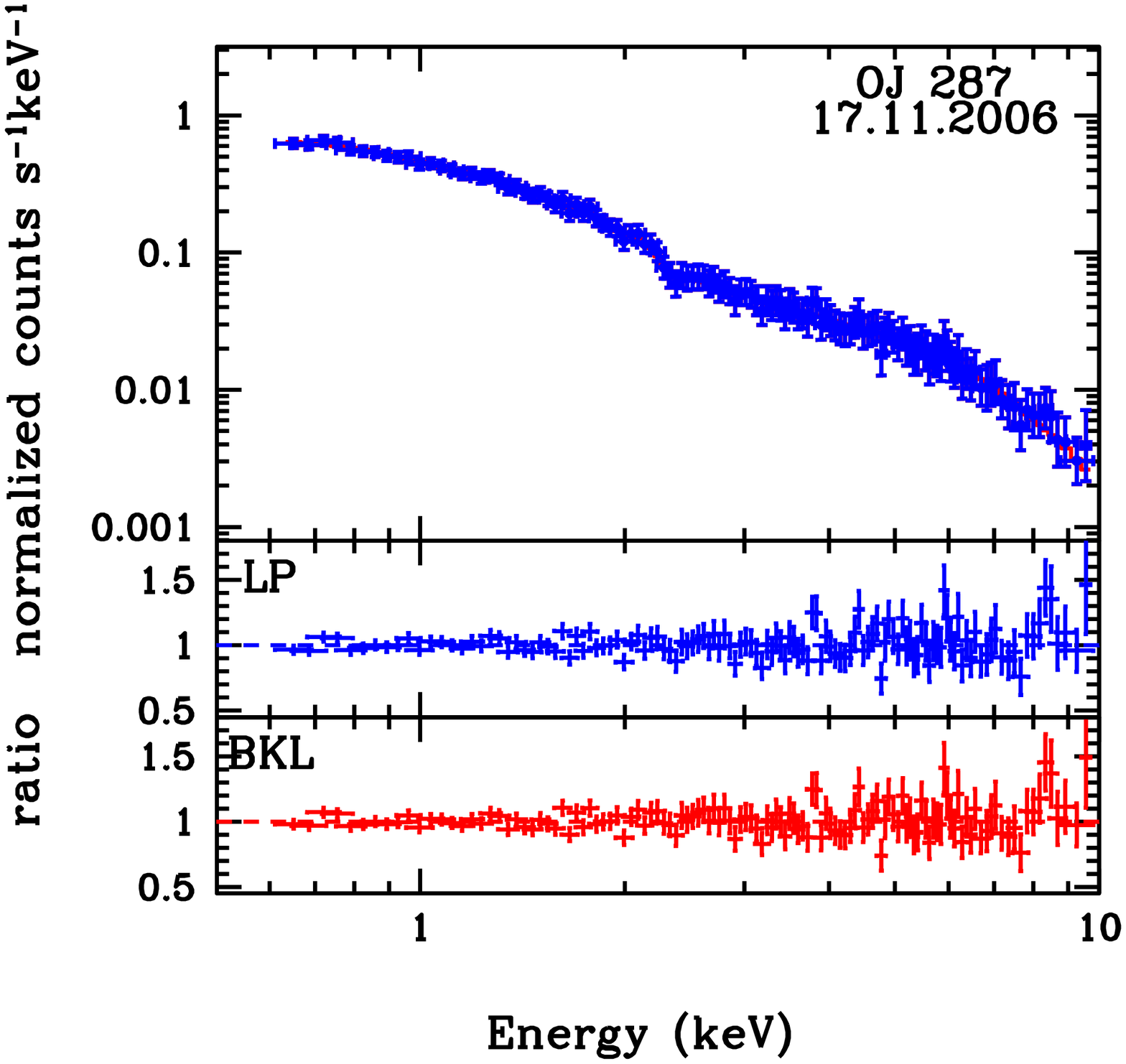}
\includegraphics[width=0.44\textwidth]{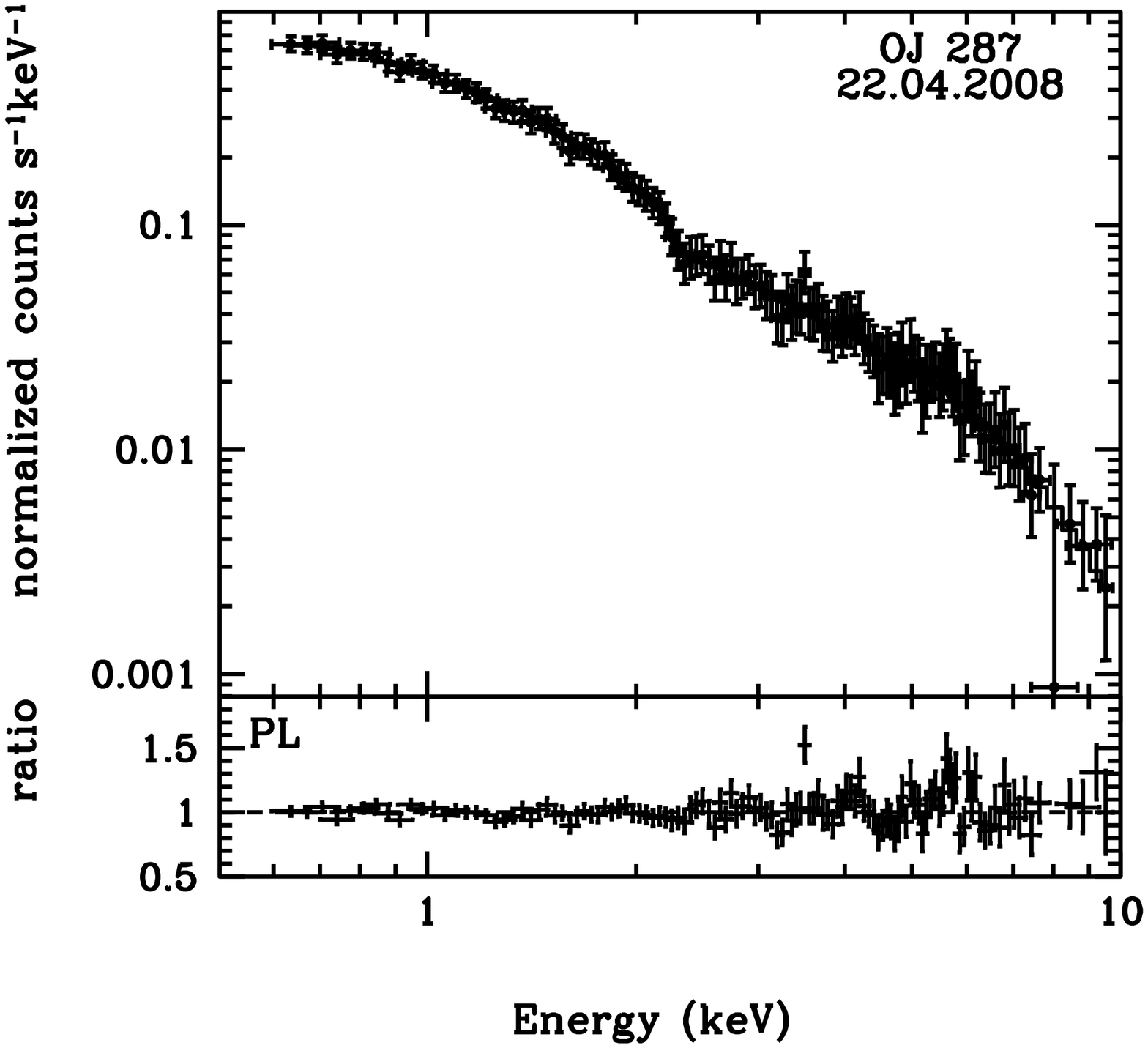}
\caption{Same as in figure 2. }
\end{figure*}

\begin{figure*}
\centering
\includegraphics[width=0.44\textwidth]{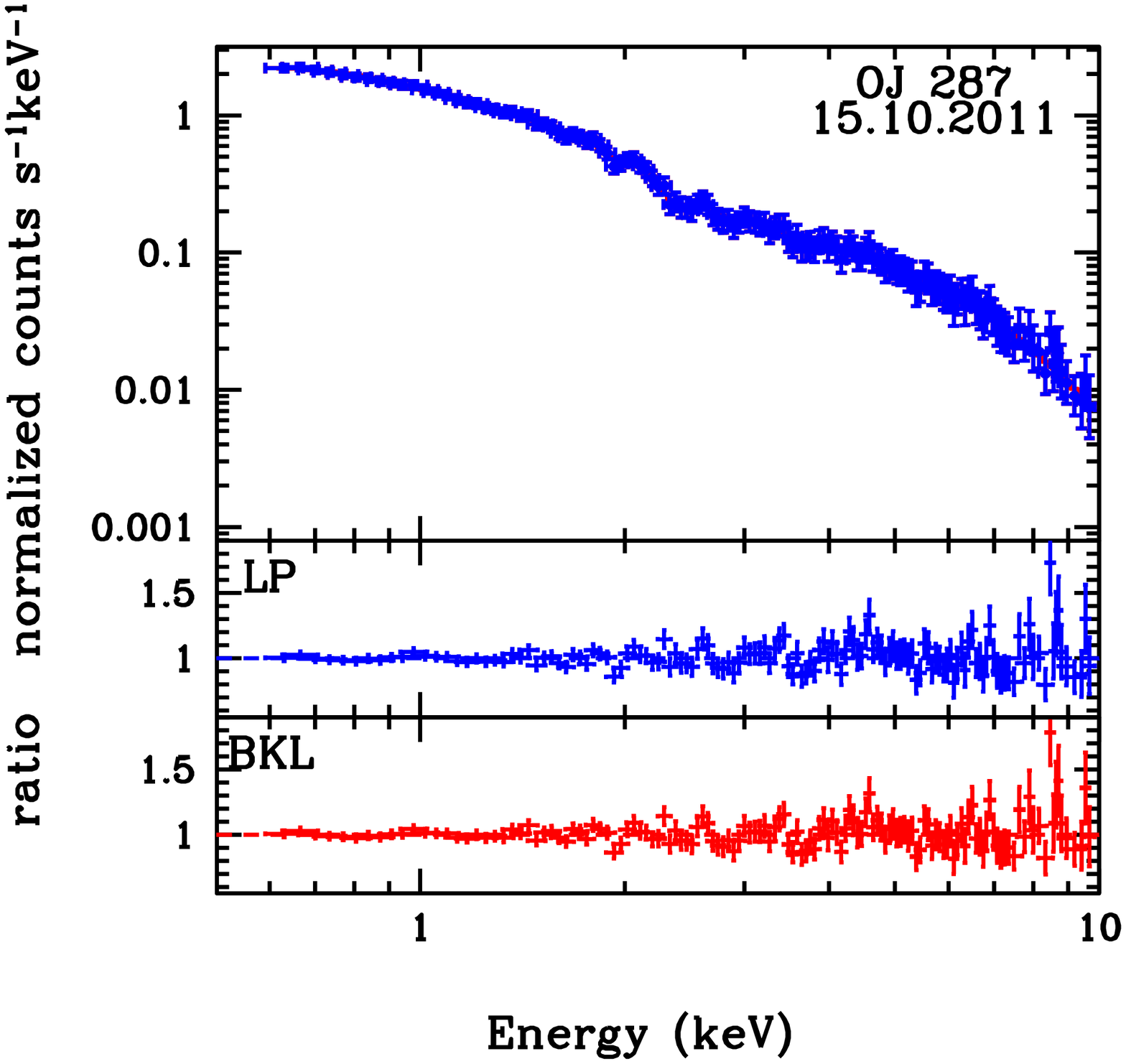}
\includegraphics[width=0.44\textwidth]{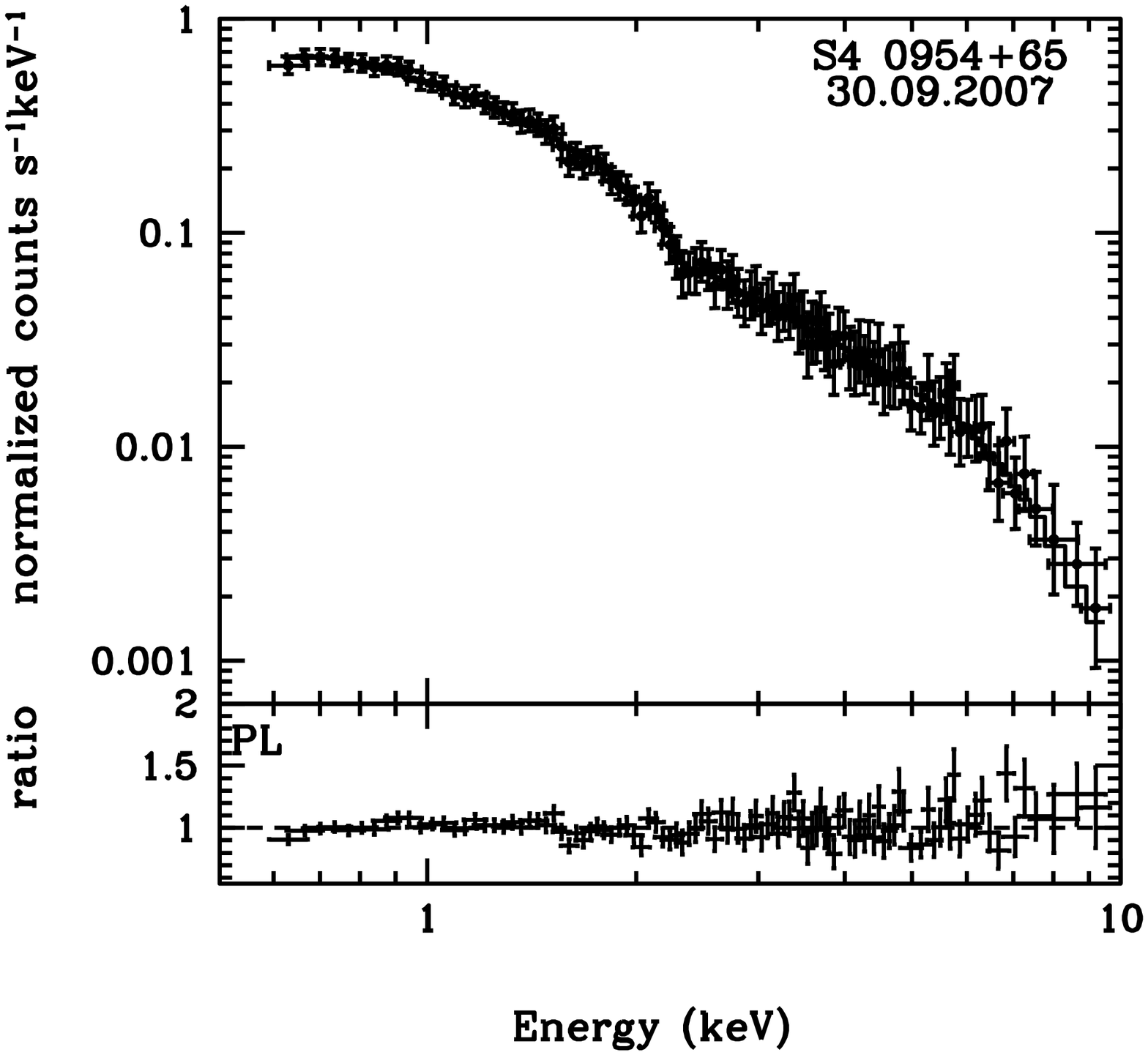}
\includegraphics[width=0.44\textwidth]{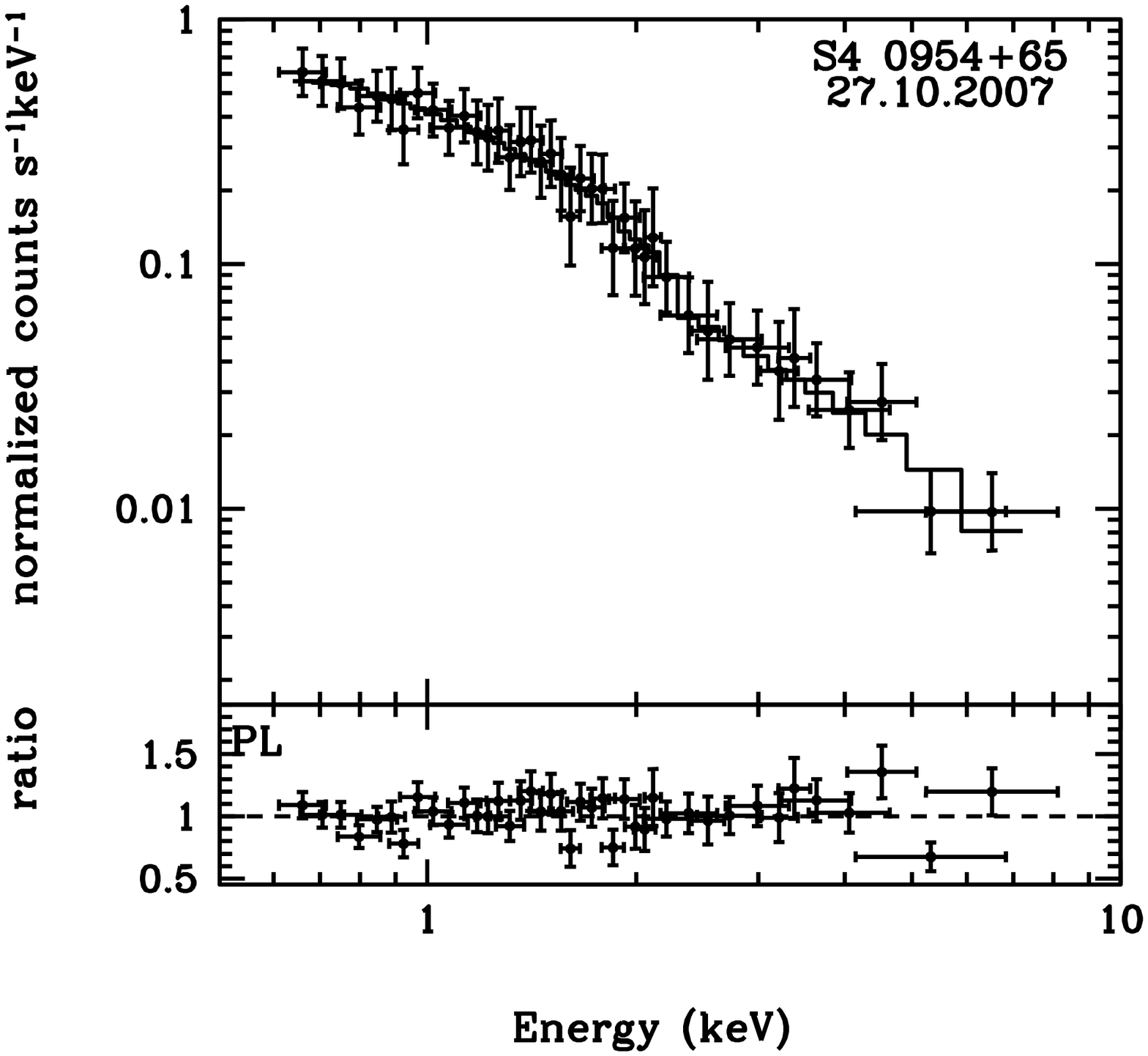}
\includegraphics[width=0.44\textwidth]{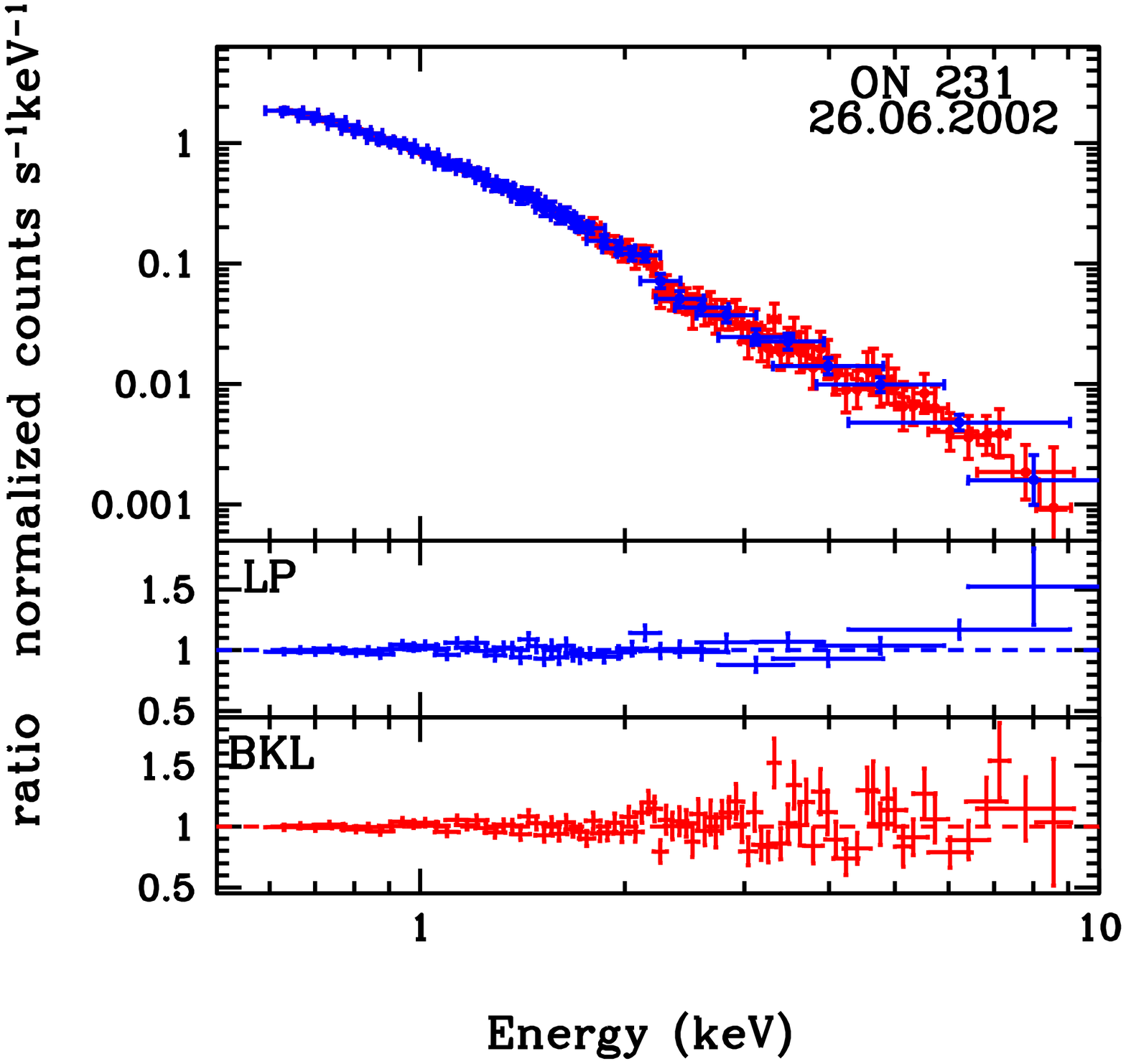}

\caption{Same as in figure 2. }
\end{figure*}

\begin{figure*}
\centering
\includegraphics[width=0.44\textwidth]{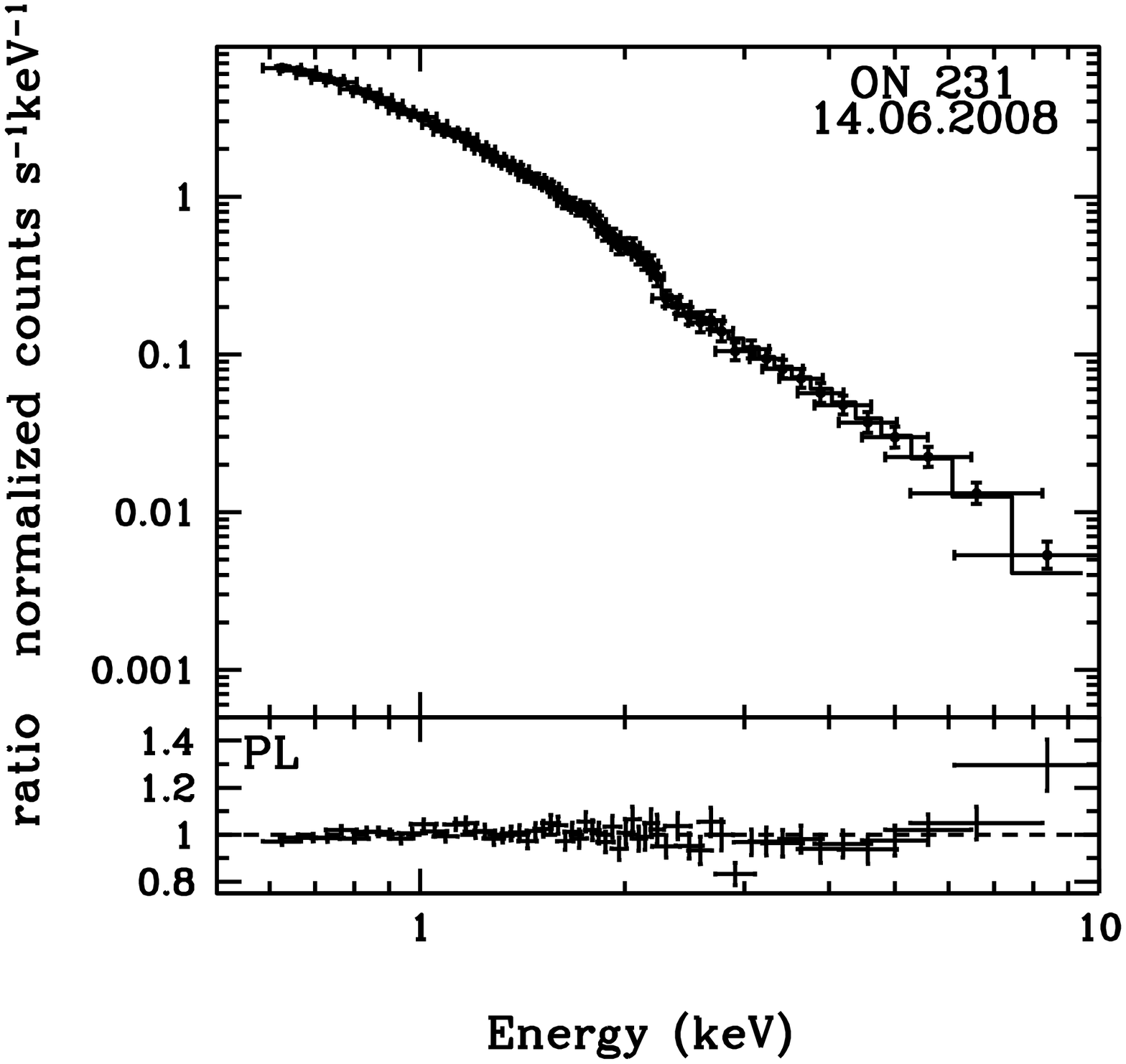}
\includegraphics[width=0.44\textwidth]{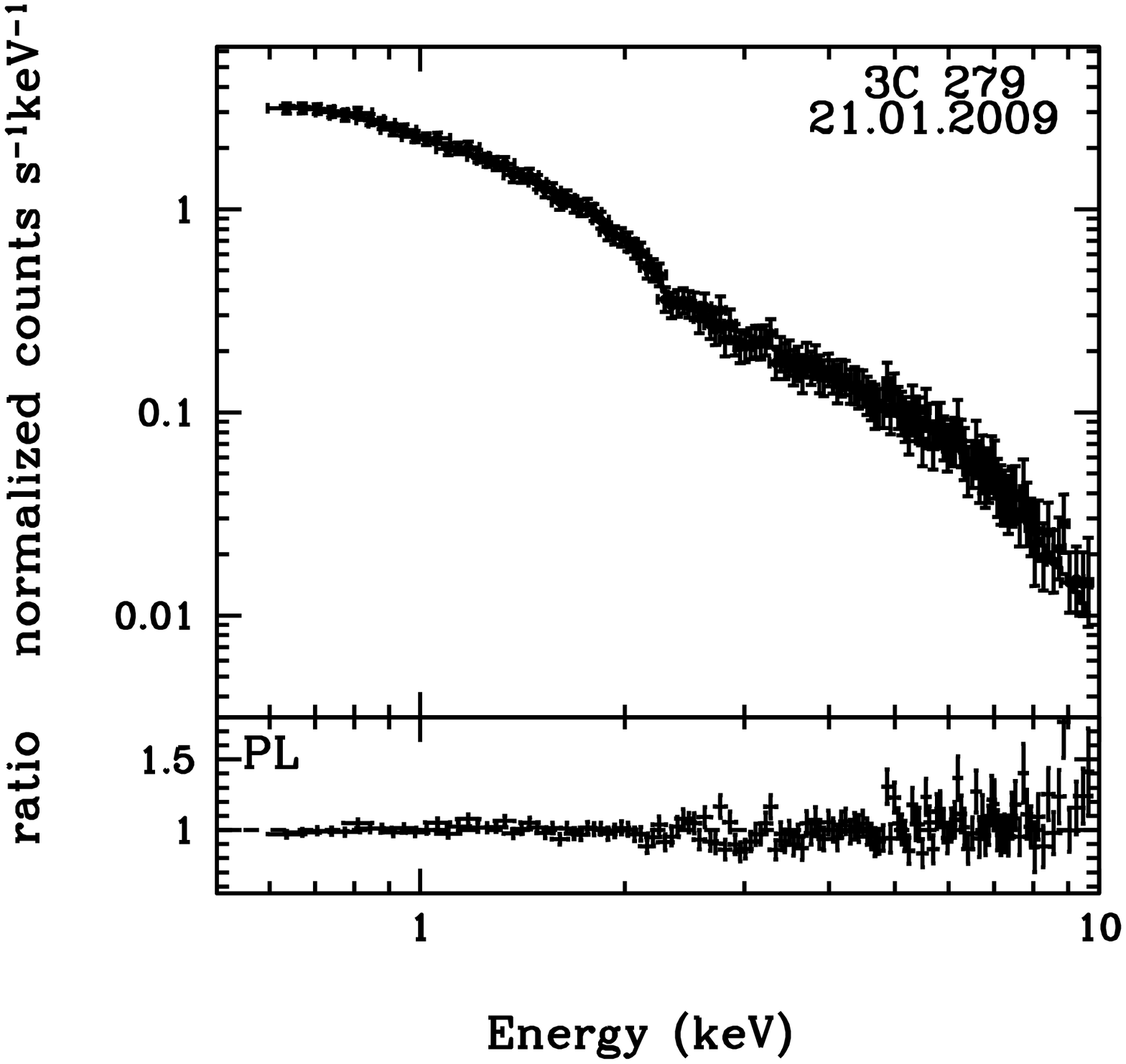}
\includegraphics[width=0.44\textwidth]{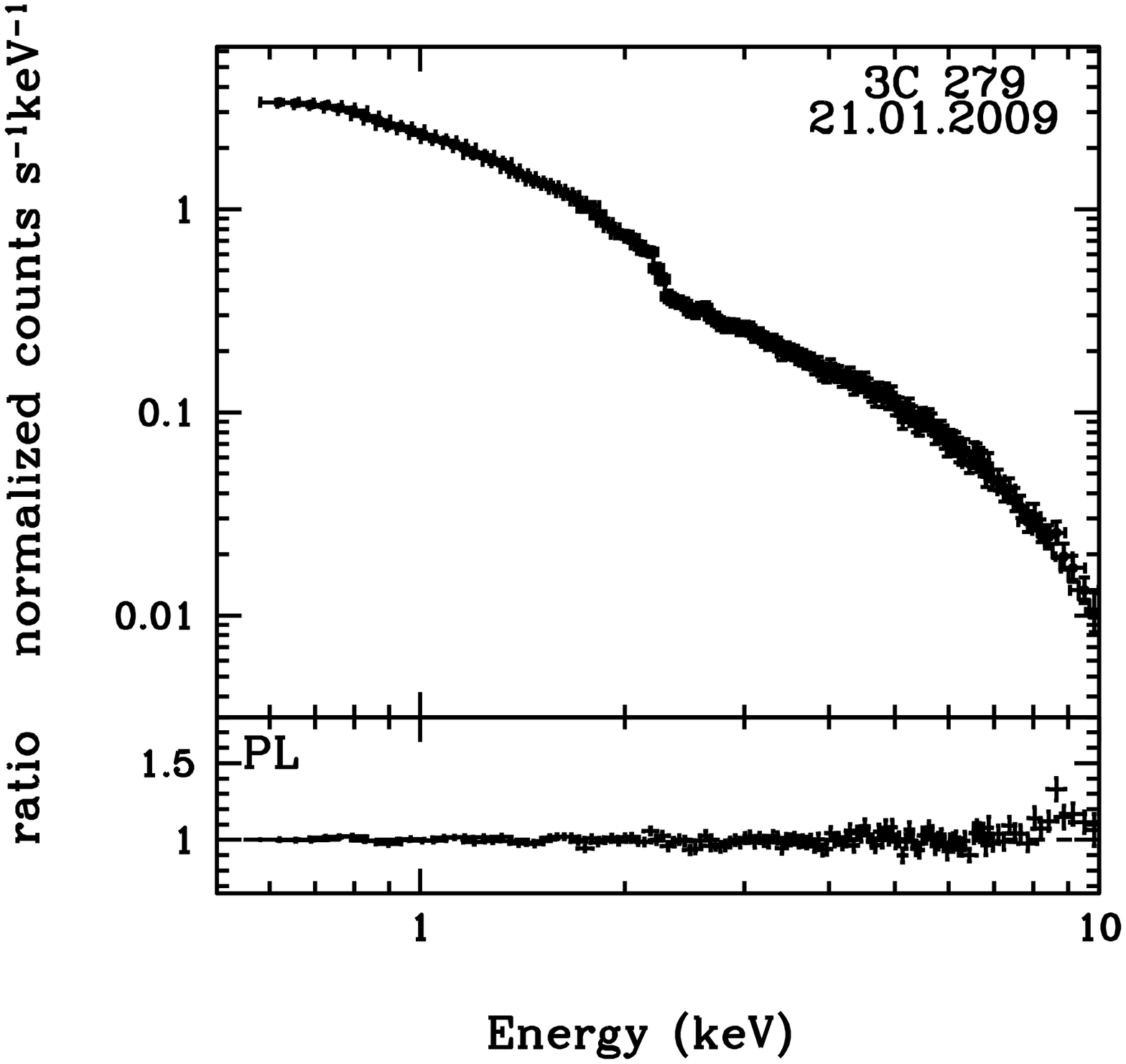}
\includegraphics[width=0.44\textwidth]{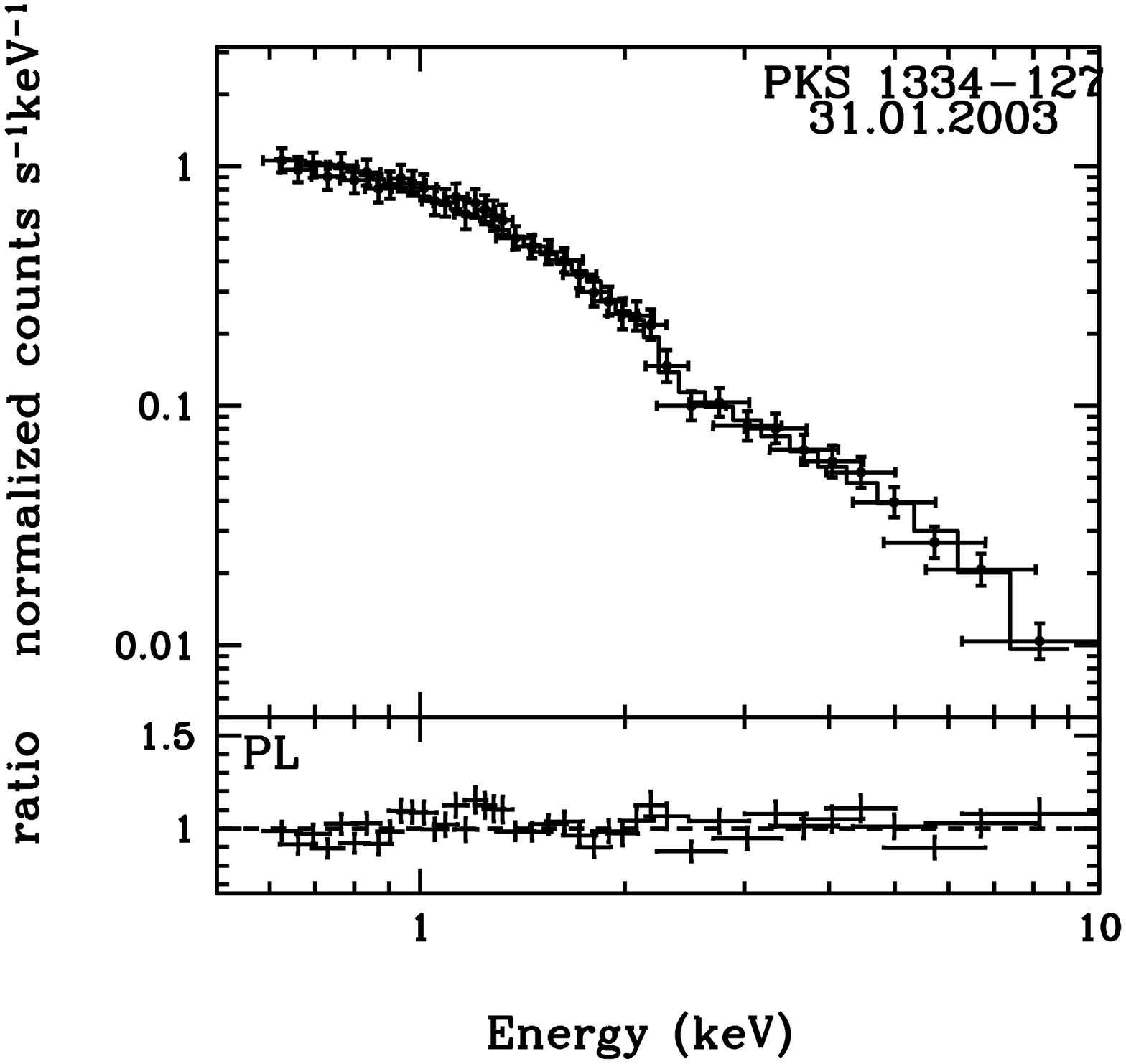}
\caption{Same as in figure 2. }
\end{figure*}

\begin{figure*}
\centering
\includegraphics[width=0.44\textwidth]{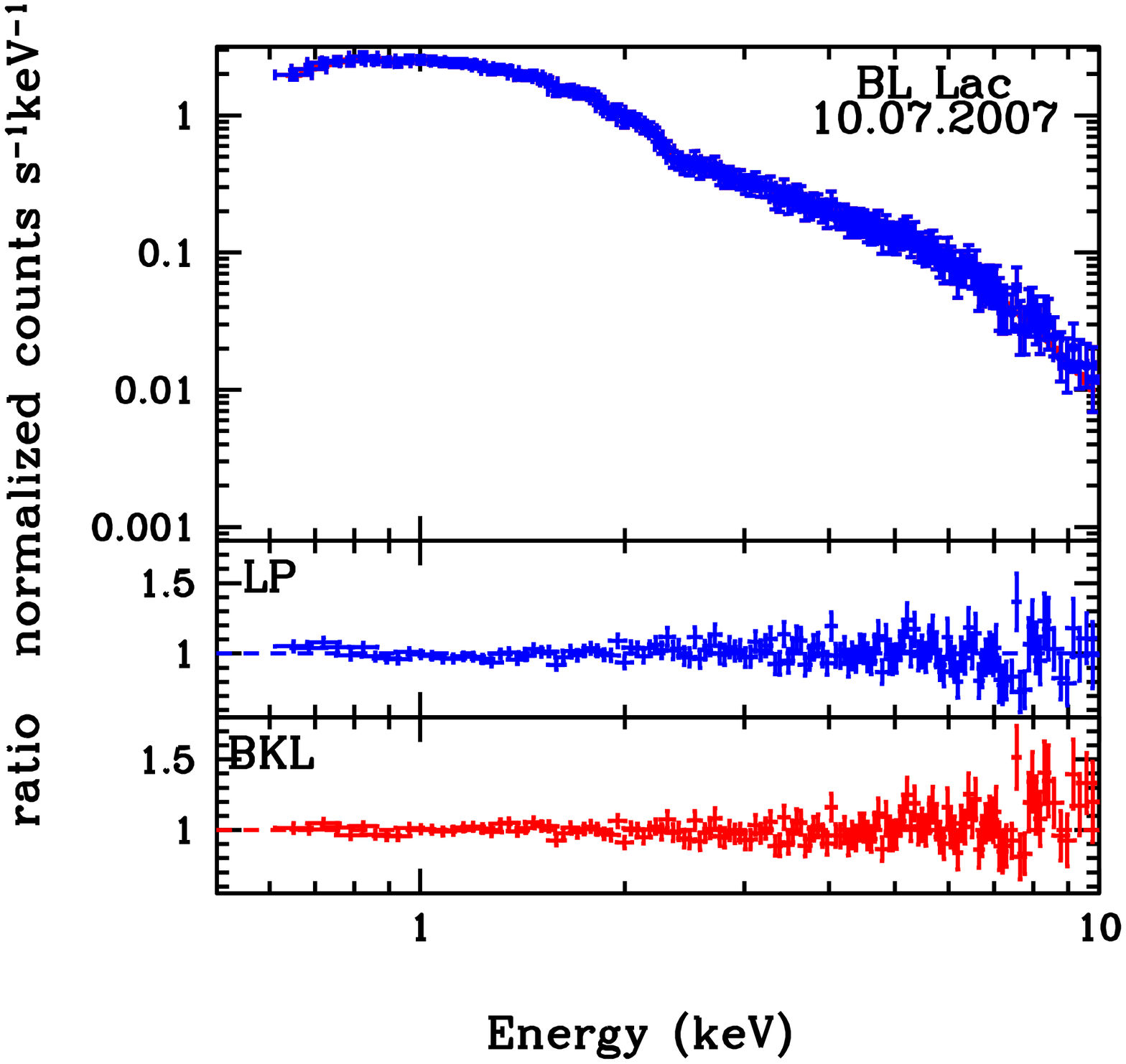}
\includegraphics[width=0.44\textwidth]{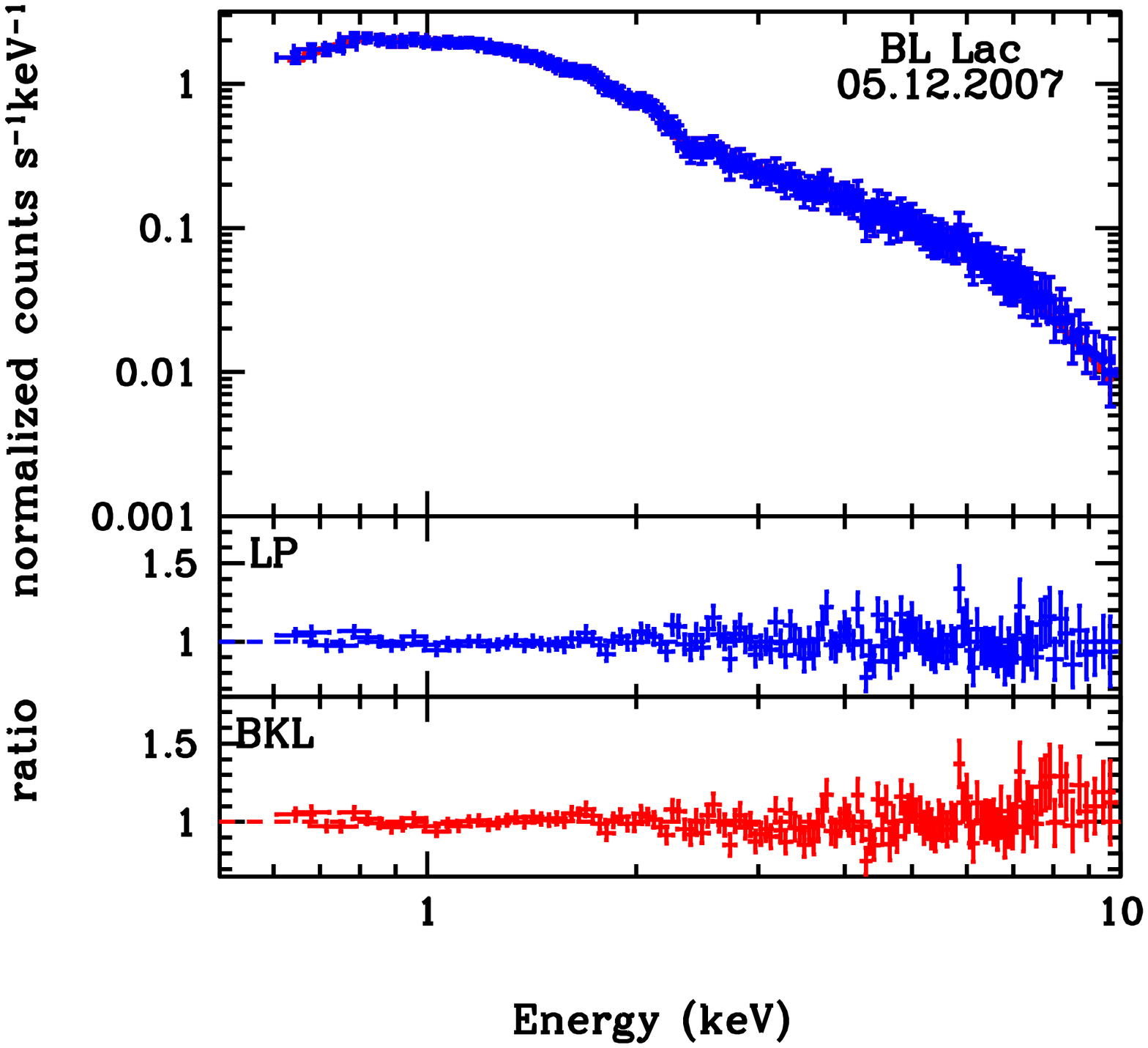}
\includegraphics[width=0.44\textwidth]{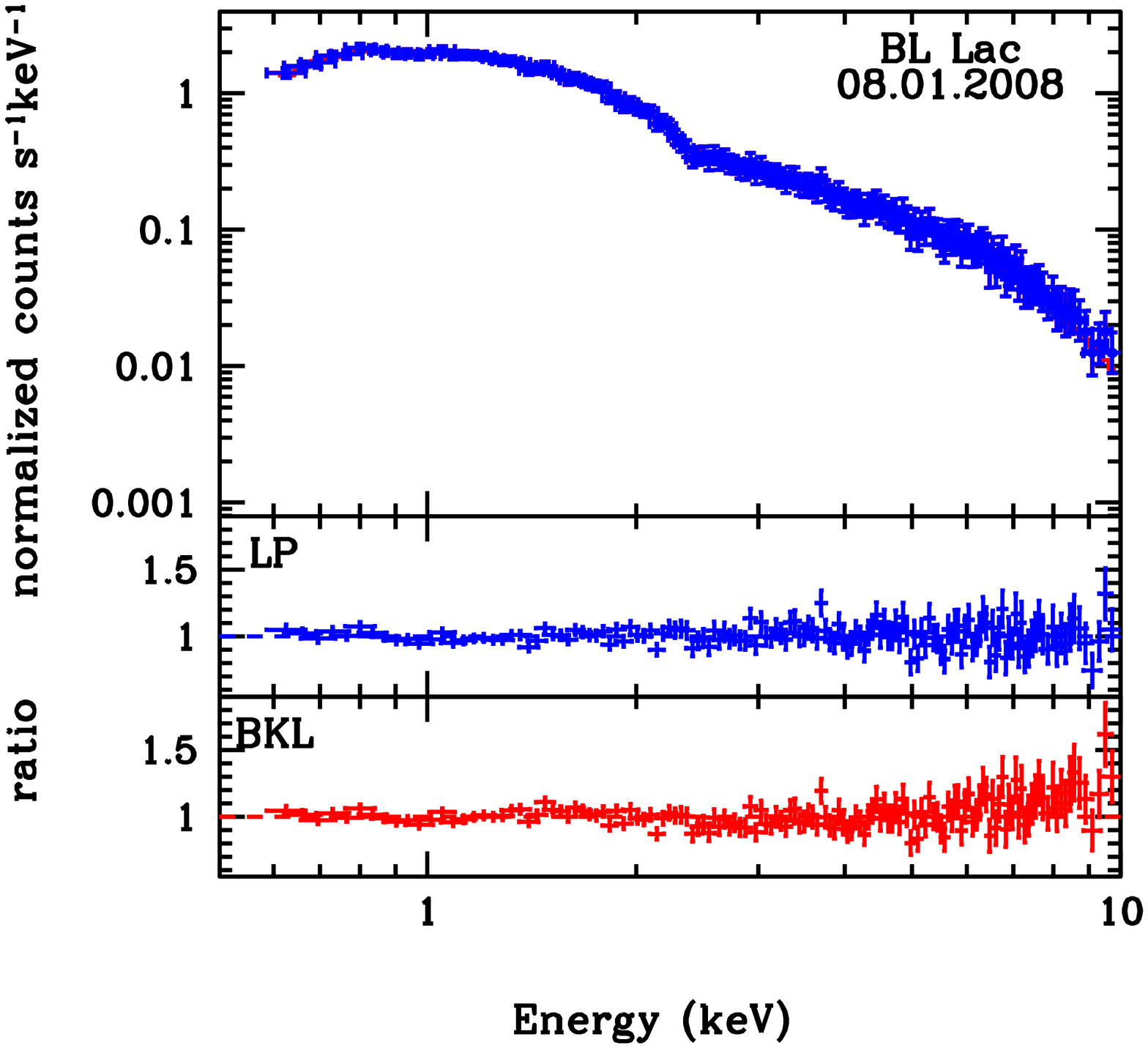}
\includegraphics[width=0.44\textwidth]{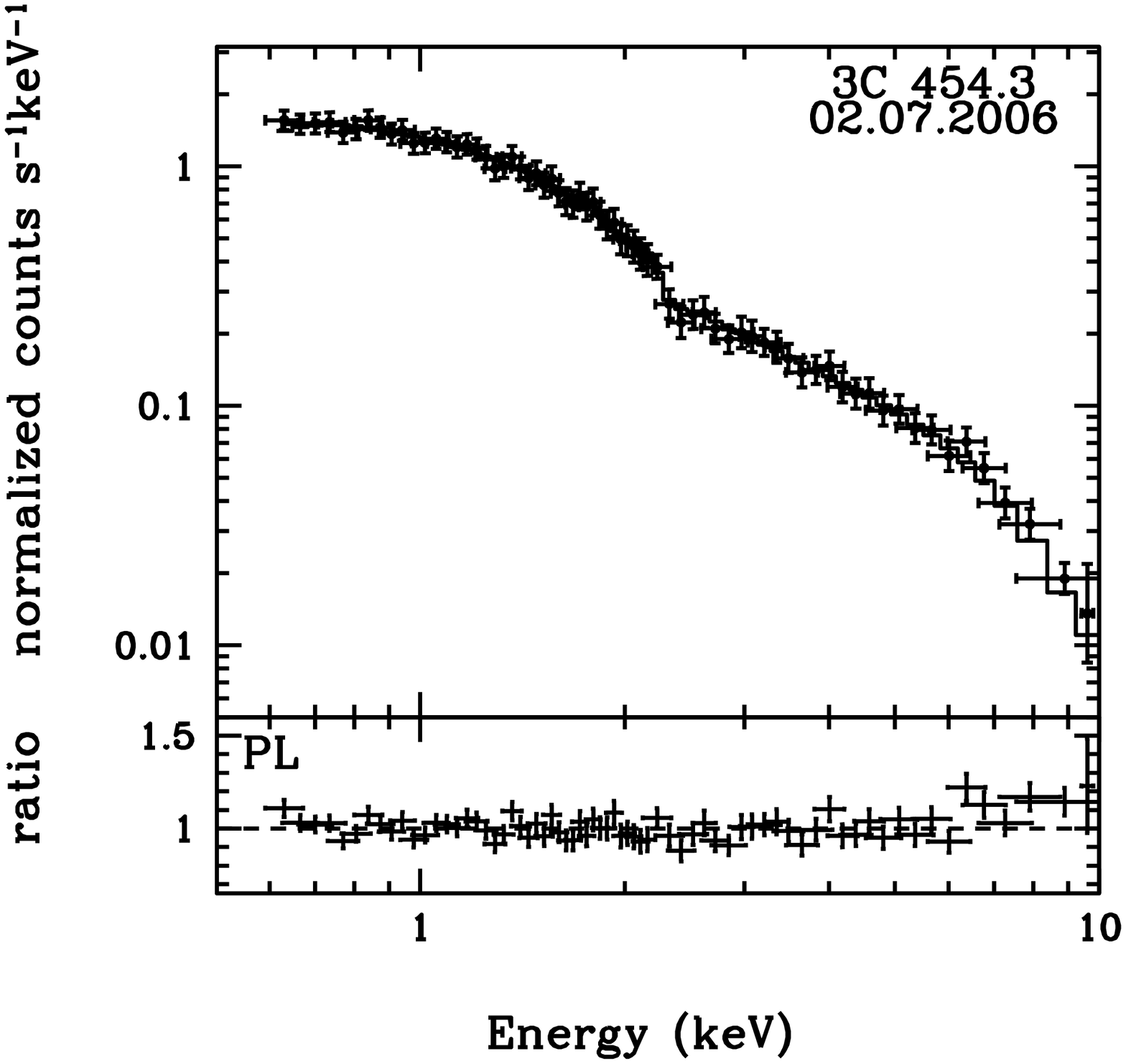}
\caption{Same as in figure 2. }
\end{figure*}

\begin{figure*}
\centering
\includegraphics[width=0.44\textwidth]{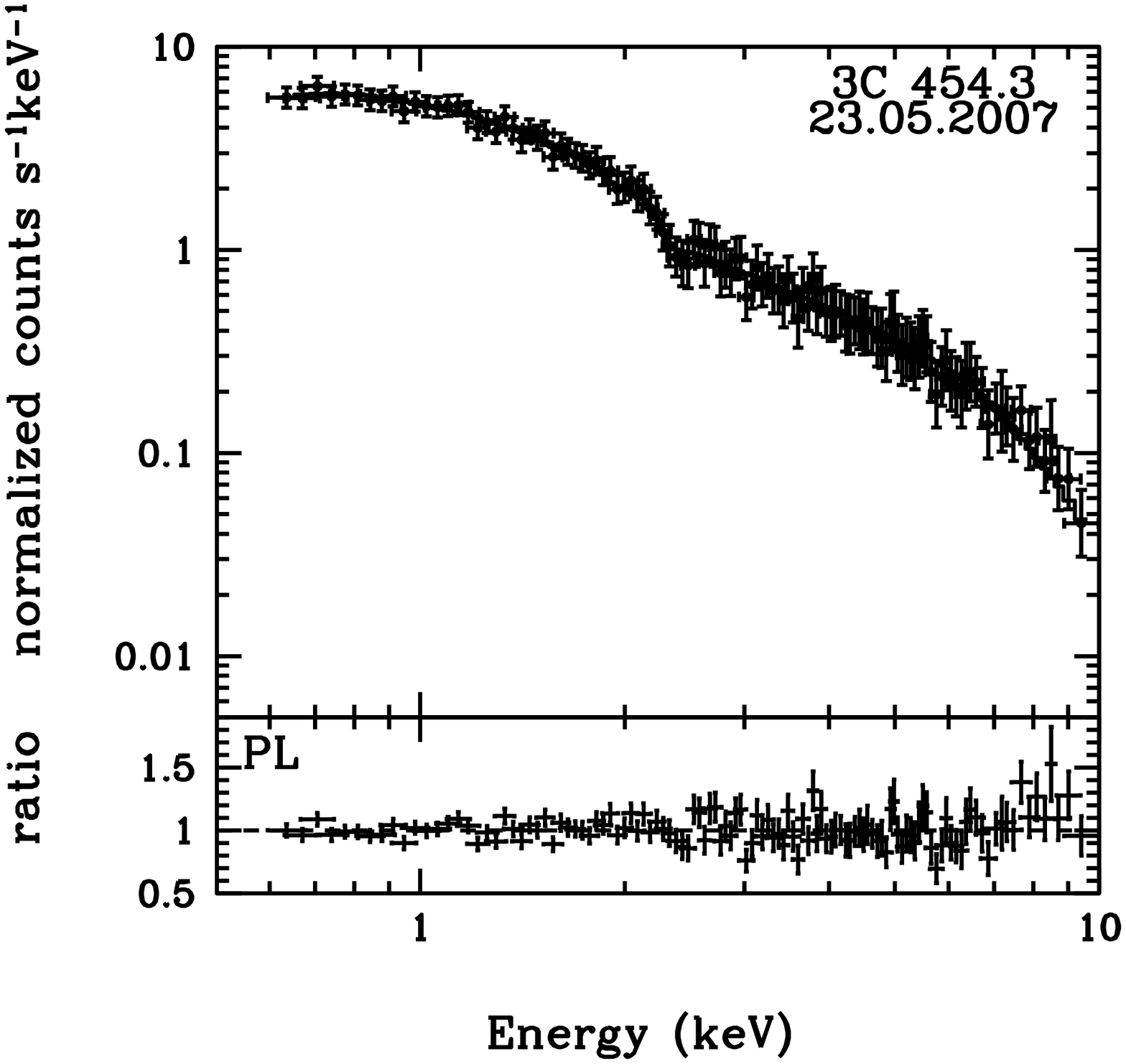}
\includegraphics[width=0.44\textwidth]{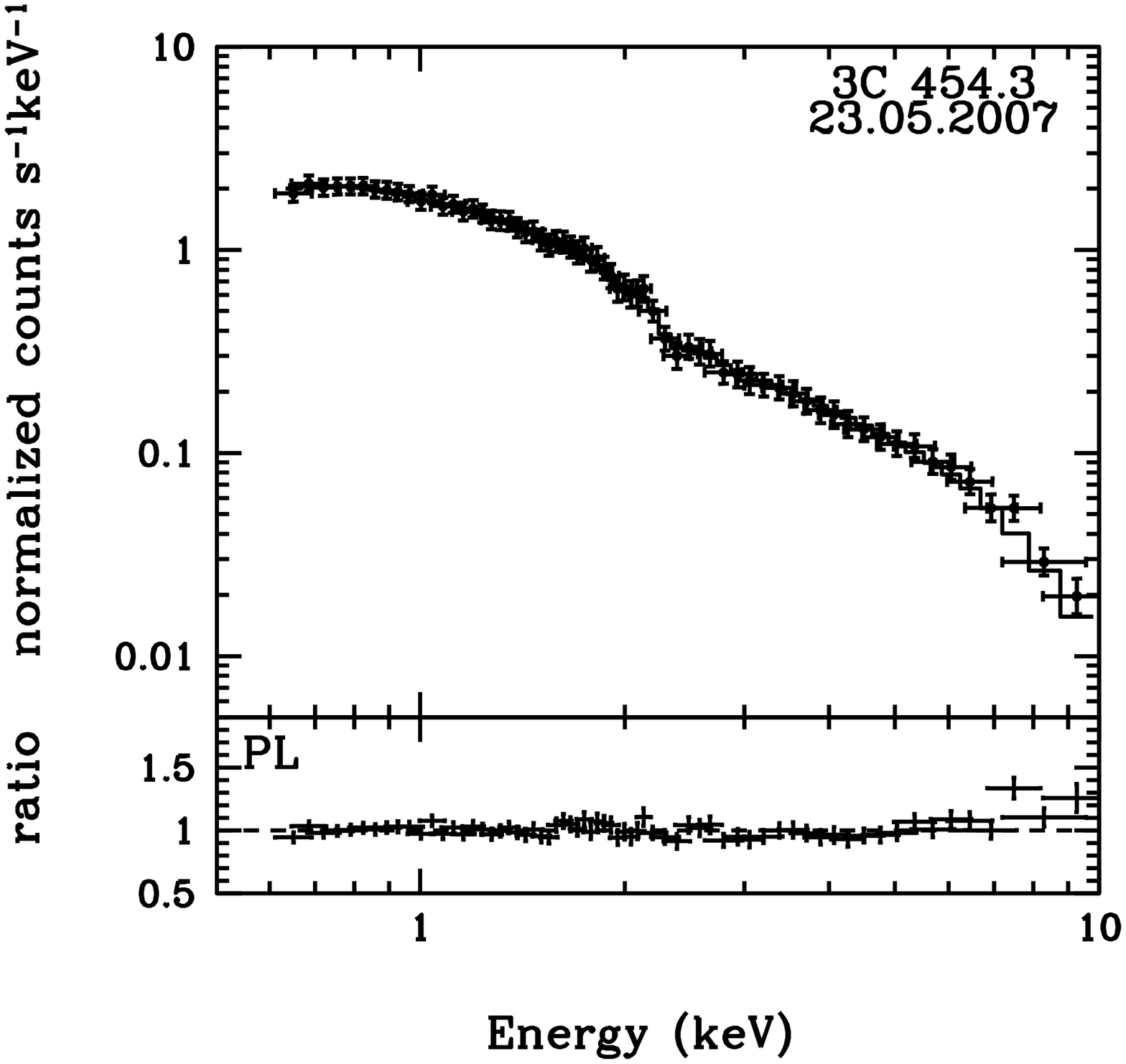}
\includegraphics[width=0.44\textwidth]{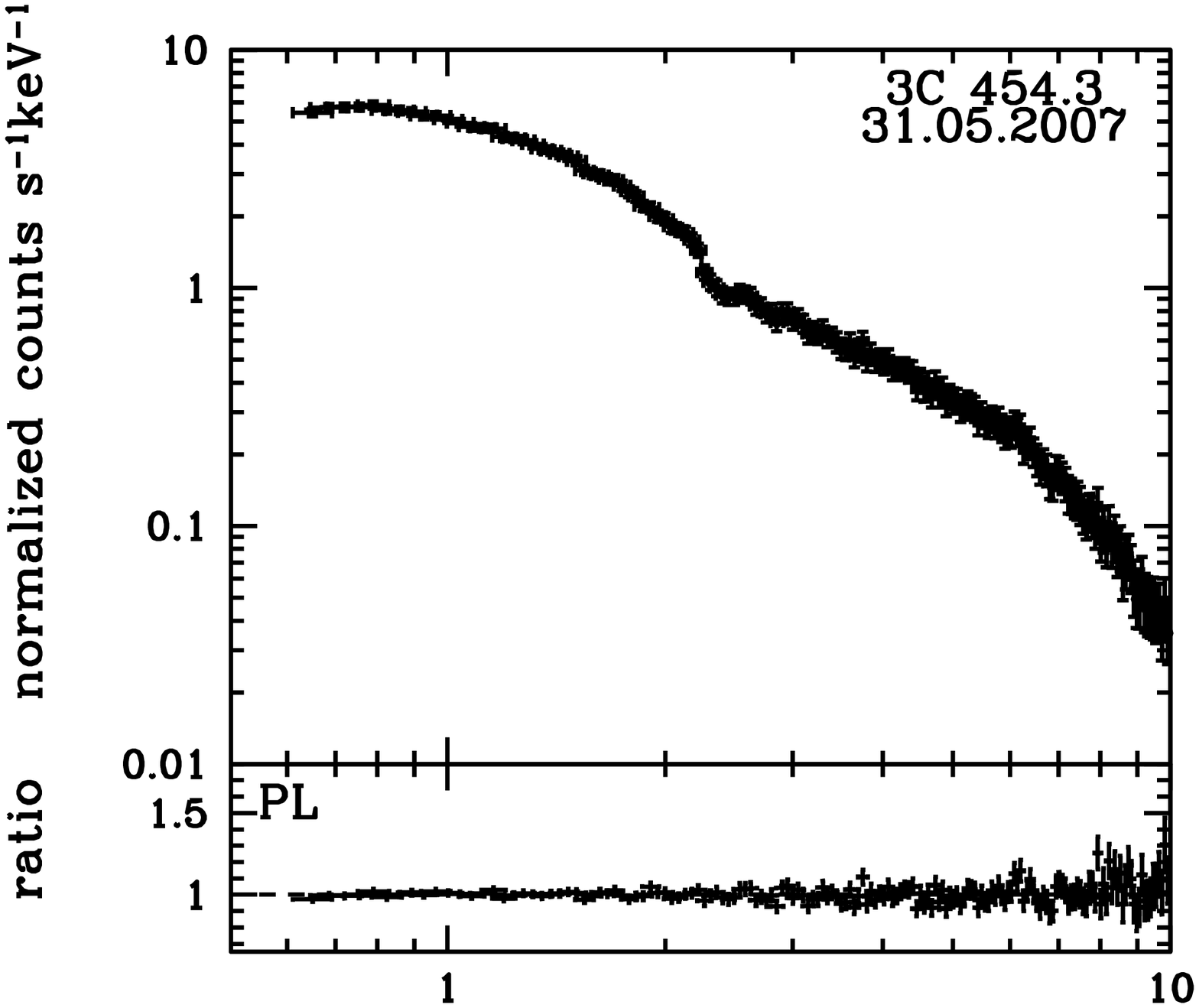}

\caption{Same as in figure 2. }
\end{figure*}

\begin{table*}
\caption{Results from the parametric PSD models fit to the LSP. Columns 1 -- 9 give the object name, the observation date,
the model (PL: power law + constant noise, BPL: bending power law + constant noise), the best-fit parameters log(A), slope $\mu$ and
the bend frequency f$_{\rm b}$, the AIC and the likelihood of a particular model.}
\centering
\scalebox{.78}{
\begin{tabular}{llllllllll}
\hline
Object & Date &Blazar & LSP       & \multicolumn{3}{c}{LSP fit parameters}  & AIC & Model      & Variability \\
       & of   &Class  & model     &        &         &                      &     & likelihood & Amplitude \\
       & obs. &           & log(A) & $\mu$   & log(f$_{\rm b}$)     &     &            & (\%) \\ \hline
TXS 0106$+$612 & 2010.02.09 &LSP & PL & -0.59 $\pm$ 0.21 & -0.01 $\pm$ 0.01 &                  & 121.62 & 1.00 & 6.24 $\pm$ 11.32 \\
               &            &    & BPL& -1.65 $\pm$ 0.18 & -3.00 $\pm$ 0.54 & -4.08 $\pm$ 0.17 & 151.87 & 2.70 $\times$ 10$^{-7}$ & \\
3C 66A         & 2002.02.05 &ISP & PL & -0.09 $\pm$ 0.01 & -0.03 $\pm$ 0.01 &                  & 205.87 & 1.00 & 5.14 $\pm$ 7.63 \\
               &            &    & BPL& -2.37 $\pm$ 0.18 & -3.00 $\pm$ 0.90 & -4.30 $\pm$ 0.18 & 223.32 & 1.62 $\times$ 10$^{-4}$ & \\
PKS 0235$+$164 & 2002.02.10*&LSP & PL & -2.57 $\pm$ 0.17 & -0.84 $\pm$ 0.45 &                  & 63.44  & 3.65 $\times$ 10$^{-1}$ & 11.76 $\pm$ 0.54 \\
               &            &    & BPL& -2.44 $\pm$ 0.18 & -1.52 $\pm$ 0.49 & -4.23 $\pm$ 0.18 & 61.42  & 1.00 & \\
               & 2004.01.18 &    & PL & -0.58 $\pm$ 0.21 & -0.01 $\pm$ 0.01 &                  & 295.67 & 1.00 & 2.13 $\pm$ 5.84\\
               &            &    & BPL& -1.94 $\pm$ 0.18 & -3.00 $\pm$ 0.87 & -4.46 $\pm$ 0.18 & 335.62 & 1.91 $\times$ 10$^{-9}$ & \\
               & 2004.08.02 &    & PL & -0.77 $\pm$ 0.16 & -0.01 $\pm$ 0.01 &                  & 118.48 & 1.00 & 3.20 $\pm$ 6.97\\
               &            &    & BPL& -1.73 $\pm$ 0.18 & -3.00 $\pm$ 0.69 & -4.06 $\pm$ 0.18 & 142.37 & 6.49 $\times$ 10$^{-6}$ & \\
               & 2005.01.28 &    & PL & -0.62 $\pm$ 0.16 & -0.01 $\pm$ 0.01 &                  & 163.07 & 1.00 & 11.70 $\pm$ 4.50\\
               &            &    & BPL& -2.02 $\pm$ 0.18 & -3.00 $\pm$ 0.81 & -4.21 $\pm$ 0.18 & 190.93 & 8.92 $\times$ 10$^{-7}$ & \\
PKS 0426$-$380 & 2012.02.11 &LSP & PL & -0.66 $\pm$ 0.16 & -0.01 $\pm$ 0.01 &                  & 201.94 & 1.00 & 5.92 $\pm$ 7.02\\
               &            &    & BPL& -1.96 $\pm$ 0.18 & -3.00 $\pm$ 0.51 & -4.30 $\pm$ 0.17 & 224.55 & 1.24 $\times$ 10$^{-5}$ & \\
PKS 0521$-$365 & 2005.08.14 &LSP & PL & -1.62 $\pm$ 0.17 & -0.49 $\pm$ 0.17 &                  & 252.86 & 1.00 & 0.80 $\pm$ 2.40\\
               &            &    & BPL& -1.62 $\pm$ 0.17 & -3.00 $\pm$ 1.89 & -4.43 $\pm$ 0.13 & 269.98 & 1.92 $\times$ 10$^{-4}$ & \\
PKS 0537$-$441 & 2010.02.27 &LSP & PL & -0.15 $\pm$ 0.12 & -0.02 $\pm$ 0.01 &                  & 291.25 & 1.00 & 3.10 $\pm$ 1.64 \\
               &            &    & BPL& -3.12 $\pm$ 0.18 & -3.00 $\pm$ 0.30 & -4.45 $\pm$ 0.19 & 320.29 & 4.96 $\times$ 10$^{-7}$ & \\
               & 2010.03.02 &    & PL & -1.29 $\pm$ 0.17 & -0.37 $\pm$ 0.13 &                  & 113.35 & 1.00 & 3.27 $\pm$ 2.61 \\
               &            &    & BPL& -1.29 $\pm$ 0.18 & -3.00 $\pm$ 0.45 & -4.05 $\pm$ 0.17 & 135.03 & 1.96 $\times$ 10$^{-5}$ & \\
               & 2010.03.04 &    & PL & -0.68 $\pm$ 0.16 & -0.01 $\pm$ 0.01 &                  & 216.82 & 1.00 & 2.57 $\pm$ 2.56 \\
               &            &    & BPL& -2.56 $\pm$ 0.18 & -3.00 $\pm$ 0.42 & -4.33 $\pm$ 0.17 & 228.46 & 2.97 $\times$ 10$^{-3}$ & \\
S5 0716$+$714  & 2007.09.24 &ISP & PL & -3.46 $\pm$ 0.60 & -1.03 $\pm$ 0.68 &                  & 227.55 & 1.00 & 17.94 $\pm$ 0.32 \\
               &            &    & BPL& -2.11 $\pm$ 1.00 & -2.03 $\pm$ 1.02 & -4.74 $\pm$ 1.00 & 270.20 & 5.50 $\times$ 10$^{-10}$ & \\
OJ 287         & 2005.04.12 &ISP & PL & -1.55 $\pm$ 0.17 & -0.53 $\pm$ 0.26 &                  & 59.22 & 1.00 & 3.66 $\pm$ 3.30 \\
               &            &    & BPL& -1.55 $\pm$ 0.13 & -3.00 $\pm$ 1.65 & -3.85 $\pm$ 0.13 & 66.15 & 3.12 $\times$ 10$^{-2}$ & \\
               & 2005.11.03 &    & PL & -2.37 $\pm$ 0.19 & -0.72 $\pm$ 0.25 &                  & 299.50 & 1.00 & 3.33 $\pm$ 1.76 \\
               &            &    & BPL& -2.73 $\pm$ 0.12 & -2.23 $\pm$ 1.22 & -4.56 $\pm$ 0.14 & 302.39 & 2.19 $\times$ 10$^{-1}$ & \\
               & 2006.11.17 &    & PL & -1.53 $\pm$ 0.17 & -0.46 $\pm$ 0.19 &                  & 432.15 & 1.00 & 1.90 $\pm$ 3.13 \\
               &            &    & BPL& -3.10 $\pm$ 0.13 & -1.13 $\pm$ 0.34 & -4.65 $\pm$ 0.18 & 438.76 & 3.69 $\times$ 10$^{-2}$ & \\
               & 2008.04.22 &    & PL & -2.37 $\pm$ 0.16 & -0.56 $\pm$ 0.25 &                  & 532.54 & 1.00 & 3.83 $\pm$ 1.72 \\
               &            &    & BPL& -2.73 $\pm$ 0.12 & -2.23 $\pm$ 1.11 & -4.73 $\pm$ 0.12 & 544.78 & 2.19 $\times$ 10$^{-3}$ & \\
               & 2011.10.15 &    & PL & -1.48 $\pm$ 0.21 & -0.35 $\pm$ 0.10 &                  & 219.80 & 1.00 & 1.48 $\pm$ 1.65 \\
               &            &    & BPL& -2.94 $\pm$ 0.11 & -1.20 $\pm$ 0.49 & -4.33 $\pm$ 0.19 & 230.49 & 4.77 $\times$ 10$^{-3}$ & \\
S4 0954$+$65   & 2007.09.30 &LSP & PL & -3.55 $\pm$ 0.21 & -1.04 $\pm$ 0.31 &                  & 260.67 & 1.97 $\times$ 10$^{-1}$ & 12.59 $\pm$ 0.85 \\
               &            &    & BPL& -2.63 $\pm$ 0.17 & -1.77 $\pm$ 0.87 & -4.62 $\pm$ 0.18 & 257.42 & 1.00 & \\
               & 2007.10.27 &    & PL & -0.61 $\pm$ 0.16 & -0.02 $\pm$ 0.01 &                  & 117.30 & 1.00 & 5.37 $\pm$ 0.39 \\
               &            &    & BPL& -1.56 $\pm$ 0.18 & -3.00 $\pm$ 0.42 & -4.05 $\pm$ 0.16 & 137.25 & 4.66 $\times$ 10$^{-5}$ & \\
ON 231         & 2002.06.26 &ISP & PL & -3.33 $\pm$ 0.80 & -0.88 $\pm$ 0.28 &                  & 520.33 & 1.00 & 52.12 $\pm$ 0.59 \\
               &            &    & BPL& -2.38 $\pm$ 1.00 & -2.21 $\pm$ 1.99 & -4.61 $\pm$ 1.00 & 632.28 & 2.54 $\times$ 10$^{-4}$ & \\
               & 2008.06.14 &    & PL & -4.09 $\pm$ 0.82 & -1.12 $\pm$ 0.31 &                  & 408.68 & 1.00 & 36.94 $\pm$ 0.39 \\
               &            &    & BPL& -2.24 $\pm$ 1.00 & -2.26 $\pm$ 2.03 & -4.45 $\pm$ 1.00 & 462.92 & 1.67 $\times$ 10$^{-12}$ & \\
3C 279         & 2009.01.21 &LSP & PL & -0.18 $\pm$ 0.18 & -0.06 $\pm$ 0.03 &                  & 257.22 & 1.00 & 1.69 $\pm$ 1.03 \\
               &            &    & BPL& -3.42 $\pm$ 0.17 & -0.43 $\pm$ 0.20 & -4.40 $\pm$ 0.19 & 268.48 & 3.59 $\times$ 10$^{-3}$ & \\
               & 2011.01.18 &    & PL & -0.91 $\pm$ 0.16 & -0.14 $\pm$ 0.06 &                  & 1246.92 & 1.00 & 1.64 $\pm$ 0.62 \\
               &            &    & BPL& -2.78 $\pm$ 0.14 & -1.49 $\pm$ 0.80 & -5.09 $\pm$ 0.18 & 1329.79 & 1.01 $\times$ 10$^{-18}$ & \\
PKS 1334$-$127 & 2003.01.31 &LSP & PL & -0.77 $\pm$ 0.16 & -0.01 $\pm$ 0.01 &                  & 124.23 & 1.00 & 3.61 $\pm$ 2.01 \\
               &            &    & BPL& -2.04 $\pm$ 0.18 & -3.00 $\pm$ 0.39 & -4.08 $\pm$ 0.16 & 136.97 & 1.71 $\times$ 10$^{-3}$ & \\
BL Lac         & 2007.07.10 &ISP & PL & -2.67 $\pm$ 0.22 & -0.71 $\pm$ 0.28 &                  & 182.54 & 1.00 & 1.71 $\pm$ 1.31 \\
               &            &    & BPL& -2.67 $\pm$ 0.12 & -2.15 $\pm$ 1.14 & -4.26 $\pm$ 0.12 & 186.37 & 1.48 $\times$ 10$^{-1}$ & \\
               & 2007.12.05 &    & PL & -2.08 $\pm$ 0.16 & -0.56 $\pm$ 0.24 &                  & 188.64 & 1.00 & 2.22 $\pm$ 1.28 \\
               &            &    & BPL& -2.08 $\pm$ 0.18 & -3.00 $\pm$ 0.72 & -4.28 $\pm$ 0.17 & 201.69 & 1.47 $\times$ 10$^{-3}$ & \\
               & 2008.01.08 &    & PL & -0.35 $\pm$ 0.18 & -0.13 $\pm$ 0.06 &                  & 227.63 & 1.00 & 1.40 $\pm$ 1.72 \\
               &            &    & BPL& -3.20 $\pm$ 0.10 & -0.97 $\pm$ 0.38 & -4.35 $\pm$ 0.19 & 235.64 & 1.82 $\times$ 10$^{-2}$ & \\
3C 454.3       & 2006.07.02 &LSP & PL & -0.65 $\pm$ 0.22 & -0.01 $\pm$ 0.01 &                  & 161.76 & 1.00 & 2.62 $\pm$ 1.67 \\
               &            &    & BPL& -1.83 $\pm$ 0.18 & -3.00 $\pm$ 0.78 & -4.20 $\pm$ 0.18 & 190.90 & 4.70 $\times$ 10$^{-7}$ & \\
               & 2006.12.18 &    & PL & -1.00 $\pm$ 0.17 & -0.29 $\pm$ 0.12 &                  & 127.93 & 1.00 & 3.36 $\pm$ 1.20 \\
               &            &    & BPL& -1.79 $\pm$ 0.17 & -2.93 $\pm$ 1.17 & -4.12 $\pm$ 0.15 & 143.41 & 4.36 $\times$ 10$^{-4}$ & \\
               & 2007.05.23 &    & PL & -1.41 $\pm$ 0.28 & -0.01 $\pm$ 0.01 &                  & 34.71 & 1.00 & 0.53 $\pm$ 4.28 \\
               &            &    & BPL& -1.41 $\pm$ 0.18 & -3.00 $\pm$ 0.66 & -3.46 $\pm$ 0.18 & 50.66 & 3.45 $\times$ 10$^{-4}$ & \\
               & 2007.05.31 &    & PL & -1.96 $\pm$ 0.21 & -0.47 $\pm$ 0.21 &                  & 291.46 & 1.00 & 1.14 $\pm$ 0.70 \\
               &            &    & BPL& -2.17 $\pm$ 0.18 & -3.00 $\pm$ 0.63 & -4.46 $\pm$ 0.17 & 309.62 & 1.14 $\times$ 10$^{-4}$ & \\ \hline
\end{tabular}}
\label{pgrambestfit2}
\end{table*}

\begin{table*}
{\caption {Best fit spectral parameters for the power law, broken power law and log parabolic model.} }
\noindent
\setlength{\tabcolsep}{0.015in}
\scalebox{.78}{
\begin{tabular}{cccccccccc} \hline \hline

Source &Blazar  &Date of          &Model   &$\Gamma_{1}$  &$\Gamma_{2}$ &E$_{b}$/$b$    &log$_{10}$Flux       &$\chi_{Red}^{2}$/dof &F(p) \\
       &Class   &observation      &        &              &             &(KeV)          &                     &       & \\ \hline

TXS0106$+$612 &LSP &2010.02.09   &PL & $  1.178_{-  0.084 }^{+  0.085}$  &   &  &$-11.895_{-  0.026 }^{+  0.025}$   &0.76/52  &  \\
              & &   &LP  & $0.783_{-0.530 }^{+0.496}$ &  &$  0.291_{-0.360 }^{+0.383}$ &$-11.911_{-0.033 }^{+0.032}$   &0.74/51 &2.41(0.127) \\
3C 66A        &ISP &2002.02.05   &PL & $  2.447_{-  0.065 }^{+  0.067}$   &  &   &$-12.209_{-  0.018 }^{+  0.017}$   &1.30/52   &\\
              &  &   &LP & $2.685_{-0.187 }^{+0.182}$    &   &$ -0.302_{-0.207 }^{+0.221}$ &$-12.185_{-0.025 }^{+0.025}$  &1.23/51 &3.96(0.052) \\

PKS 0426$-$380$^{*}$ &LSP &2012.02.11   &PL & $  1.905_{-  0.065 }^{+  0.067}$  &  & &$-12.231_{-  0.024 }^{+  0.024}$   &0.96/54            &\\
             & &   &LP  & $2.338_{-0.187 }^{+0.183}$     &   &$ -0.473_{-0.181 }^{+0.182}$ &$-12.184_{-0.030 }^{+0.029}$   &0.68/53  &23.23(1.25$\times10^{-5}$)\\
             & &   &BPL & $  2.111_{-  0.110 }^{+  0.152}$  &$  1.511_{-  0.233 }^{+  0.201}$  &$  2.109_{-  0.536 }^{+  0.740}$  &$-12.186_{-  0.030 }^{+  0.027}$    &0.63/52 &\\
PKS 0537$-$441$^{*}$ &LSP &2010.02.27$^{*}$   &PL   &$  2.009_{-  0.018 }^{+  0.018}$ & & &$-11.328_{-  0.006 }^{+  0.006}$  &1.57/ 131    &\\
              &  &           &LP   &$  2.310_{-  0.054 }^{+  0.053}$ & &$ -0.320_{-  0.052 }^{+  0.053}$ &$-11.304_{-  0.007 }^{+  0.007}$  &0.87/ 130 &106.40(1.36$\times10^{-18}$)\\
              &  &           &BPL  &$  2.128_{-  0.031 }^{+  0.070}$ &$  1.752_{-  0.075 }^{+  0.107}$  &$  2.283_{-  0.665 }^{+  0.384}$ &$-11.307_{-  0.007 }^{+  0.007}$    &0.93/ 129   &\\
             &  &2010.03.02   &PL   &$  1.970_{-  0.029 }^{+  0.029}$ & & &$-11.331_{-  0.010 }^{+  0.010}$  &1.19/ 100   &\\
              &  &           &LP   &$  2.189_{-  0.095 }^{+  0.094}$ & &$ -0.226_{-  0.091 }^{+  0.093}$    &$-11.315_{-  0.012 }^{+  0.012}$   &1.04/99 &
15.42(1.59$\times10^{-4}$)\\
              &  &           &BPL  &$  2.040_{-  0.046 }^{+  1.005}$ &$  1.759_{-  0.263 }^{+  0.185}$  &$  2.707_{-  1.818 }^{+  1.674}$ &$-11.317_{-  0.012 }^{+  0.012}$  &1.07/  98   &\\
            &  &2010.03.04$^{*}$   &PL   &$  1.957_{-  0.022 }^{+  0.023}$ & & &$-11.333_{-  0.008 }^{+  0.008}$  &1.29/ 137  &\\
            &  &             &LP   &$  2.246_{-  0.066 }^{+  0.065}$ & &$ -0.315_{-  0.065 }^{+  0.067}$    &$-11.304_{-  0.010 }^{+  0.010}$ &0.88/ 136  &64.83(3.65$\times10^{-13}$)\\
            &  &             &BPL  &$  2.073_{-  0.037 }^{+  0.125}$ &$  1.716_{-  0.072 }^{+  0.127}$  &$  2.156_{-  0.845 }^{+  0.275}$ &$-11.307_{-  0.010 }^{+  0.009}$  &0.91/ 135 &\\
S5 0716$+$714$^{*}$ &ISP &2007.09.24   &PL & $  2.178_{-  0.010 }^{+  0.011}$  &  & &$-11.060_{-  0.003 }^{+  0.003}$   &5.64/161     &      \\
              & &  &LP & $2.628_{-0.028 }^{+0.028}$        &  &$ -0.521_{-0.029 }^{+0.029}$ &$-11.010_{-0.004 }^{+0.004}$  &1.09/160  &673.06(3.313$\times$10$^{-59}$)\\
              & &  &BPL& $  2.364_{-  0.020 }^{+  0.021}$ &$  1.824_{-  0.036 }^{+  0.033}$   &$  1.904_{-  0.103 }^{+  0.122}$ &$-11.019_{-  0.004 }^{+  0.004}$   &1.62/ 159  &\\
OJ 287$^{*}$   &ISP     &2005.04.12   &PL & $  1.579_{-  0.043 }^{+  0.044}$  &   & &$-11.429_{-  0.018 }^{+  0.017}$ &0.79/79      &      \\
              & &  &LP & $1.736_{-0.159 }^{+0.156}$        & &$ -0.150_{-0.142 }^{+0.146}$ &$-11.417_{-0.021 }^{+0.020}$ &0.76/78  &4.118(0.046)\\
            &  &2005.11.03$^{*}$   &PL & $  1.824_{-  0.022 }^{+  0.022}$  &  &  &$-11.548_{-  0.008 }^{+  0.008}$ &1.40/126     &      \\
            &  &   &LP  & $2.075_{-0.068 }^{+0.067}$       &  &$ -0.255_{-0.064 }^{+0.065}$ &$-11.528_{-0.010 }^{+0.009}$ &1.09/125  &36.835(1.42$\times10^{-8}$)\\
            &  &   &BPL & $  2.011_{-  0.053 }^{+  0.059}$ &$  1.691_{-  0.040 }^{+  0.039}$   &$  1.539_{-  0.179 }^{+  0.181}$  &$-11.531_{-  0.009 }^{+  0.009}$ &1.03/ 124  &\\
            &  &2006.11.17$^{*}$   &PL & $  1.752_{-  0.017 }^{+  0.017}$  &  & &$-11.592_{-  0.007 }^{+  0.006}$ &1.11/139     &      \\
            &  &   &LP & $1.906_{-0.057 }^{+0.057}$        &  &$ -0.152_{-0.053 }^{+0.054}$ &$-11.581_{-0.007 }^{+0.007}$ &0.97/138 &21.062(9.896$\times10^{-6}$) \\
            &  &   &BPL& $  1.801_{-  0.025 }^{+  0.047}$  &$  1.579_{-  0.089 }^{+  0.109}$  &$  3.076_{-  1.339 }^{+  0.703}$  &$-11.581_{-  0.007 }^{+  0.007}$  &0.97/ 137  &\\
            &  &2008.04.22    &PL & $  1.755_{-  0.017 }^{+  0.017}$  &  & &$-11.583_{-  0.007 }^{+  0.007}$ &1.24/136     &      \\
            &  &    &LP & $1.787_{-0.059 }^{+0.058}$        &  &$ -0.032_{-0.055 }^{+0.056}$ &$-11.580_{-0.008 }^{+0.008}$ &1.24/135  &1(0.319)\\
            &  &2011.10.15$^{*}$   &PL& $  1.769_{-  0.013 }^{+  0.013}$  &  & &$-11.068_{-  0.005 }^{+  0.005}$ &1.36/144     &      \\
            &  &   &LP & $1.904_{-0.042 }^{+0.042}$        &  &$ -0.133_{-0.039 }^{+0.039}$ &$-11.059_{-0.005 }^{+0.005}$ &1.15/143  &27.296(6.054$\times10^{-7}$)\\
            &  &   &BPL& $  1.858_{-  0.038 }^{+  0.039}$ &$  1.700_{-  0.039 }^{+  0.026}$   &$  1.664_{-  0.337 }^{+  0.559}$ &$-11.060_{-  0.005 }^{+  0.005}$  &1.14/ 142   &\\
S4 0954$+$65 &LSP  &2007.09.30   &PL & $  1.967_{-  0.020 }^{+  0.020}$  &  &  &$-11.517_{-  0.007 }^{+  0.007}$ &1.03/123   &\\
             & &   &LP & $1.966_{-0.067 }^{+0.067}$    &  &$  0.001_{-0.068 }^{+0.069}$ &$-11.517_{-0.009 }^{+0.009}$ &1.03/122 &1(0.319)\\
             & &2007.09.30   &PL & $  1.939_{-  0.062 }^{+  0.064}$  &  &  &$-11.584_{-  0.022 }^{+  0.022}$ &0.90/  48   &\\
             & &   &LP & $1.826_{-0.235 }^{+0.228}$   & &$  0.118_{-0.230 }^{+0.241}$ &$-11.593_{-0.029 }^{+0.028}$ &0.91/47  &0.473(0.495)\\
ON 231$^{*}$ &ISP  &2002.06.26$^{*}$   &PL & $  2.766_{-  0.022 }^{+  0.022}$  &  &  &$-11.466_{-  0.005 }^{+  0.005}$ &1.15/101    &\\
             & &   &LP & $2.837_{-0.063 }^{+0.062}$ &  &$ -0.095_{-0.037 }^{+0.039}$ &$-11.461_{-0.007 }^{+0.007}$ &1.02/100  &12.745(5.503$\times10^{-4}$)\\
             & &   &BKL&$2.784_{-0.023 }^{+0.023}$  &$2.032_{-2.203 }^{+0.331}$  &$4.328_{-0.511 }^{+1.873}$ &$-11.452_{-0.008 }^{+0.008}$ &1.03/99 &6.883(0.001)\\
             & &2008.06.14   &PL & $  2.743_{-  0.013 }^{+  0.013}$  &  &  &$-11.064_{-  0.003 }^{+  0.003}$ &0.97/126     &\\
             & &    &LP& $2.691_{-0.039 }^{+0.039}$   & &$ 0.069_{-0.049 }^{+0.049}$ &$-11.067_{-0.004 }^{+0.004}$ &0.93/125  &5.37(0.022)\\
3C 279      &LSP  &2009.01.21   &PL & $  1.775_{-  0.012 }^{+  0.012}$  &  &  &$-10.916_{-  0.005 }^{+  0.005}$ &1.12/ 145   &\\
            &  &   &LP & $1.807_{-0.041 }^{+0.041}$ &   &$ -0.032_{-0.039 }^{+0.039}$ &$-10.914_{-0.005 }^{+0.005}$ &1.11/144  &2.306(0.131)\\
            &  &2011.01.18   &PL & $  1.764_{-  0.005 }^{+  0.005}$  &  &  &$-10.892_{-  0.002 }^{+  0.002}$ &1.28/ 166    &\\
            &  &   &LP & $1.796_{-0.017 }^{+0.017}$     & &$ -0.033_{-0.016 }^{+0.017}$ &$-10.889_{-0.002 }^{+0.002}$ &1.22/165  &9.164(0.0029)\\
PKS 1334$-$127 &LSP  &2003.01.31   &PL & $  1.810_{-  0.025 }^{+  0.025}$  &  &  &$-11.354_{-  0.009 }^{+  0.009}$ &1.11/ 112    &\\
               & &   &LP & $1.959_{-0.083 }^{+0.082}$    & &$ -0.149_{-0.078 }^{+0.079}$ &$-11.344_{-0.011 }^{+0.011}$ &1.03/111  &9.699(0.0023)\\
3C 454.3      &LSP  &2006.07.02   &PL & $  1.566_{-  0.016 }^{+  0.016}$  &  &  &$-10.996_{-  0.007 }^{+  0.006}$ &0.96/137   &\\
              &  &   &LP & $1.687_{-0.060 }^{+0.059}$   &   &$ -0.112_{-0.053 }^{+0.053}$  &$-10.988_{-0.007 }^{+0.007}$ &0.88/136 &13.455(3.49$\times$10$^{-4}$)\\
              &  &2007.05.23   &PL & $  1.580_{-  0.019 }^{+  0.019}$    &  & &$-10.400_{-  0.008 }^{+  0.008}$ &1.10/129   & \\
              &  &   &LP & $1.567_{-0.073 }^{+0.072}$    &  &$  0.012_{-0.065 }^{+0.066}$  &$-10.401_{-0.009 }^{+0.009}$ &1.11/128  &-0.162(999)\\
              &  &2006.12.18   &PL & $  1.635_{-  0.017 }^{+  0.017}$    & &  &$-10.892_{-  0.007 }^{+  0.006}$ &1.14/136   & \\
              &  &   &LP & $1.706_{-0.061 }^{+0.061}$    &   &$ -0.067_{-0.055 }^{+0.056}$  &$-10.887_{-0.008 }^{+0.007}$ &1.12/135  &3.429(0.066)\\
              &  &2007.05.31   &PL & $  1.598_{-  0.006 }^{+  0.006}$   & &  &$-10.421_{-  0.002 }^{+  0.003}$ &0.89/164  & \\
              &  &   &LP & $1.602_{-0.024 }^{+0.023}$    &  &$ -0.004_{-0.021 }^{+0.021}$  &$-10.421_{-0.003 }^{+0.003}$ &0.89/163 &1(0.319)\\
BL Lac$^{*}$  &ISP &2007.07.10$^{*}$  &PL & $  2.163_{-  0.014 }^{+  0.015}$  &  &  &$-10.727_{-  0.004 }^{+  0.004}$ &2.15/147   &\\
              &  &   &LP & $2.549_{-0.048 }^{+0.048}$  & &$ -0.365_{-0.043 }^{+0.043}$ &$-10.705_{-0.004 }^{+0.005}$ &0.95/146  &186.7(6.786$\times$10$^{-28}$)\\
              &  &   &BPL& $  2.574_{-  0.081 }^{+  0.092}$  &$  2.054_{-  0.023 }^{+  0.022}$   &$  1.256_{-  0.090 }^{+  0.113}$ &$-10.708_{-  0.004 }^{+  0.004}$  &0.88/ 145  &\\
              &  &2007.12.05$^{*}$   &PL & $  2.159_{-  0.016 }^{+  0.016}$  &  &  &$-10.833_{-  0.004 }^{+  0.004}$ &2.21/ 143  &\\
              &  &   &LP & $2.601_{-0.051 }^{+0.051}$  & &$ -0.418_{-0.046 }^{+0.046}$ &$-10.806_{-0.005 }^{+0.005}$ &0.80/142  &253.0(2.53$\times$10$^{-31}$)\\
              &  &   &BPL& $  2.455_{-  0.077 }^{+  0.067}$  &$  1.998_{-  0.050 }^{+  0.031}$   &$  1.555_{-  0.148 }^{+  0.291}$ &$-10.811_{-  0.005 }^{+  0.005}$  &0.92/ 141   &\\
              &  &2008.01.08$^{*}$   &PL & $  2.108_{-  0.014 }^{+  0.014}$  &  &  &$-10.812_{-  0.004 }^{+  0.004}$ &3.05/ 150   &\\
              &  &   &LP & $2.592_{-0.046 }^{+0.045}$  & &$ -0.455_{-0.040 }^{+0.040}$ &$-10.783_{-0.004 }^{+0.004}$ &1.00/149  &308.5(4.0$\times$10$^{-38}$)\\
              &  &   &BPL& $  2.483_{-  0.130 }^{+  0.066}$ &$  1.952_{-  0.071 }^{+  0.025}$  &$  1.433_{-  0.101 }^{+  0.438}$  &$-10.789_{-  0.004 }^{+  0.004}$  &1.21/ 148   &\\  \hline
\end{tabular}}

$\Gamma_{1}$: Low energy spectral index; $E_{b}$: Break Energy; $b$: curvature; Flux in ergs/sec/cm$^{2}$; $\Gamma_{2}$: High  energy spectral index;
$\chi_{Red}^{2}$: Reduced $\chi^{2}$; dof: degree of freedom; $F$: statistics value for $F$-test and $p$: is its probability;
* : Sources with significant negative curvature ($>$99\%).
\end{table*}

\section{ Sample Selection and Data Analysis}
The sample of the blazars are selected from the catalogue of TeV sources (TeVCat\footnote{(TeVCat online catalogue provided by Scott Wakely 
\& Deirdre Horan (http://tevcat.uchicago.edu/)}). For our sample, we selected all ISPs and LSPs from the XMM-Newton catalogue and 
considered all observations since its launch. The blazar sample and their observation log is provided in Table 1.

The blazars in our sample are observed by the European Photon Imaging Camera (EPIC) on board the XMM-Newton satellite (Jansen et al.\ 2001). The EPIC is composed of three co-aligned X-ray telescopes which simultaneously observe a source by accumulating photons in the three CCD-based instruments: the twins MOS 1 and MOS 2 and the pn (Turner et al.\ 2001; Str\"{u}der et al.\ 2001). The EPIC instrument provides imaging and spectroscopy in the energy range from 0.2 to 15 keV with a good angular resolution (PSF = 6 arcsec FWHM) and a moderate spectral resolution ($E/\Delta E \approx 20-50$). We consider here only the EPIC-pn data as it is most sensitive and  less affected by the photon pile-up effects.

We used the XMM-Newton Science Analysis System (SAS) version 14.0.0 for the light curve extraction and spectral analysis. The observation summary or Observation Data File (ODF) and the calibration index file (CIF) are generated using updated calibration data files or Current Calibration Files (CCF) following ``The XMM$-$Newton ABC Guide'' (version 4.6, Snowden et al. 2013). XMM$-$Newton EPCHAIN pipeline is used to generate the event files. In order to identify intervals of flaring particle background, we extracted the high energy (10 keV $< E <$ 12 keV) light curve for the full frame of the exposed CCD and found background flares in few light curves. The flaring portions are removed in affected light curves. Pile up effects are examined for each observation by using the SAS task EPATPLOT. We found that the observations are not affected by the pile-up effects. We read out source photons recorded in the entire 0.3 $-$ 10 keV energy band, using a circle varying between 30--40 arcsec
radius centered on the source. These radii have been chosen to sample most of the point spread function according to the observing mode. Background photons were read out from a circular region with an area comparable to the source region, located about 180 arcsec off the source on the same chip set. Redistribution matrices and ancillary response files were produced using the SAS tasks {\it  rmfgen} and {\it arfgen}. The pn spectra were created by the SAS tool {\it XMMSELECT} and grouped to have at least 30 counts in each energy bin to ensure the validity of $\chi^{2}$ statistics. The X-ray spectra are expected to be significantly affected by the instrumental uncertainties at energies below 0.5 keV, hence we consider only the 0.6 $-$ 10 keV energy band for our studies. 

\begin{table*}
{\caption {Best fit spectral parameters for the power law, broken power law and log parabolic model for those sources where intrinsic absorption
is required or free $N_{H}$ provided better fit.} }
\noindent
\scalebox{.78}{
\begin{tabular}{ccccccccccc} \hline \hline

Source  &Blazar &Date of         &Model   &$\Gamma_{1}$  &$\Gamma_{2}$ &$E_{b}$/b    &log$_{10}$Flux    &$N_{1}$    &$\chi_{Red}^{2}$/dof &F(p) \\
        &Class  &observation     &        &              &             &(KeV)        &                  &$10^{22}cm^{2}$     &                 & \\ \hline

PKS 0235$+$164$^{*}$ &LSP &2002.02.10$^{*}$   &PL &$  2.306_{-  0.027 }^{+  0.027}$ & & &$-11.047_{-  0.005 }^{+  0.005}$ &$  0.102_{-  0.022 }^{+  0.022}$ &1.32/ 132     &\\
            &  &   &LP   &$2.763_{-0.153 }^{+0.153}$  & &$ -0.344_{-0.112 }^{+0.113}$ &$-11.040_{-0.005 }^{+0.005}$ &$0.239_{-0.051 }^{+0.051}$ &1.14/131 &21.842(7.253$\times$10$^{-6}$)\\
            &  &   &BPL  &$2.387_{-0.038 }^{+0.045}$  &$  1.931_{-  0.160 }^{+  0.155}$  &$  4.029_{-  0.772 }^{+  0.612}$ &$-11.038_{-  0.006 }^{+  0.005}$  &$ 0.141_{-  0.026 }^{+  0.028}$ &1.10/ 130 &\\
            &  &2004.01.18   &PL &$  1.607_{-  0.043 }^{+  0.044}$ & & &$-11.538_{-  0.011 }^{+  0.011}$ &$  0.155_{-  0.048 }^{+  0.050}$ &1.01/ 127 &\\
            &  &   &LP &$ 1.706_{-  0.282 }^{+  0.281}$ & &$ -0.069_{-  0.193 }^{+  0.195}$ &$-11.536_{-  0.012 }^{+  0.012}$ & $  0.190_{-  0.108 }^{+  0.110}$ &1.02/ 126  &-0.245(999) \\
            &  &2004.08.02   &PL &$  1.590_{-  0.064 }^{+  0.066}$  & & &$-11.544_{-  0.017 }^{+  0.016}$ &$  0.201_{-  0.074 }^{+  0.077}$ &0.89/  89 &\\
            &  &   &LP  &$  1.556_{-  0.436 }^{+  0.438}$ & &$  0.024_{-  0.300 }^{+  0.303}$ &$-11.545_{-  0.019 }^{+  0.018}$ &$  0.189_{-  0.163 }^{+  0.169}$ &0.90/  88 &0.011(0.916) \\
            &  &2005.01.28   &PL &$  1.702_{-  0.092 }^{+  0.096}$  & & &$-11.921_{-  0.023 }^{+  0.022}$ &$  0.129_{-  0.098 }^{+  0.103}$ &0.98/  70 &     \\
            &  &   &LP &$  2.129_{-  0.622 }^{+  0.622}$ &  &$ -0.305_{-  0.433 }^{+  0.441}$ &$-11.911_{-  0.027 }^{+  0.026}$ & $  0.274_{-  0.231 }^{+  0.241}$ &0.97/  69 &1.721(0.194)\\
PKS 0521$-$365$^{a}$ &LSP &2005.08.14   &PL &$  1.771_{-  0.023 }^{+  0.023}$ & & &$-10.729_{-  0.005 }^{+  0.005}$ &$0.036_{-  0.008 }^{+  0.008}$ &0.99/ 138 &\\
               &   &          &LP &$1.792_{-0.162 }^{+0.162}$ & &$ -0.015_{-0.113 }^{+0.113}$ &$-10.728_{-0.012 }^{+0.012}$ &$  0.038_{-0.021 }^{+0.022}$ &0.99/137 &\\
\\ \hline

\end{tabular}}

$\Gamma_{1}$: Low energy spectral index; $E_{b}$: Break Energy; $b$: curvature; Flux in ergs/sec/cm$^{2}$; $\Gamma_{2}$: High  energy spectral index;
$\chi_{Red}^{2}$: Reduced $\chi^2$; $N_{1}$: intrinsic absorption or absorption due to the intervening medium;
 dof: degree of freedom; $F$: statistics value for $F$-test and $p$: is its probability; * : Source with significant negative curvature ($>$99\%).
a: Best fit for free $N_{H}$.
\end{table*}

\subsection{Timing Analysis}
\subsubsection{Excess Variance and Variability Amplitude}

We calculate the excess variance (e.g. Edelson et al.\ 2002; Vaughan et al.\ 2003), which is an estimator of the intrinsic source variance over and above the underlying noise. The variance, after subtracting the excess contribution from the measurement errors is
\begin{equation}
\sigma_{\rm{NXS}}^{2} = S^{2} - \overline{\sigma_{\rm{err}}^2},
\end{equation}
where $\overline{\sigma_{\rm{err}}^{2}}$ is the mean square error,
\begin{equation}
\overline{\sigma_{\rm{err}}^{2}} = \frac{1}{N}\sum_{i=1}^{N} \sigma_{{\rm
err}, i}^{2}.
\end{equation}
The normalized excess variance is given by
$\sigma_{\rm{NXS}}^{2}=\sigma_{\rm{XS}}^{2}/\bar{x}^{2}$ and the fractional root mean square (rms) variability
amplitude ($F_{\rm{var}}$;
Edelson, Pike \& Krolik 1990; Rodriguez-Pascual et al.\ 1997) is 
\begin{equation}
F_{\rm{var}} = \sqrt{ \frac{S^{2} -
\overline{\sigma_{\rm{err}}^{2}}}{\bar{x}^{2}}}.
\end{equation}
and the error on the fractional amplitude is
\begin{eqnarray}
{err(F_{\rm{var}}) =
\frac{1}{2 F_{\rm{var}}} err(\sigma_{\rm{NXS}}^{2}) = }
\nonumber \\
{\qquad
\sqrt{ \left\{ \sqrt{\frac{1}{2N}} \cdot\frac{ \overline{\sigma_{\rm{err}}^{2}}
}{  \bar{x}^{2}F_{\rm{var}} }  \right\}^{2}
+
\left\{ \sqrt{\frac{\overline{\sigma_{\rm{err}}^{2}}}{N}}
\cdot\frac{1}{\bar{x}}  \right\}^{2}  }.
}
\end{eqnarray}

We calculated $F_{\rm{var}}$ for all of our light curves and the results are presented in Table 2.

\subsubsection{Lomb-Scargle periodogram}

The Lomb-Scargle periodogram \cite[][]{1976Ap&SS..39..447L,1982ApJ...263..835S} is used to determine the power spectral density (PSD) generated by a physical process manifested through the observed light curve. It is an effective measure of variability in un-evenly sampled light curves including those with short data gaps such as those encountered here. The periodogram evaluated from a light curve $x(t_k)$ spanning $N$ points is determined by a least squares fit to a mean subtracted time series using
\begin{equation}
x(t_k) = C_1 \sin 2 \pi f_j (t_k-\tau)+C_2 \cos 2 \pi f_j (t_k-\tau)
\end{equation}
and is given by \cite[e.g.][]{1986ApJ...302..757H}
\begin{align}
P (f_j)&= \frac{1}{2 \sigma^2} \left[\frac{(\sum^{N}_{k=1} (x(t_k)-\bar{x}) \cos 2 \pi f_j (t_k-\tau))^2}{\sum^{N}_{k=1} \cos^2 2 \pi f_j (t_k-\tau)}\right . \nonumber \\ 
&+ \left . \frac{(\sum^{N}_{k=1} (x(t_k)-\bar{x}) \sin 2 \pi f_j (t_k-\tau))^2}{\sum^{N}_{k=1} \sin^2 2 \pi f_j (t_k-\tau)}\right] ,
\end{align}
where $\tau$ is a time shift parameter and is given by
\begin{equation}
\tan{(4 \pi f_j \tau)}=\frac{\sum^{N}_{k=1} \sin(4 \pi f_j t_k)}{\sum^{N}_{k=1} \cos(4 \pi f_j t_k)}
\end{equation}
where $\bar{x}$ is the mean of the light curve and $P (f_j)$ is evaluated at frequencies $f_j = j/(t_N-t_1)$ where $j = 1, 2, .., N/2$ and $(t_N-t_1)$ is the total duration of the observation. The periodogram is evaluated using the algorithm presented in \cite{1989ApJ...338..277P} in order to achieve fast computational speeds.

We performed a linear interpolation of the light curves to populate small ($<$ 200 - 300 s; 2-3 points) data gaps and sample the light 
curves at regular intervals $\Delta t = 100$ s. The LSP is the Fourier periodogram for an evenly sampled light curve which leads to a proper 
estimation of the underlying PSD as noted in \cite{2017ApJ...837..127G}, where it was inferred that the LSP of unevenly sampled light curves 
tended to be dominated by higher temporal frequencies thus yielding systematically flatter PSD slopes.
We use two competing parametric models to infer the PSD shape, a power law model
\begin{equation}
I(f_j) = A f^{\mu}_j+C,
\end{equation}
 with amplitude $A$, slope $\mu$, and a constant Poisson noise $C$, and a bending power law model
\begin{equation}
I(f_j) = A f^{-1}_j \left(1+(f_j/f_b)^{-\mu-1}\right)^{-1}+C,
\end{equation}
 with amplitude $A$, slope $\mu$, bend frequency $f_{b}$, and a constant Poisson noise $C$. By maximizing the likelihood function
\begin{equation}
\mathcal{L} (\theta_k) = \prod^{N/2}_{j = 1} \frac{1}{I(f_j,\theta_k)} e^{-P(f_j)/I(f_j,\theta_k)},
\end{equation}
we determine the parameters of the model $\theta_k$. The Akaike informaion criteria (AIC) is then calculated for each competing model and the best fit model is that with higher likelihood of describing the data effectively. The best fit model parameters are employed in the  simulation of 1000 random realizations of light curves with similar statistical and spectral properties as the original light curve, including the same sampling pattern, using an algorithm similar to that prescribed in \cite{1995A&A...300..707T}. The periodogram of each of these simulated light curves is determined using at the same sampling frequencies as that used for the original light curve and at each frequency bin, we construct an empirical distribution function to determine the 3$-\sigma$ model error at each ordinate. Details of the fitting procedure, model selection using the AIC and the light curve simulations procedure are presented in \cite{2015MNRAS.452.2004M,2016MNRAS.456..654M} and references therein. For a light curve populated by random Gaussian noise, its periodogram ordinates are expected to be $\chi^2_2$ distributed \cite[e.g.][]{1986ApJ...302..757H}. The residuals of the best fit PSD are thus expected to be $\chi^2_2$ distributed. From the cumulative distribution function of the $\chi^2_2$ distribution, and after accounting for the number of frequencies sampled and the model errors inferred from the simulations described above, we set a significance threshold of 99 \%, used to infer the presence of any statistically significant quasi-periodic oscillations in the data and to infer the statistical significance of each periodogram peak.

\subsection{Spectral Analysis}

We fit each spectra using following models: 
\begin{enumerate}
\item Power law model defined by $k~E^{\Gamma}$ with fixed Galactic absorption. It is characterized by a normalization $k$ and spectral index $\Gamma$. 
\item Logarithmic parabola model defined by $k~E^{(-\alpha+\beta log(E/E_{1})}$ (e.g. Landau et al. 1986; Massaro et al. 2004; 2008) with fixed Galactic absorption. It is characterized by a normalization $k$, spectral index $\alpha$, transition energy $E_{1}$ and curvature parameter $\beta$. We fix $E_{1} = 0.6$ keV, the lowest probed energy without loss of generality as the particular choice does not drastically affect the results owing to its logarithmic dependance. 
\item Broken power law model defined by $k~E^{\Gamma_{1}}$ for $E < E_{\rm break}$ and $k~E^{\Gamma_{2}}$ otherwise, with fixed Galactic absorption. This model is also characterized by a normalization $k$, two spectral indices $\Gamma_{1}$ and $\Gamma_{2}$, and a break energy $E_{\rm break}$. This model is used only to fit those spectra when there is significant concave curvature ($>$99\% significance) using the log parabolic model and is used to infer the break energy $E_{\rm break}$ in the spectrum, which is the point at which synchrotron component turns over to inverse Compton component. 
\end{enumerate}

Some FSRQs in our sample are at high redshift and can have an intervening absorption due to gravitational microlensing and damped Lyman-$\alpha$ 
systems i.e. AO 0235$+$164, 3C 454.3 (e.g. Foschini et al. 2006, Raiteri et al. 2008). For these blazars, all models are fit after including an 
absorber along the line of sight in addition to the Galactic absorption density. 

The spectra are fit using the XSPEC software package version 12.8.1. The XSPEC routine ``cflux'' is used to obtain unabsorbed flux and its error. The Galactic absorption is taken from the survey by Willingale et al. (2013) which includes both the atomic gas column density $N_{HI}$ and the molecular hydrogen column density $N_{H_{2}}$. The $N_{HI}$ is adopted from the Leiden Argenine Bonn Survey (Kalberla et al. 2005) which is obtained by merging two surveys: the Leiden/Dwingeloo Survey (Hartmann \& Burton 1997) and the Instituto Argentino de Radioastronoma Survey (Arnal et al. 2000; Bajaja et al. 2005). The $N_{H_{2}}$ is estimated using the maps of dust infrared emission by Schlegel, Finkbeiner \& Davis (1998) and the dust-gas ratio by Dame, Hartmann \& Thaddeus (2001). 

\section{Results and inferences}

\subsection{Temporal Variability}

The light curves of the blazars are analyzed to infer their normalized  excess variance and their PSD shape with the results being summarized in 
Table 1. The Lomb-Scargle periodogram analysis for some of the sources including the best fit model are plotted in Fig. 1. The variability 
amplitude shows a large spread, ranging between $0.80 \pm 2.4$ \% (PKS 0521--365) and $52.12 \pm 0.59$ \% (ON 231: 26-06-2002) with a median of 
$5.76 \pm 2.26$ indicating a low to moderate overall variability with a few outliers. This is consistent with expectations for LSP and ISP 
sources, due to their X-ray emission mainly originating from the inverse Compton scattering of seed photons by the lower energy tail of the 
electron population and thus indicating lower variability in the $0.6 - 10$ keV band.

From the parametric model fit to the periodogram, the power law model is inferred to be the best fit PSD shape in 29/31 light curves with 
the power law index in the range $-0.01$ to $-1.12$ and many light curves being white noise dominated. This indicates that the inferred 
variability here probes only the noise floor as opposed to intrinsic source based variability which often leads to indices in the range $-1.5--2.5$
 (e.g. Mohan et al. 2014, 2015, 2016, Agarwal et al. 2017; Goyal et al. 2017 and references therein). The bending power law is the best fit model 
in the light curves of PKS 0235$+$164: 10-02-2002 and PKS 0528$+$134: 14-09-2009. However, as the bend frequency indicates that it is close to the 
edge of the observation duration, these results are similar to the power law PSD model. There are no inferred statistically significant quasi-periodic 
oscillations in any of the light curves. Thus, no conclusion can be drawn from the PSD slopes alone in relation to the typical distance of 
variability origin or the emission mechanism, necessitating the analysis of the spectra to clear these uncertainties.

\subsection{Spectral Variability}
The spectral parameters of power law, log parabolic and broken power law model fits are presented in Table 3--4. The preferred models for each
 spectra are shown in Fig 2--7. The goodness of fit of power law and log parabolic model are compared using the F-test (Bevington \& Robinson 2003) 
and values are quoted in Table 3--4. More than half of the observation (18/31) are described well with the power law model and the remaining 
13/31 are well described by the log parabolic model. In 7 (PKS 0235+164, PKS 0426-380, PKS 0537-441, S5 0716+714, OJ 287, ON 231
and BL Lacertae) of the 14 blazars, a significant negative curvature is inferred. The spectra of such sources are also fit with a broken 
power law to constrain the $E_{\rm break}$.  \\

\begin{figure*}
\centering
\includegraphics[width=0.44\textwidth]{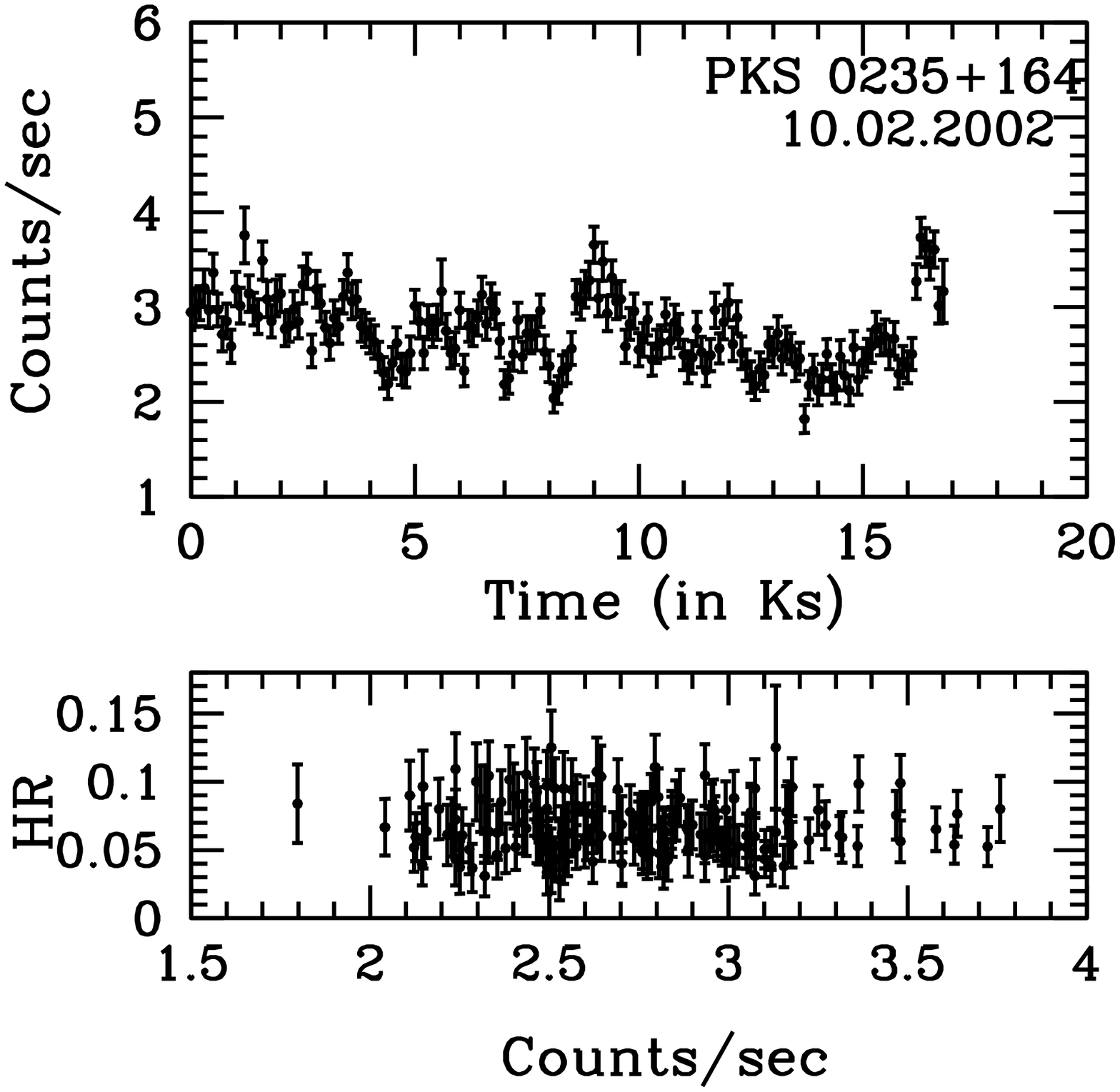}
\includegraphics[width=0.44\textwidth]{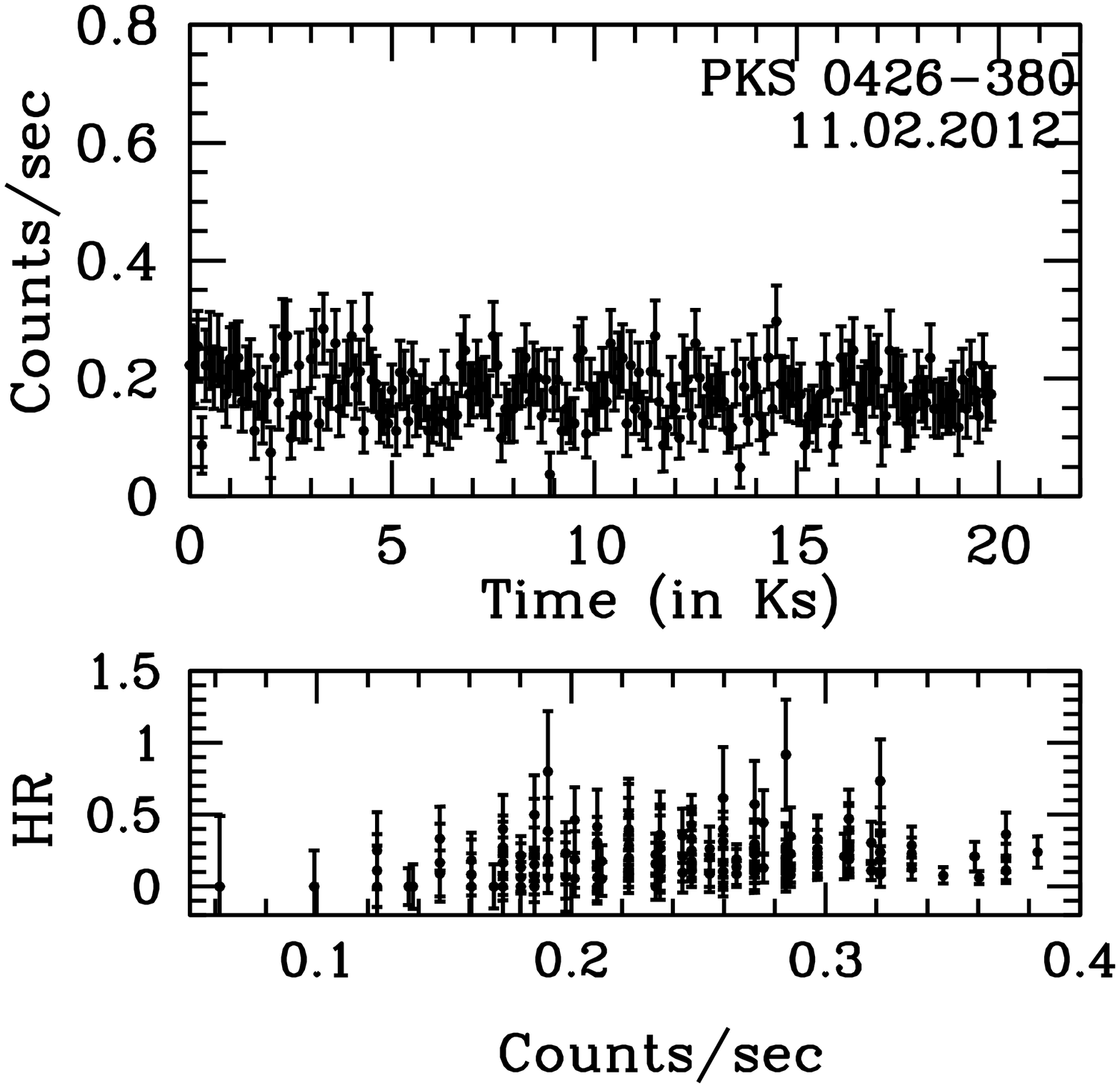}
\includegraphics[width=0.44\textwidth]{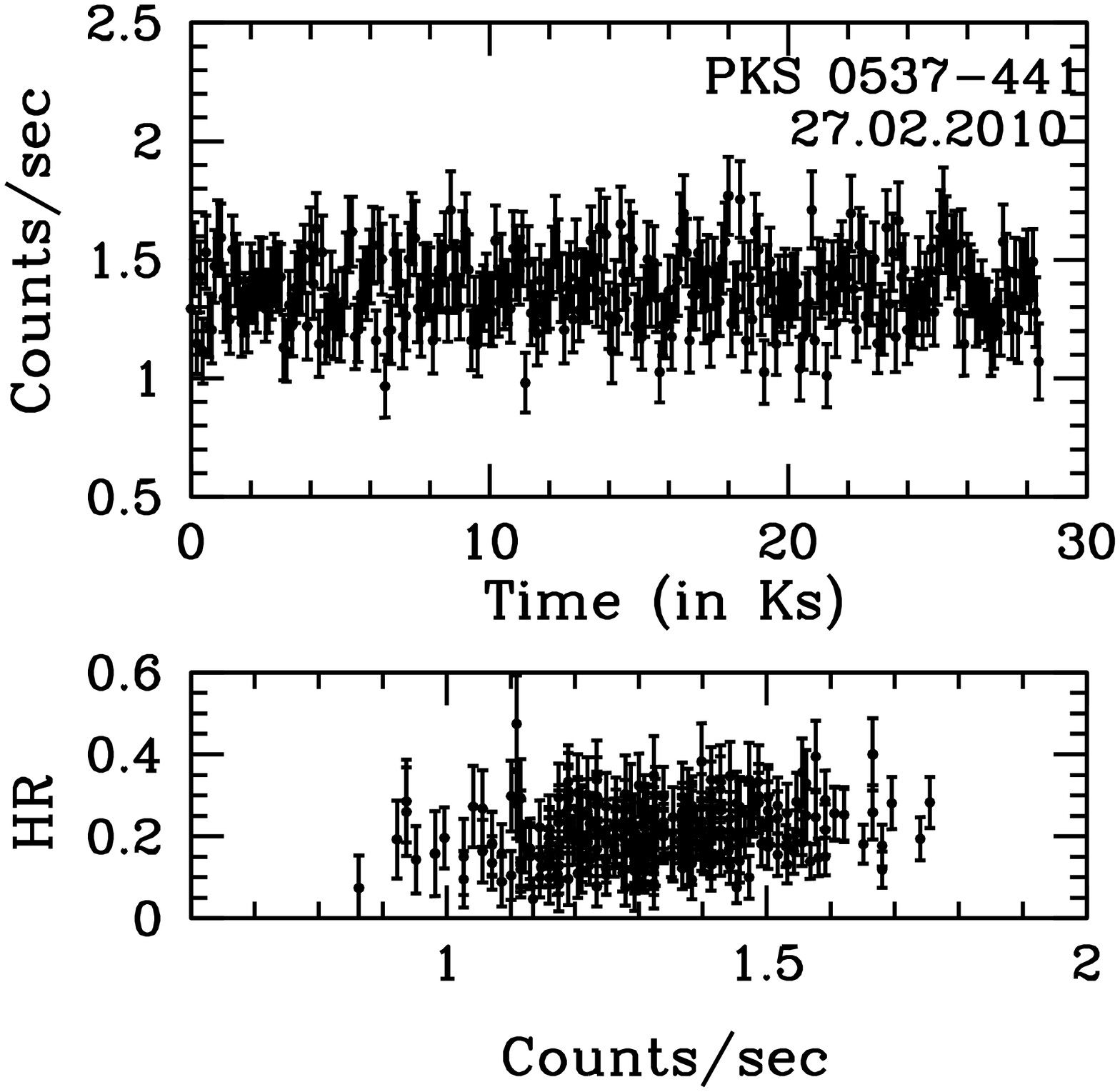}
\includegraphics[width=0.44\textwidth]{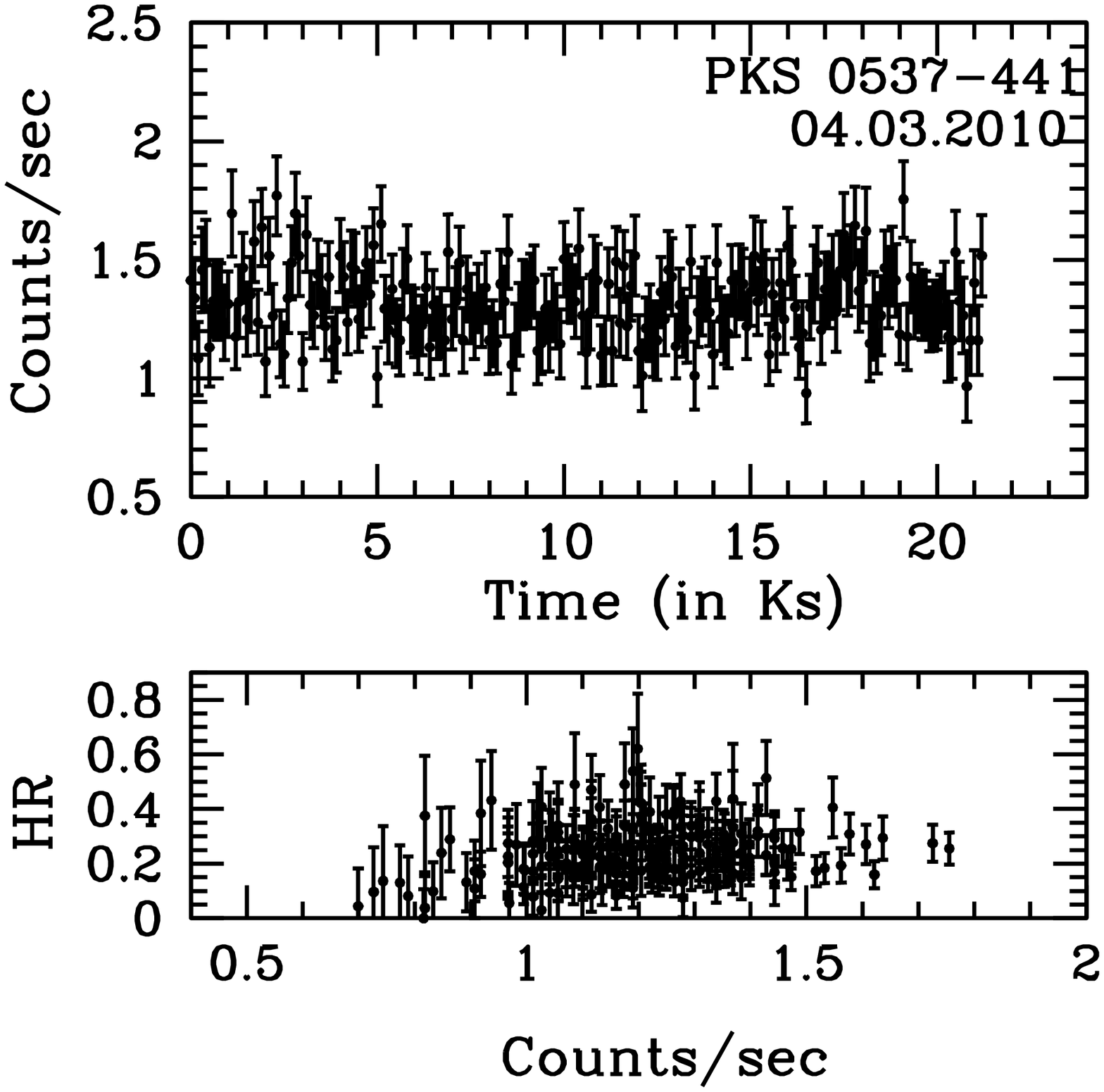}
\caption{Light curves (in the range of 0.6--10 KeV) of those blazars where we found significant spectral flattening.}
\end{figure*}

\begin{figure*}
\centering
\includegraphics[width=0.44\textwidth]{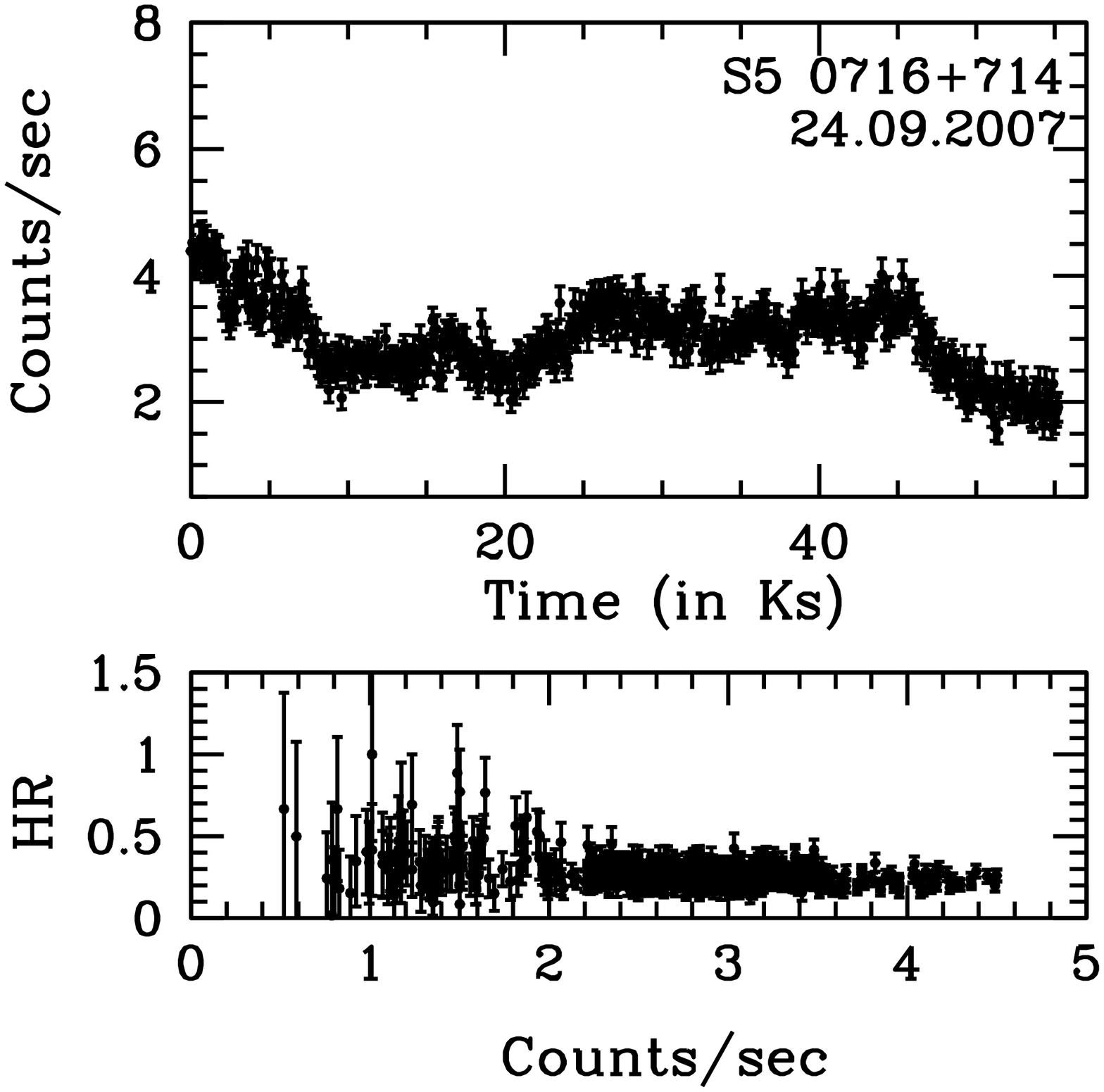}
\includegraphics[width=0.44\textwidth]{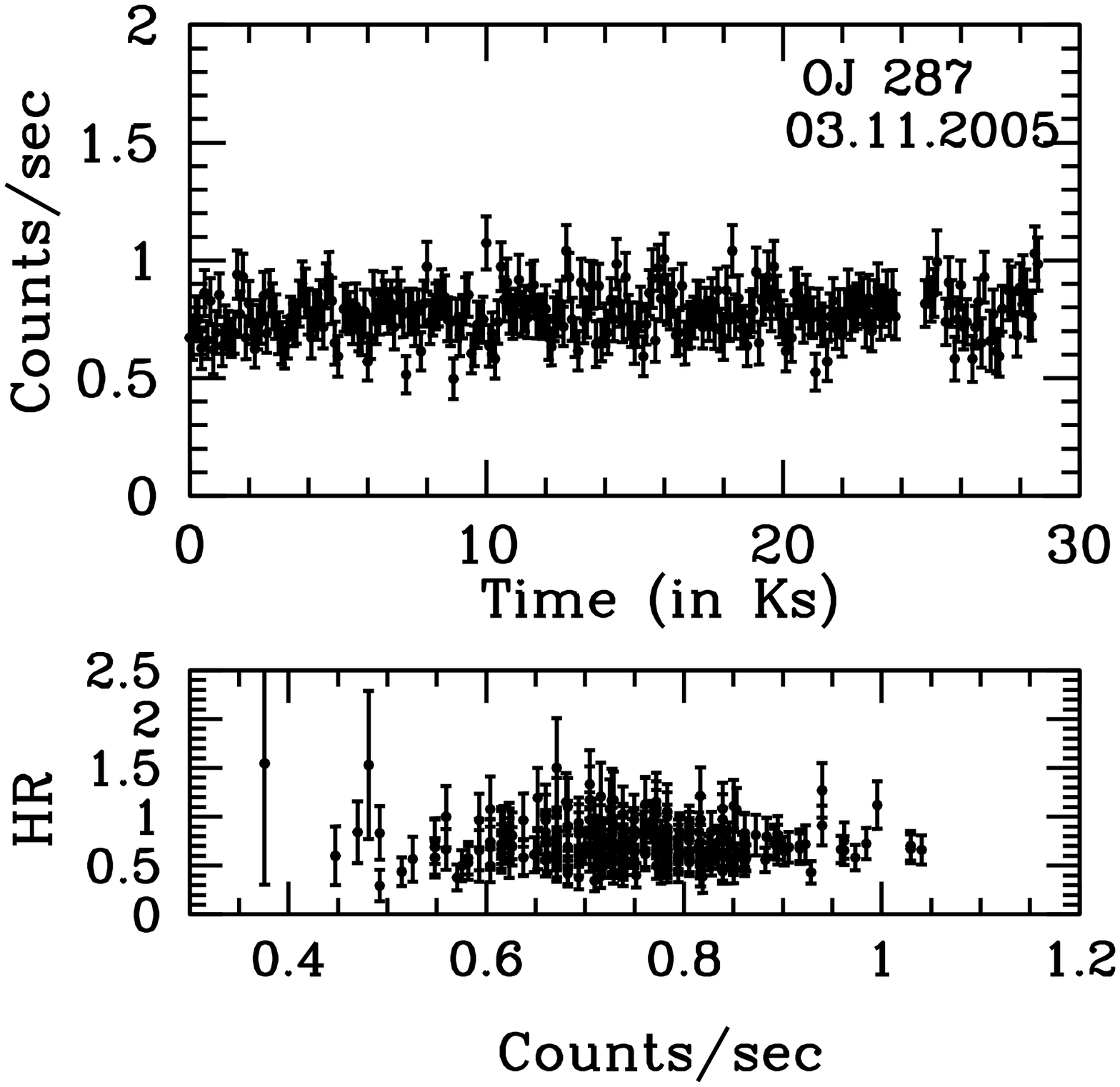}
\includegraphics[width=0.44\textwidth]{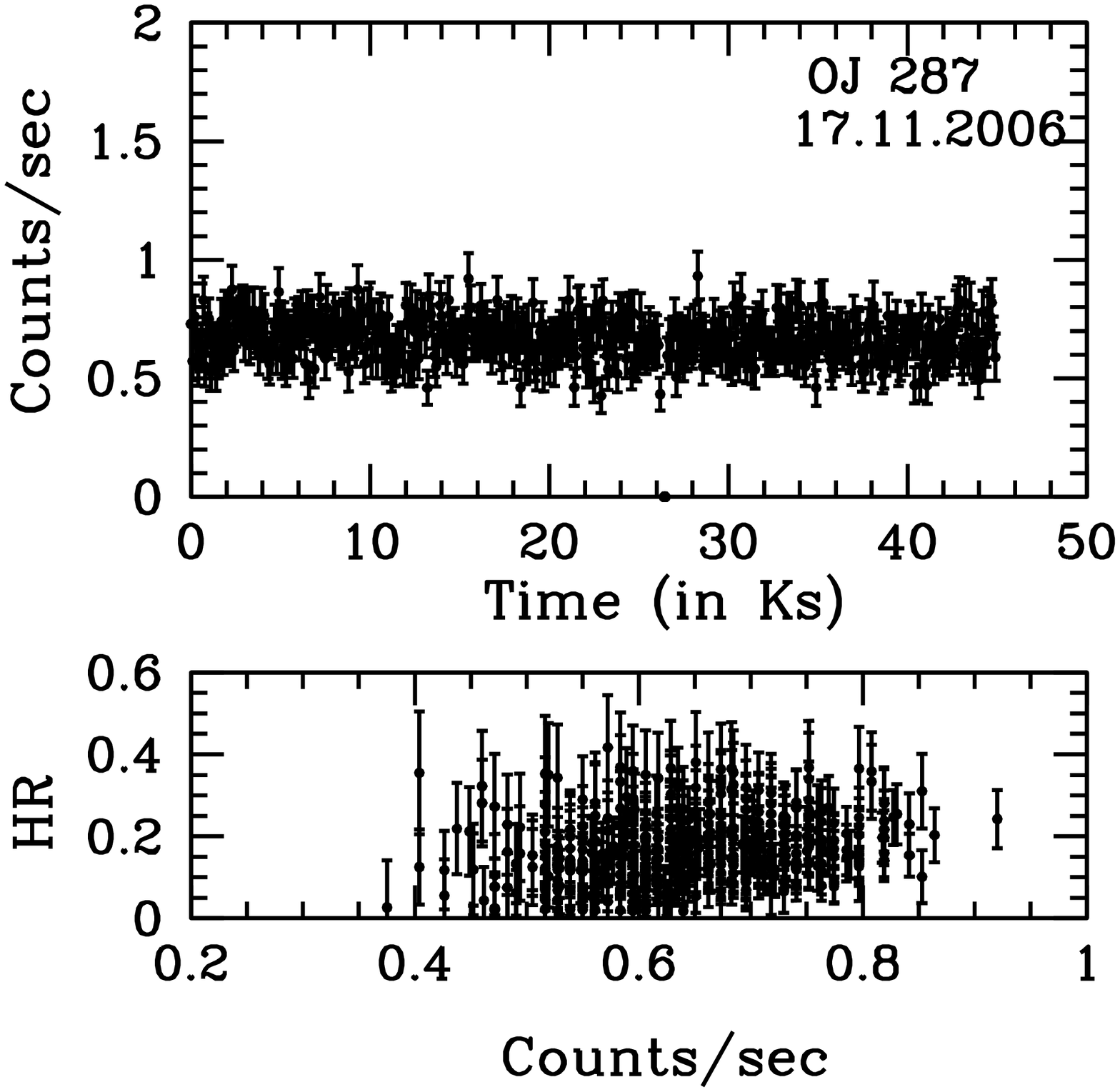}
\includegraphics[width=0.44\textwidth]{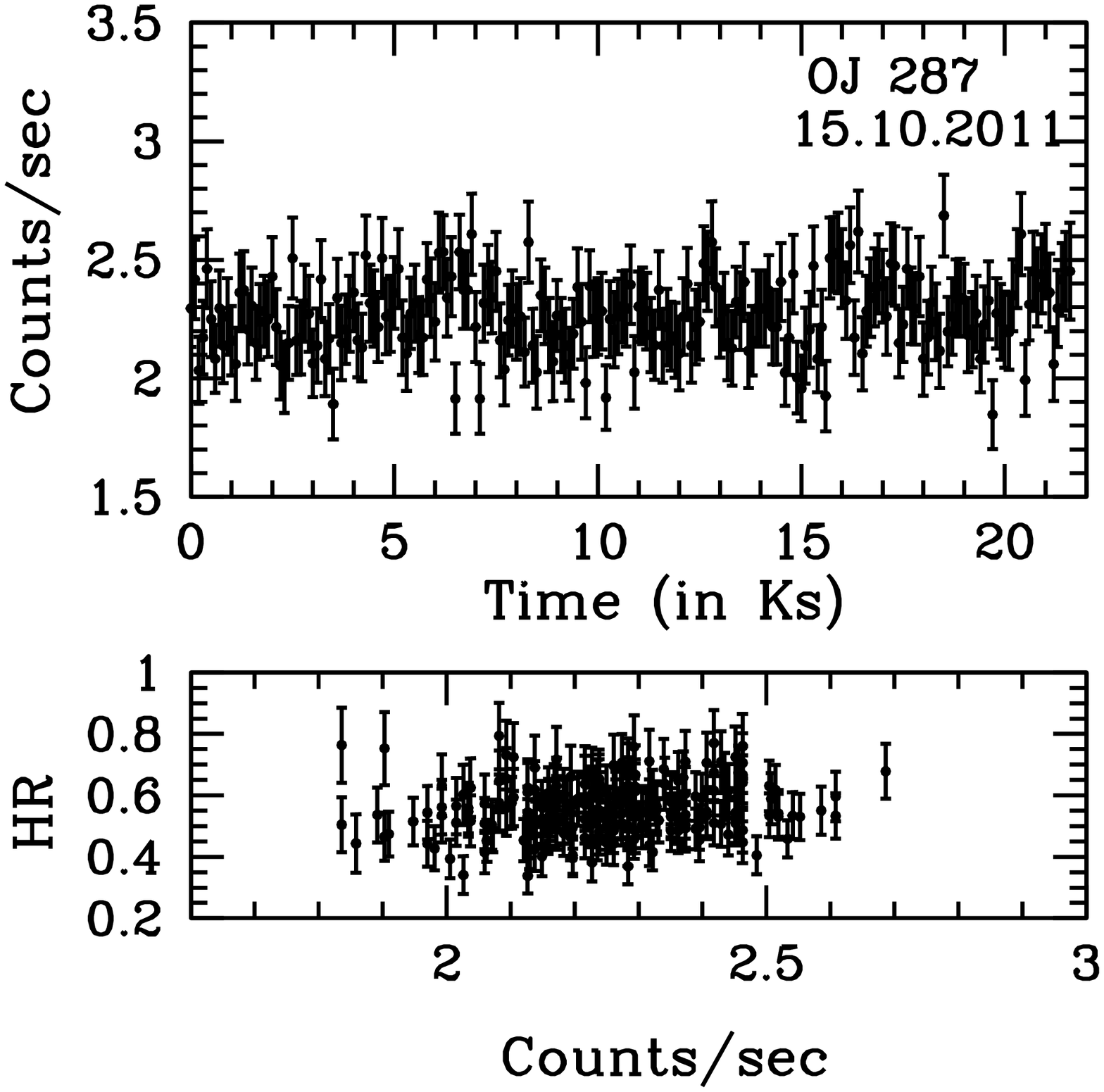}
\caption{Same as in figure 10.}
\end{figure*}

\begin{figure*}
\centering
\includegraphics[width=0.44\textwidth]{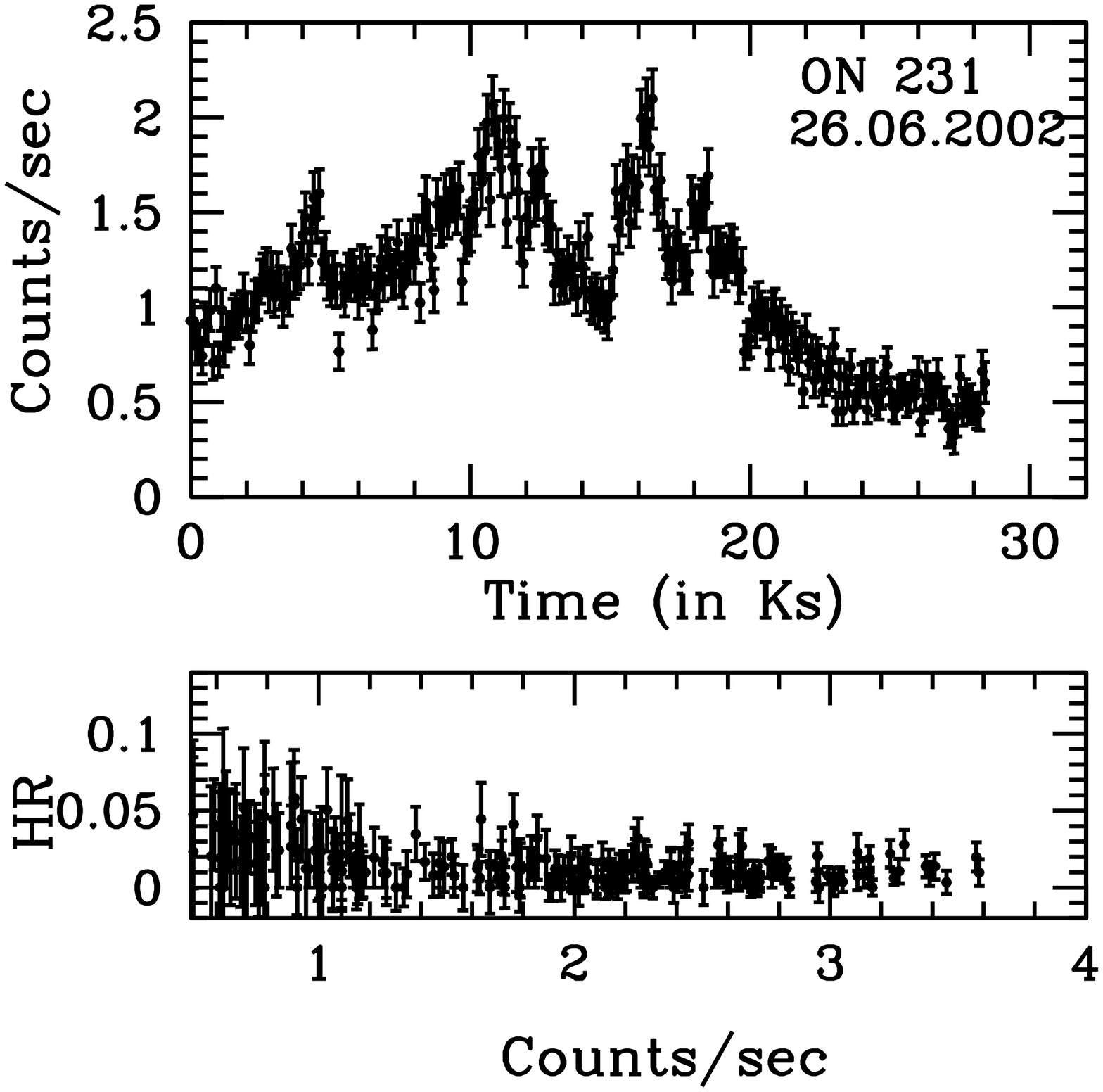}
\includegraphics[width=0.44\textwidth]{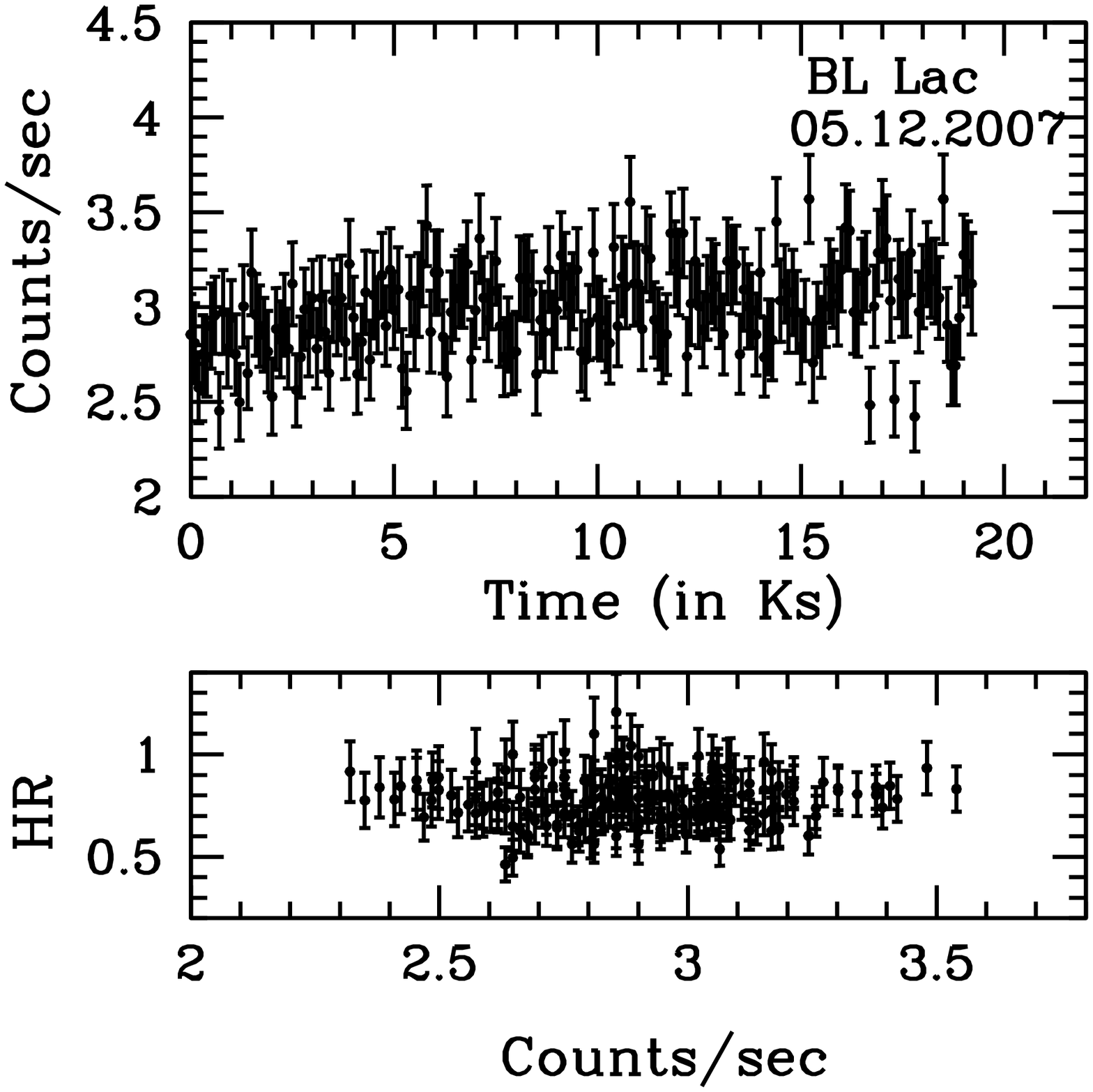}
\includegraphics[width=0.44\textwidth]{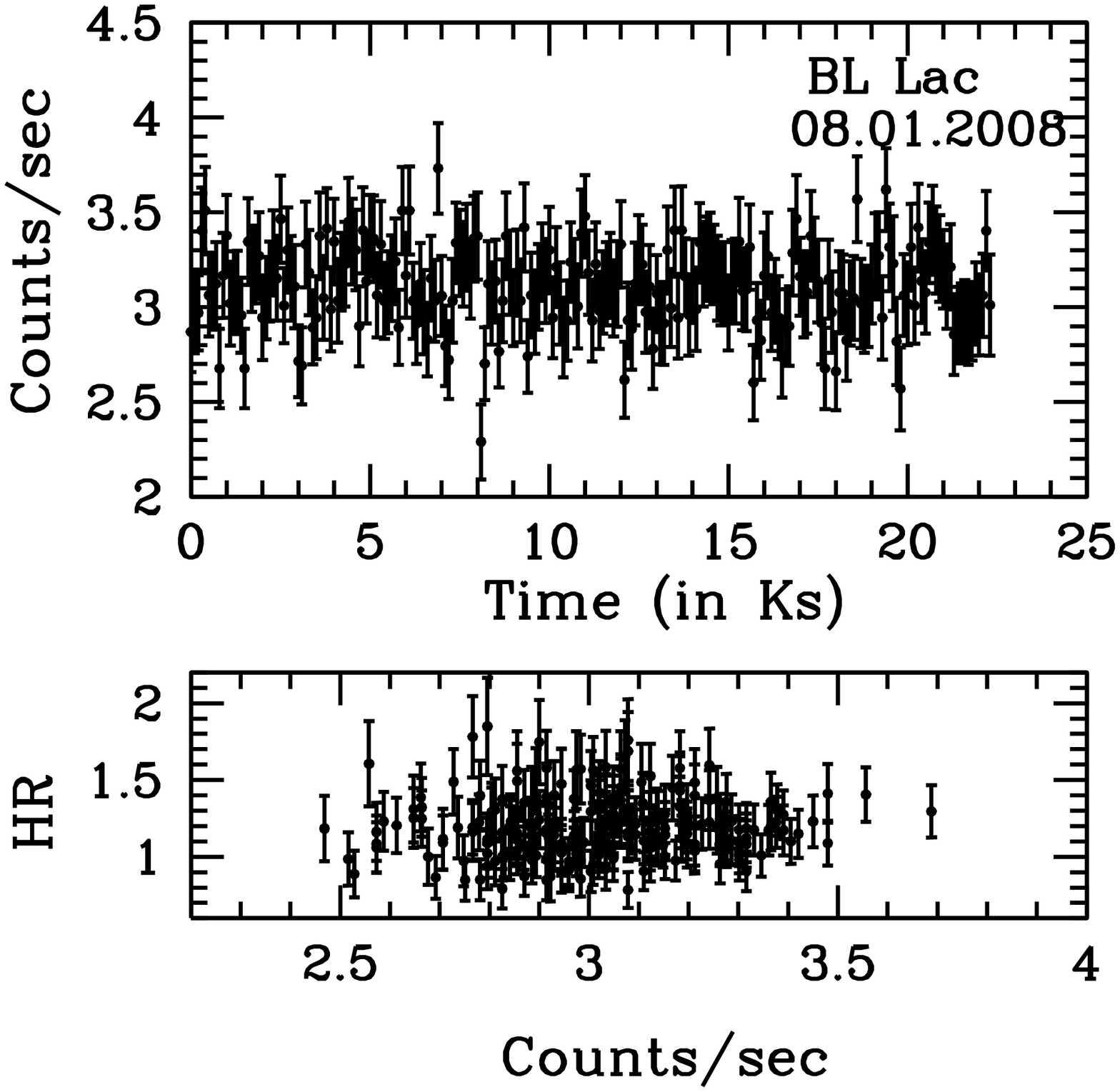}
\includegraphics[width=0.44\textwidth]{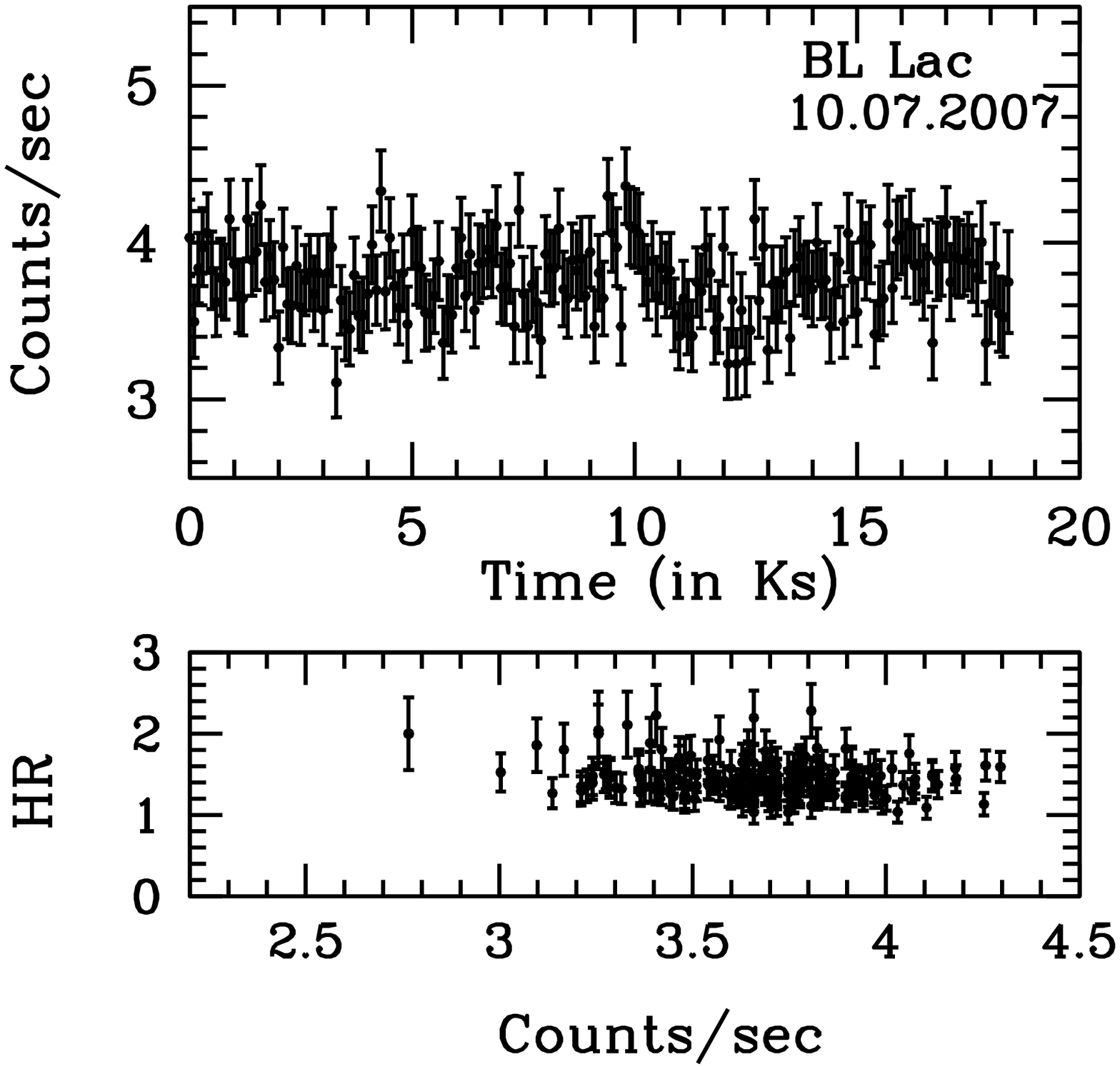}
\caption{Same as in figure 10.}
\end{figure*}

\begin{figure*}
\centering
\includegraphics[width=0.44\textwidth]{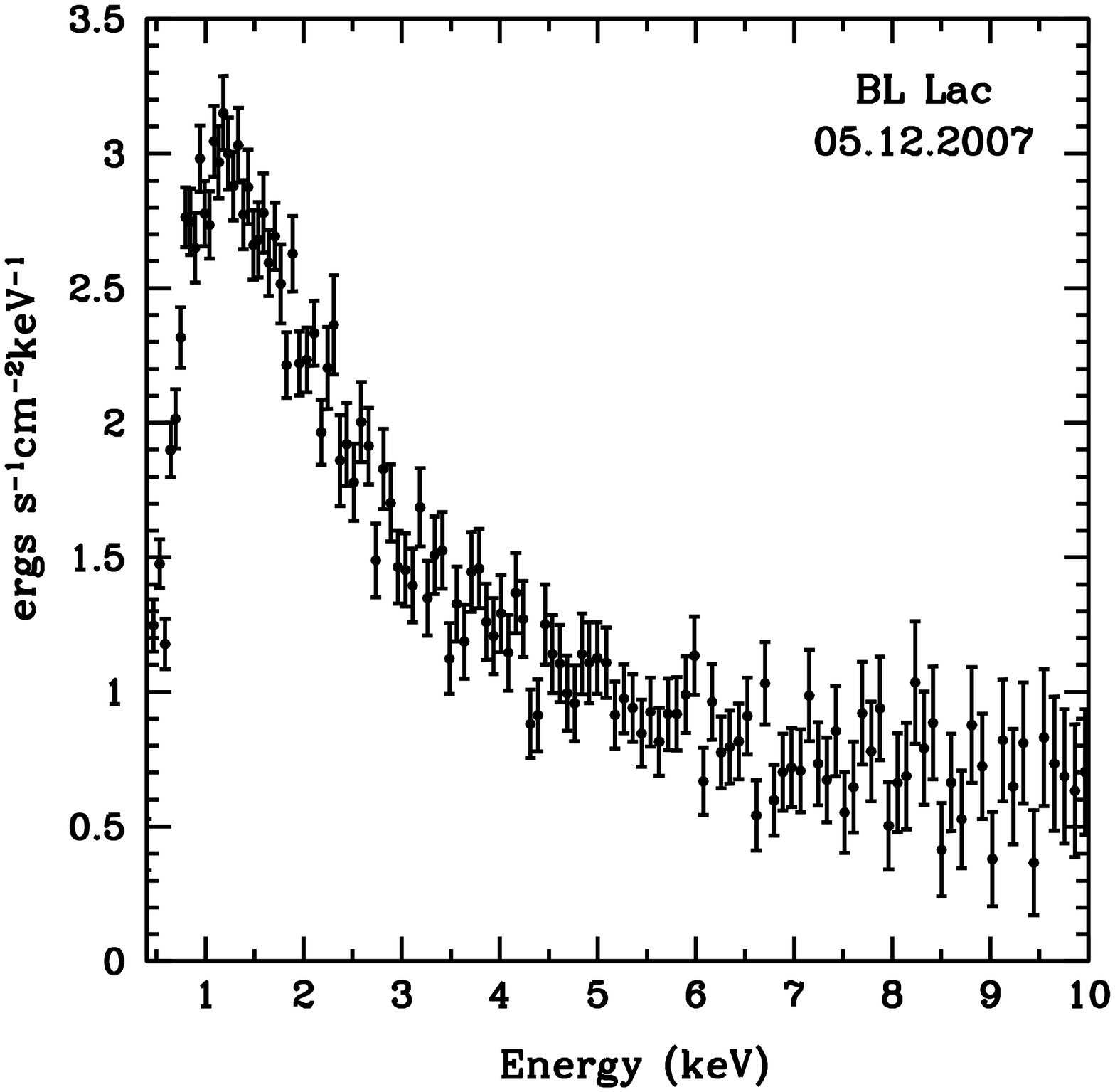}
\includegraphics[width=0.44\textwidth]{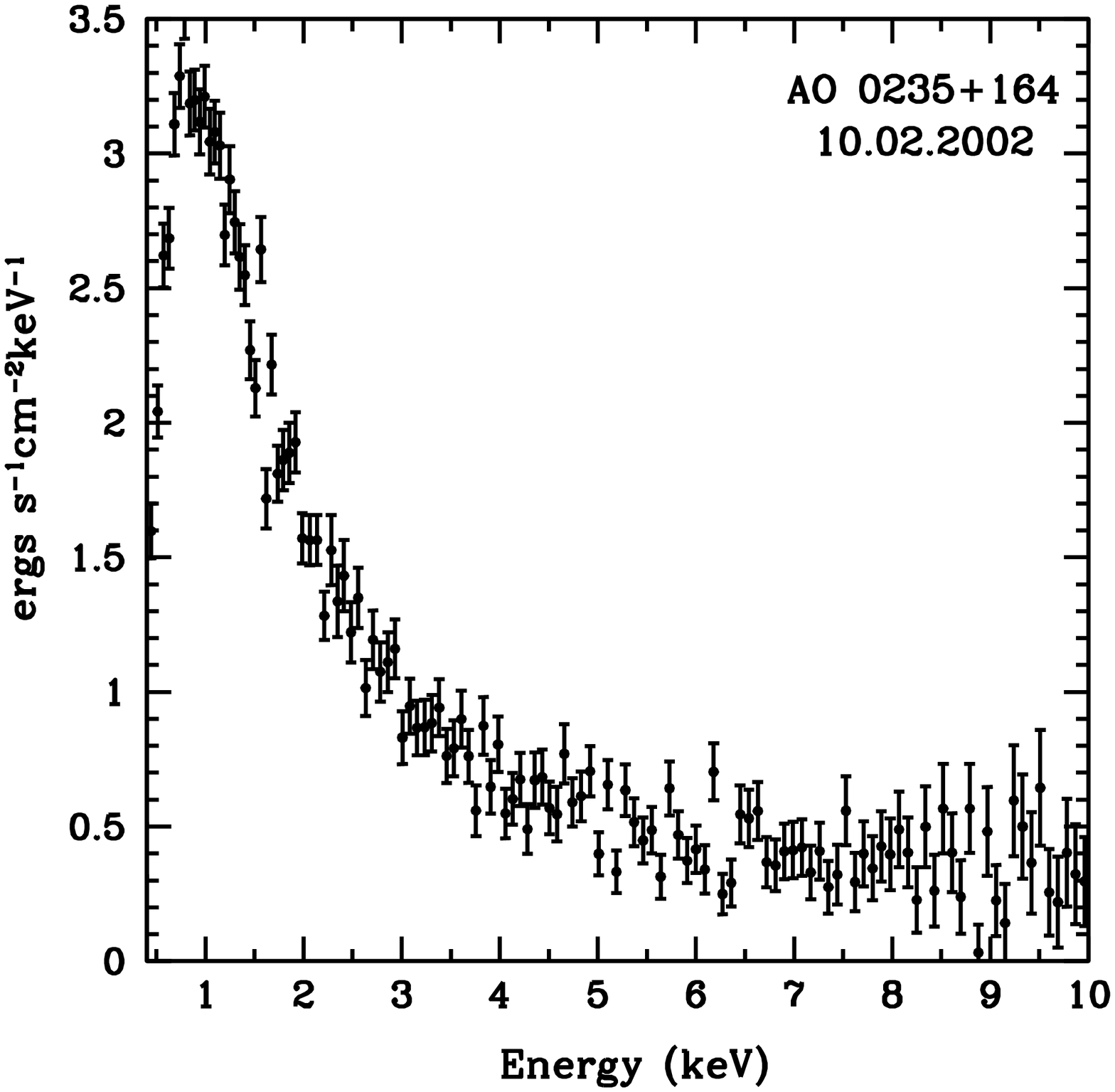}
\caption{Examples of SED of AO 0235$+$164 and BL Lac which shows negative curvature.}
\end{figure*}

\subsection{Notes on individual blazars} 

{\bf TXS 0106$+$612:}  This LSP was first discovered by Gregory \& Taylor (1981) in a radio survey
of the Galactic plane. This source is a TeV blazar (e.g. Tsujimoto et al. 2015). The X-ray spectrum of this blazar is 
better described with a power law with fixed $N_{H}$ 
rather than log parabolic model.  Also, we fit the spectra with free $N_{H}$ but did not get any improvement in the 
$\chi^{2}$ values. We did not find any significant curvature in the studied energy range.   

{\bf 3C 66A:} This source is a well known ISP (\textbf{e.g.} Aliu et al. 2009). In previous studies, significant negative curvature is found 
in the SED using {\it Swift}/XRT observations (i.e. Wierzcholska \& Wagner 2016) with break energy of $\sim$3.4 keV.
 We examined the spectra of this blazar using XMM$-$Newton 
 pointing and found that it is well described with power law model. 
The spectral index ($\Gamma$=2.45) is well matched with the previous studies. However, we did not find significant spectral curvature for this source. 
Similar results are found by Donato et al. (2005) where they did not find any significant curvature using BeppoSAX observations and found the spectrum 
to be well fitted by the power law model with $\Gamma$=2.26. We also model our spectrum by making $N_{H}$ as free parameter but
$\chi^{2}$ values did not improve significantly.

{\bf PKS 0235$+$164:} This source is a BL Lac object and a TeV $\gamma$-ray emitting blazar (e.g. Tsujimoto et al. 2015). Donato et al. (2005) 
studied the 
X-ray spectrum of this blazar 
using BeppoSAX observations and found power law to be well fitted to the spectra. Raiteri et al. (2006) analyzed XMM$-$Newton and Chandra
observations during the period 2000--2005 and found log-parabolic model to be superior than power law model.
The source has an intervening system along the line of sight at z=0.524 which likely absorbs the soft X-ray spectrum (Raiteri et al. 2006).
This intervening medium has been measured by ROSAT and ASCA obtaining a value of $N_{H}$=2.8 $\times10^{21}$ $cm^{-2}$ (Madejski et al. 1996)
and $\sim$2.4--2.6 $\times10^{21}$ $cm^{-2}$ using XMM$-$Newton observations (Raiteri et al. 2005; 2006; Foschini et al. 2006). 
We analyzed four pointings of XMM$-$Newton satellite during the period 2002--2005. During the observation performed in 2002, the source was 
in outburst state (Raiteri et al. 2005). The spectrum is found to be well fitted with log parabolic model with significant concave curvature
with break energy at around 4 keV in 2002. The source is in faint state during the observation performed in the period 2004--2005 and we 
found the spectra to be well described with a power law model with $\Gamma$=1.5--1.7.   

{\bf PKS 0426$-$380:} The source is classified as a FSRQ (i.e. Ghisellini et al. 2011; Sbarrato et al. 2012). Tanaka et al. (2013)
reported the discovery of the Very High Energy $\gamma$-ray emission from this blazar. We analyzed the XMM$-$Newton pointing of the blazar 
in 2012 and found it to be well described with a log-parabolic model which showed significant concave spectrum. The spectrum is also 
fitted with the broken power law and the upturn break energy is found to be at around 2.1 keV. We fit the spectra by making 
$N_{H}$ as free parameter but did not get any significant improvement in $\chi^{2}$ values. This the first occasion where the spectral
flattening is found for this blazar.

{\bf PKS 0521$-$365:} The X-ray spectrum of this blazar is studied by Donato et al. (2005) and found to be well characterized by
simple power law with $\Gamma$ $\sim$1.7. Foschini et al. (2006) also fit the spectrum with broken power law and found
break energy at $\sim$1.5 keV. In our studies, X-ray spectrum analyzed for one pointing of XMM$-$Newton is well fitted by the power 
law when we fixed $N_{H}$. This spectrum shows significant improvement in the model when we kept $N_{H}$ as free parameter.
$N_{H}$ is found to be 3.8 which is very close to the Galactic absorption column density [$10^{20}$ cm$^{-2}$] from Dickey \& Lockman (1990).
Since, this blazar is a FSRQ, we also fit these two observations using power law plus black body and power law plus bremsstrahlung but 
did not find any improvement on the power law model.   

{\bf PKS 0537$-$441:} This blazar is a FSRQ and a TeV source (Tsujimoto et al. 2015). 
The spectrum is well fitted by simple power law by Donato et al. (2005) with spectral index of 1.8.
We observed three pointings of XMM$-$Newton in 2010 and found the X-ray spectrum of this blazar to be well described by log-parabolic model.
There is significant spectral flattening with negative curvature. All the spectra are also fitted with the broken power law model with
 break energy at around 2.1--2.7 keV.

{\bf S5 0716$+$714:} This is a well known ISP (Giommi et al. 2008). The X-ray spectrum of this source is very well studied
in literature by Tagliaferri et al. (2003); Ferrero et al. (2006); Donato et al. (2005); Wierzcholska \& Wagner (2016)
and has shown an upturn in the SED at the break energy varying between $\sim$ 1.5--2.73 keV. We studied one pointing of XMM$-$Newton
in 2007 and found that the spectra is well described with a log parabolic model. The curvature is significantly negative. We fit the spectrum 
with the broken power law to locate the break energy at $\sim$1.9. The spectral indices ($\Gamma=$2.63) are consistent
with the previous studies. Fitting parameters are not significantly altered by making $N_{H}$ as free parameter.  

{\bf OJ 287:} The X-ray spectrum of this blazar is analyzed by Donato et al. (2005) in two occasions and found the spectra were
well characterized by simple power law with $\Gamma$=1.6--1.9. Massaro et al. (2003) analyzed the X-ray spectra of this source
during its optical bright state in 2001 and found the spectrum to be described by power law model. Seta et al. (2009) analyzed
Suzaku observations of the blazar in April and November 2007, in quiescent and flaring states,  respectively. They found the simple
power law to describe the spectra with photon indices $\Gamma$=1.65 and 1.50 in quiescent and flaring states,  respectively.
In more recent studies Siejkowski \& Wierzcholska (2017) found flat spectrum of OJ 287, which can be interpreted as an indication of 
the spectral upturn located in the X-ray regime.
We analyzed five pointings of XMM$-$Newton and found one observation in April 2005 and April 2008 to be well fitted by simple power law. 
Other three pointings in November 2005, November 2006 and October 2011 have shown significant negative spectral curvature with 
break energy varying between 1.5--3.0 keV. We did not find any significant improvement by keeping $N_{H}$ as free parameter in the models.

{\bf S4 0954$+$65:} This source is a TeV blazar. Tanaka et al. (2016) analyzed Swift/XRT observations of this source in optical
flaring state and found softening of the X-ray spectrum, with a photon index of $\Gamma$ $\sim$1.7 (compared to the earlier outburst
with $\Gamma$ $\sim$1.38) possibly indicating a modest contribution of synchrotron emission by the highest-energy 
electrons superposed on the inverse Compton component. We analyzed the X-ray spectra of this blazar on two occasions in 2007 and 
found them to be well described with simple power law model with photon index $\Gamma$ $\sim$1.96. Free $N_{H}$ in the fitting models 
did not alter the results. 

{\bf ON 231: }  This blazar has shown significant concave curvature in literature (i.e. Donato et al.
2005; Tagliaferri et al. 2000; Wierzcholska \& Wagner 2016). Wierzcholska \& Wagner (2016) fitted the Swift/XRT observation of this source
using the broken power law and found a break at 2.01 keV. Donato et al. (2005) found the break energy to be at $3.09_{-1.092 }^{+0.557}$. 
We analyzed the spectra of this source on four occasions and found
significant curvature in one of the observations performed in 2002. We found a break energy of $4.33_{-1.87 }^{+0.51}$ keV which is in 
consistent with previous studies within errorbars. We did not find any significant improvement by keeping $N_{H}$ as free parameter in the models.    
   
{\bf 3C 279:} This blazar is very well studied FSRQ and is classified as an LSP (Ackermann et al. 2011). In previous studies by Donato et al. 
(2005), its spectra were fitted by power law and broken power law  and break energy is found at low energy ($\sim$ 0.5--0.7) keV.
Wierzcholska \& Wagner (2016) studied the source and found it to be well described by the log parabola model with positive curvature.
We analyzed the spectra of 3C 279 in two occasions in 2009 and 2011 using XMM$-$Newton. In both the pointings, we did not find any significant 
curvature and F-test gave reasonable fit using power law model. The photon indices are varying between 1.77--1.80 which are 
in accordance with previous studies.

{\bf PKS 1334$-$127:} This FSRQ is a $\gamma$-ray loud blazar studied by Foschini et al. (2006) using XMM$-$Newton observations 
and found it to be well fitted with power law model with $\Gamma$=1.8 and Galactic absorption equal to 6.7$\pm$0.9 in 0.4--10 keV energy band. 
We also analyzed this spectrum in the (0.6--10) keV and found it to be well described by the power law model with $\alpha$=1.8.
We did not find any significant improvement by keeping $N_{H}$ as free parameter in our fitting parameters. 

{\bf BL Lac}: The X-ray spectra of this source is very well studied with log-parabolic as well as with power law model in previous studies. 
Tanihata et al. (2000) found that a soft steep component dominates below 1 keV and hard component dominates otherwise. Ravasio et al. (2002)
 observed BL Lac in two observations and found significant negative curvature above 5-6 keV in July 1999. However, December 1999 
observation was well fitted by a simple power law. Donato et al. (2005) also found significant concave curvature after the break 
energy of $\sim$1.71 keV. Massaro et al. (2008) also preferred log-parabolic model for the BeppoSAX observations of this source. 
Raiteri et al. (2009) analyzed the XMM$-$Newton observations of this source during the period 2007--2008 in its low state and found significant 
concave curvature in all of the observations. Wierzcholska \& Wagner (2016) also found the signature of
negative curvature in the BL Lac spectrum and found the break energy at $\sim$1.1 keV. We analyzed three pointings of XMM$-$Newton and found
that the spectra shows an upturn at the break energy of around $\sim$1.25--1.55 keV.  Fitting parameters are not significantly altered 
by making $N_{H}$ as free parameter. 

{\bf 3C 454.3:} This source is studied by Donato et al. (2005) and found that it is well described by simple power law with $\Gamma$=1.34.
XMM$-$Newton observations are analyzed by Raiteri et al. (2007); 2008 in its post outburst and outburst states, respectively. We analysed four 
observations of XMM$-$Newton and found that all are well described by the power law model. The spectral indices varies between 1.5--1.7
which is in accordsnce with the previous findings. 

\subsection{Blazars with concave spectrum}

In 7 (PKS 0235+164, PKS 0426-380, PKS 0537-441, S5 0716+714, OJ 287, ON 231 and BL Lacertae) of 14 blazars in our sample, we infer a significant 
negative curvature suggesting that both synchrotron and inverse Compton components are located in the X-ray regime. The spectra of these sources 
are also fit by a broken power law model to infer the $E_{\rm break}$, with two energy bands for each source. The soft energy band is in the energy 
range 0.6 keV--$E_{\rm break}$, representing the synchrotron component and the hard energy band is in the range $E_{break}$--10 keV, representing 
the inverse Compton component. Though the power law PSD slopes based on the 0.6--10 keV band integrated light curves for these sources are dominated by 
white noise, if the underlying physical mechanism as inferred from the spectral analysis shows a demarcation between the synchrotron and inverse Compton 
components, it is expected that the variability in the above defined soft and hard energy bands must show clear differences. The $F_{\rm var.}$ is then 
calculated for both segments of each of the light curves for the sources displaying a spectral break to help distinguish their origin. Below, we 
determine the spectral and temporal variability of the soft and the hard components separately for all those sources and the figures for these blazars 
are provided in Fig 10--12. 

{\bf AO 0235$+$164:} The light curve of this source has shown pronounced variability in one of the occasion in 10.02.2002 with 
$F_{var}$ of 11.76\% and is shown in fig 10. The hardness ratio versus counts/sec is also showed in lower panel. We did not find any correlation between 
these two. To separate the synchrotron and inverse Compton components, we divided the light 
curve in two bands i.e. soft (0.6--4.03 keV) and hard (4.03--10 keV) energy bands. The fractional variability amplitude for the
soft and hard energy bands are 11.94$\pm$0.56\% and 8.88$\pm$5.45\%, respectively. It can be seen that the synchrotron components is highly
variable and contributing mostly to the total emission. Hard IC component appears to be almost stable.   

{\bf PKS 0426$-$380:} The source has not shown any variability during the observation. A weak positive correlation is found between HR and 
counts/sec with (Pearson correlation coefficient, $r$=0.20, $p$=0.007). We separated the soft 
(0.6-2.1 keV) and hard (2.1-10 keV) energy components and found that the hard energy band representing the IC component showed higher fractional 
variability amplitude with $F_{var}$=21.94$\pm$6.25 as compared to stable synchrotron component (1.82$\pm$15.15).


{\bf PKS 0537$-$441:} The source has not shown significant variability in X-ray bands and are shown in fig 10 along with their respective
HR versus counts/sec. We found weak positive correlation between HR versus counts/sec for both the pointings i.e. 27.02.2010 ($r$=0.239, 
$p$=4.49$\times10^{-5}$) and 04.03.2010 ($r$=0.222, $p$=6.6$\times10^{-4}$). It indicates that both spectra are showing bluer-when-brighter trend.
We divided the light curve of 27.02.2010 into (0.6--2.28) and (2.28--10 keV), 
respectively. The hard energy band representing the IC component is highly variable ($F_{var}$=11.47$\pm$3.77) in this case. 
Synchrotron component is almost constant. Also, in the other observation performed on 04.03.2010, we divided the light curve 
into soft (0.6--2.16 keV) and hard (2.16--10 keV) energy bands. The synchrotron component has shown more pronounced variability with 
$F_{var}$=7.10$\pm$1.52 as compared to stable IC component. For the above two observations, $E_{break}$ is consistent within the errorbars.

{\bf S5 0716$+$714:} This blazar has shown significant variability in the X-ray band with $F_{var}$=17.94\%.
We found a negative correlation between HR versus counts/sec ($r$=-0.397, $p <$2.2$\times10^{-16}$)  which indicates redder-when-brighter trend. 
 We divided the light curve in soft (0.6--1.9 keV) and hard (1.9--10 keV) energy bands.  For this sources, synchrotron 
component reveals more pronounced variability with $F_{var}$=38.03$\pm$0.39 as compared to lesser variable IC component (28.40$\pm$1.10). 

{\bf OJ 287:} This blazar has shown significant variability in one occasion of 17.11.2006 with $F_{var}$=7.98$\pm$0.96.
We did not find any correlation between HR versus counts/sec during two pointings of 03.11.2005 and 15.10.2011. However, positive
correlation of $r$=0.256, $p$=4.078$\times10^{-8}$ is found during the observation dated 12.11.2006.
We divided the light curve into soft (0.6--3.08 keV) and hard (3.08--10 keV) and found that IC component (17.84$\pm$3.90) has higher variability
amplitude as compared to the synchroton component (7.16$\pm$1.23). Similarly, during the pointing on 15.10.2011, $E_{break}$ shifts to
1.66 keV and IC component shows more pronounced variability of 5.47$\pm$1.48 as compared to the flatter synchrotron component
(2.88$\pm$1.42). However, the case is very different on 03.11.2005 where  $E_{break}$ is consistent with the  $E_{break}$ of pointing
on 15.10.2011, but synchrotron component is highly variable (7.20$\pm$1.98) as compared to almost constant IC component 3.17$\pm$5.77.

{\bf ON 231:} This ISP has shown strong variability during the pointing on 26.06.2002. The HR versus counts/sec is also shown in fig 8 and we 
found strong negative correlation ( $r$=-0.414,  $p$=3.739$\times10^{-11}$) which represents redder-when-brighter trend. 
In the spectral fitting, we found a break at 4.1 keV
and divided the spectra into soft (0.6--4.3) keV and hard (4.3--10) keV energy bands, respectively. It has been found that the hard
energy component has more variability amplitude ($F_{var}$=65.72$\pm$3.43) as compared to the synchrotron component (52.86$\pm$3.61).

{\bf BL Lacertae:} This blazar has not shown variability in all of the three pointings it was observed with XMM$-$Newton.
Significant correlation is not found between HR versus counts/sec for two pointings of 05.12.2007 and 08.01.2008. However,
significant negative correlation ($r$=-0.468, $p$=1.9$\times10^{-11}$) is found for the pointing dated on 10.07.2007.
We divided the light curve observed on 10.07.2007 into soft (0.6-1.26 keV) and hard (1.26--10 keV) energy bands and found that synchrotron 
component is highly variable with $F_{var}$=8.92$\pm$0.99 than the constant IC component ($F_{var}$=1.87$\pm$2.13). We divided
the pointing on 05.12.2007 at $E_{break}$=1.56 and we found that both, synchrotron as well as IC components are stable with
$F_{var}$=2.09$\pm$2.49 and 2.69$\pm$2.61, respectively. Third observation is divided at $E_{break}$=1.43 and found the IC component
to be more variable ($F_{var}$=4.19$\pm$1.29) as compared to flat synchrotron component ($F_{var}$=1.18$\pm$4.79).  



\begin{table*}
{\caption {Fractional Variability Amplitude of soft and hard bands} }
\noindent
\begin{tabular}{lllllll} \hline \hline

Source &Blazar &Date of         &$F_{var}$ &$F_{var}$  &$E_{break}$ & PSD slope \\
       &Class  &observation     &(Soft)    &(Hard)  & Slope & $\mu$\\ \hline

PKS 0235$+$164 &LSP &2002.02.10 &11.94$\pm$0.56  &8.88$\pm$5.45 &$4.03_{-0.77 }^{+0.61}$ & -0.84$\pm$0.45 \\
PKS 0426$-$380$^{*}$ &LSP &11.02.2012 &1.82$\pm$15.15  &21.94$\pm$6.25 &$2.11_{-0.54}^{+0.74}$ & -0.01$\pm$0.01\\
PKS 0537$-$441$^{*}$ &LSP &2010.02.27 &0.87$\pm$6.90   &11.47$\pm$3.77 &$2.28_{-0.67}^{+0.38}$ & -0.02$\pm$0.01\\
               & &2010.03.04 &7.10$\pm$1.52   &6.49$\pm$9.54 &$2.16_{-0.85}^{+0.28}$ & -0.01$\pm$0.01\\
S5 0716$+$714  &ISP &2007.09.24 &38.03$\pm$0.39  &28.40$\pm$1.10 &$1.90_{-0.10}^{+0.12}$ & -1.03$\pm$0.68\\
OJ 287$^{*}$   &ISP &2005.11.03 &7.20$\pm$1.98   &3.17$\pm$5.77  &$1.54_{-0.18}^{+0.81}$& -0.72$\pm$0.25\\
               & &2006.11.17 &7.16$\pm$1.23   &17.84$\pm$3.90 &$3.08_{-1.34}^{+0.70}$ & -0.46$\pm$0.19\\
               & &2011.10.15 &2.88$\pm$1.42   &5.47$\pm$1.48 &$1.66_{-0.34}^{+0.56}$ & -0.35$\pm$0.10\\
ON 231         &ISP &2002.06.26 &52.86$\pm$3.61  &65.72$\pm$3.43 &$4.3_{-0.51}^{+1.87}$ & -0.88$\pm$0.28\\
BL Lac$^{*}$   &ISP &2007.07.10 &8.92$\pm$0.99   &1.87$\pm$2.13 &$1.26_{-0.09}^{+0.11}$ & -0.71$\pm$0.28\\
               & &2007.12.05 &2.09$\pm$2.49   &2.69$\pm$2.61 &$1.56_{-0.15}^{+0.29}$ & -0.56$\pm$0.24\\
               & &2008.01.08 &1.18$\pm$4.79   &4.19$\pm$1.29 &$1.43_{-0.10}^{+0.48}$ & -0.13$\pm$0.06\\ \hline              
\end{tabular} \\  
$^{*}$: Sources which have not shown significant $F_{var}$ for (0.6--10) keV energy band and hence cannot be considered
strong cases. 
\end{table*}


\section {Discussion and Conclusions}

{We study a sample of 14 blazars (5 ISPs and 9 LSPs) in the 0.6 -- 10 keV X-ray energy band spanning 31 observation epochs.} 
The analyses includes a timing study using the Lomb-Scargle periodogram to infer the power spectral density shape and possible 
quasi-periodic oscillations in their light curves and the fitting of the 0.6--10 keV spectrum with parametric models depending on 
source properties. We did not find any characteristic timescale in any light curve. The PSD shape is also well fitted with a power 
law with no temporal breaks or quasi periodic components. This is expected for this class of blazars as their synchrotron component 
peaks in optical/UV and hence are highly variable in these bands. X-ray emission generally comes from less variable and flatter 
inverse Compton component (e.g. Gupta et al. 2016). The inferred fractional variability amplitude $F_{var}$=52\% for one pointing of 
ON 231 which is the highest in the present sample. In 20/31 of the pointings, we found small amplitude flickering ($F_{var}<$5\%) in the X-ray light curves. In six of the observations, we found significant ($>$3 $\sigma$) variability amplitude in the range varying from 5--52 \%. This includes one pointing of PKS 0235$+$164, S5 0716$+$714, two poinings of S4 0954$+$65 and two pointings of ON 231. Spectral analysis is performed to check whether power law or log parabolic model well describes the LBL and IBL spectra. In 18/31 epochs, the spectra are well described by the power law model. The remaining 13/31 are well described by the log parabolic model. The spectral indices $\Gamma$ for our sample of blazars lie in the range 1.2--2.7. The results are consistent with the previous studies where for LSPs and ISPs, flatter X-ray spectral slopes varying between 1.5--2 are found (e.g. Donato et al. 2005; Massaro et al. 2008; Wierzcholska \& Wagner 2016).

 In seven of these blazars (PKS 0235+164, PKS 0426-380, PKS 0537-441, S5 0716+714, OJ 287, ON 231 and BL Lacertae), we found significant 
concave or negative curvature. The spectral fits with their residuals are shown in figure 2--9. It is clear from the figure that there 
are discrepancies around 1--3 keV with significant spectrum flattening. As the synchrotron peak lies in the optical/IR for LSPs and ISPs, 
it is expected that the transition from the synchrotron emission to inverse Compton occurs in the X-ray bands. Hence, the concave curvature 
could be interpreted as the beginning of a spectral upturn from the steep component which is the high energy tail of the synchrotron 
component to more flatter low-energy side of the inverse Compton component. The inferred break energy for these sources are consistent 
within error bars and do not vary significantly with flux changes.
Similar spectral upturns are found in previous studies also i.e. 3C 66A (Donato et al. 2005, Wierzcholska \& Wagner 2016); S5 0716$+$714 (Giommi et al. 1999; Tagliaferri et al.2003; Donato et al. 2005; Ferrero et al. 2006; Wierzcholska \& Siejkowski 2015); ON 231 (Tagliaferri et al. 2000); BL Lacertae (Tanihata et al. 2000; Ravasio et al. 2002);  AO 0235$+$164 (Raiteri et al. 2006); OQ 530 (Tagliaferri et al. 2003); 4C $+$21.35 (Wierzcholska \& Wagner 2016). For the well known HBL Mrk 421, Kataoka \& Stawarz (2016) found that the X-ray observations from NuStar Satellite is dominated by the highest-energy tail of the synchrotron continuum till 20 keV and the variable excess hard X-ray emission is related to inverse Compton emission in its very low state. 
All these sources are TeV blazars and have shown variability in $\gamma$-bands (Abdo et al. 2010). 

Differences in the fractional variability amplitude for the soft and hard components is additionally used to demarcate the emission mechanisms. 
Four sources do not show significant variability based on their 0.6 -- 10 keV band integrated light curves (marked in Table 5). Though, on five of the twelve occasions, a more pronounced variability is inferred in the soft energy band i.e the high energy tail of the synchrotron component. In the remaining seven occasions (including five blazars PKS 0426$-$380, PKS 0537$-$441, OJ 287, ON 231 and BL Lacertae), the variability is more pronounced in the hard energy band i.e. inverse Compton component. The study of Wiercholska \& Wagner (2016) also found more pronounced variability of inverse Compton component for the blazar BL Lacertae. For the ISP S5 0716$+$714, this component appears to show pronounced variability with $F_{var}$=28.40\% and the synchrotron component also highly variable with $F_{var}$=38.03\%. 
It must be noted however that as the spectral slopes after $E_{\rm break}$ in some cases (PKS 0537$-$441: 2010.02.27, OJ 287: 2006.11.17 and 
ON231: 2002.06.26) are still soft, in addition to a possible onset of transition from synchrotron to inverse Compton emission, this may also be 
due to a hardening of synchrotron emission from the highest energy electrons which is expected in physical scenarios including a dominant inverse 
Compton cooling in the Klein-Nishina regime \cite[e.g.][]{2005MNRAS.363..954M} or a flattening intrinsic to the mechanisms causing the electron 
acceleration. Such scenarios can also produce an increased $F_{\rm var}$.

 We infer a negative spectral curvature in seven sources (Table 5), four of which are IBLs (BL Lacs) and the remaining three, LSPs (both BL Lacs and FSRQ). Their high energy emission (X-ray--$\gamma$ ray) in the leptonic scenario is produced by inverse-Compton scattering of lower energy seed photons.
The seed photons can originate from the synchrotron emitting electrons (synchrotron self-Compton, SSC process) or external sources (external Compton, EC process), including the accretion disk, broad line region clouds, dust torus or the cosmic microwave background (e.g. Sikora, Begelman \& Rees 1994; Dermer, Sturner \& Schlickeiser 1997). From a systematic study of blazar broadband spectra, Ghisellini et al. (1998) suggest that along the sequence HSP--LSP--FSRQ, there exists an increasing contribution to the energy density from external radiation fields in turn leading to an increasing incidence of Compton cooling. 
The high energy spectrum of FSRQs thus contains a prominent contribution from EC processes (e.g. Mukherjee et al. 1999; Hartman et al. 2001) while that of HSPs can be sufficiently fit with SSC models (Pian et al. 1998; Bottcher et al. 2002). The ISPs and LSP BL Lacs being intermediate between these classes then require to account for a non-neglible contribution from EC processes to reproduce the high energy spectrum (Madejski et al. 1999; Bottcher \& Bloom 2000). Owing to this, we incorporate all these contributory sources in the following model to estimate physical properties of the emission region.

The SSC and EC emission processes may be disentangled by appealing to clear differences in the variability pattern between the hard and soft X-ray bands. 
 The hard band variability can be similar to or less than the soft band in the SSC process and can be larger than the soft band in the EC process if it contains imprints of the rapid microvariability (hour timescales) from the accretion flow (e.g. inner accretion disk, corona and the disk-jet transition region). 
However, in the strong cooling regime, changes in the seed photons may not necessarily cause rapid changes in the EC emission. Hence, no strict conclusion 
may be drawn from the current small sample of sources indicating different $F_{\rm var}$. For both the SSC and EC scenarios, the size of the region along the jet participating in the scattering is
\begin{equation}
\Delta r \leq \frac{c \delta \tau}{1+z},
\end{equation}
where $c$ is the speed of light, $\delta$ is the Doppler factor, $\tau$ is the variability timescale and $z$ is the redshift. For the range of 
$z = 0.07 - 1.11$ corresponding to BL Lac and PKS 0426$-$380 respectively (Table 1), using a $\delta \sim 6$ obtained from studies of relativistic beaming 
in BL Lac objects \citep{2009PASJ...61..639F,2009A&A...494..527H}, and $\tau \sim 1$ day (as a majority of the light curves do not show clear trends 
and are consistent with Poisson noise), $\Delta r \leq (0.002 - 0.005)$ pc. As this small region is along the beamed jet, we can estimate the distance 
to this region $r$ from the central black hole by assuming a conically shaped jet with half opening angle $\theta_0 \sim 1/\Gamma$ where $\Gamma$ is 
the bulk Lorentz factor and $r \sim \Gamma \Delta r$. For $\Gamma \sim 10$ \cite[e.g.][]{2009A&A...494..527H}, $r \leq (0.02 - 0.05)$ pc $\sim (249 - 492)
 r_S$ where $r_S = G M_\bullet/c^2$ is the Schwarzschild radius for a black hole of mass $M_\bullet$, taken to be $\sim 10^9 M_\odot$. In canonical 
models of the AGN jet structure \cite[e.g.][]{2008Natur.452..966M}, the region at $\sim 10^2 - 10^3~r_S$ hosts helical magnetic fields which 
can lead to the appearance of quasi-periodic flux and polarization variability for favourable viewing angles due to beamed emission
 \cite[e.g.][]{2015ApJ...805...91M,2016MNRAS.463.1812M}.

In the EC scenarios relating to strong variability, we can roughly estimate the relevant scattering energies involved. For a thin accretion disk 
emitting black body radiation (e.g. Shakura \& Sunyaev 1973), the temperature structure in scaled units is
\begin{equation}
k T \sim 54~\left(\frac{\dot{m}}{m_9 \tilde{r}^3} \left(1-\left(\frac{1}{\tilde{r}}\right)^{1/2}\right)\right)^{1/4}~{\rm eV}
\end{equation}
where $m_9 = M_\bullet/(10^9 M_\odot)$, $\tilde{r} = r/r_S$ and $\dot{m} = \dot{M}/\dot{M}_{\rm Edd.}$ is the accretion rate $\dot{M}$ scaled by 
the Eddington accretion rate $\dot{M}_{\rm Edd.} = L_{\rm Edd.}/(\eta c^2)$ with $L_{\rm Edd.} = 1.3 \times 10^{47} m_9$ and $\eta = 0.007$. 
For $\tilde{r} \sim 3 - 50$ and $\dot{m} = 0.1 - 0.3$ \cite[expected range for blazars, e.g.][]{2009MNRAS.399.2041G}, $k T \sim (1.7 - 15.5)$ eV, 
corresponding to optical/ultra-violet emission. For a corona (composed of electrons) in virial equilibrium,
\begin{equation}
k T \sim \frac{m_e c^2}{2 \tilde{r}} = \frac{0.511}{2 \tilde{r}}~{\rm MeV} 
\end{equation}
where $m_e = 0.511$ MeV/c$^2$ is the electron rest mass. For $\tilde{r} \sim 3 - 50$, the accretion energy per electron $k T \sim (5.1 - 85.2)$ keV. If the lower energy optical/ultra-violet disk based photons scatter off sufficiently dense regions of the corona, this can result in upscattering of soft X-ray photons to higher energy further downstream in the jet by relativistic electrons.

The relative strength of the emission processes can be tested and physical parameters of the pc-scale jet roughly estimated using a toy model involving synchrotron and inverse Compton emission. The synchrotron and Compton power losses $L_S$ and $L_C$ (effectively the luminosity) during the emission and scattering from a relativistic electron 
respectively are in the ratio
\begin{equation}
\frac{L_S}{L_C} = \frac{U_B}{U_{\rm ph.}},
\end{equation}
where $U_B = B^2/(8 \pi)$ and $U_{\rm ph.}$ are the magnetic field and radiation energy densities respectively. We can infer the magnetic field strength as
\begin{equation}
B = \left(8 \pi U_{\rm ph.} \frac{L_S}{L_C}\right)^{1/2},
\end{equation}
once we estimate luminosities $L_S$ and $L_C$ and the energy density in the radiation field. It is assumed that the electrons emitting synchrotron 
traverse a region in the jet with a strong magnetic field such as that argued above (in the region $\sim 10^2 - 10^3~r_S$). Further, if we assume 
that the majority of scattered Compton energy density is from these synchrotron emitting electrons being Doppler beamed along the observer 
line of sight $U_{\rm ph.,S}$ and an EC process composed of energy densities from the accretion disk, broad line region clouds surrounding the disk and the dust torus \cite[e.g.][]{2009MNRAS.397..985G,2016ApJ...830...94F},
\begin{align}
U_{\rm ph.} &= U_{\rm ph.,S}+ U_{\rm ph.,D}+ U_{\rm ph.,BLR}+ U_{\rm ph.,T}\\ \nonumber
&= \frac{1}{4 \pi c} \left\{\frac{L_S}{\delta^4 \Delta r^2}+\Gamma^2 \left(\frac{L_D}{r^2_D}+\frac{L_{\rm BLR}}{r^2_{\rm BLR}}+\frac{L_T}{r^2_T}\right)\right\},
\end{align}
where $L_{\rm D,BLR,T}$ and $r_{\rm D,BLR,T}$ are the accretion disk, broad line region and torus luminosities and radii respectively. The luminosity $L_S$ is scaled by the Doppler factor $\delta$ and the luminosities $L_{\rm D,BLR,T}$ by the Lorentz factor for an energy density as measured in the jet co-moving frame \cite[e.g.][]{2014Natur.515..376G}. 

For a radiatively efficient accretion disk,
\begin{equation}
L_D =\eta \dot{M} c^2 = (9.1 \times 10^{44}~{\rm erg/s})~\dot{m} m_9,
\end{equation}
and $r_D$ is the radius of the accretion disk participating in the contribution to the EC scattered seed photons. An approximation for $r_D$ can be 
the distance at which the disk transitions from being gravitationally bound to the central supermassive black hole to being self gravitational 
(King 2016) in which case
\begin{equation}
r_D = (6.46 \times 10^{16}~{\rm cm})~\dot{m}^{-8/27} m^{1/27}_9,
\end{equation}
which is typically $0.01 - 0.1$ pc. 
If we assume that all radiation emitted from the disk participates in ionizing the BLR clouds and results in a re-processed emission from the BLR, the emergent luminosity $L_{\rm BLR}$ is similar to $L_D$ owing to ionization timescale being much lesser compared to the scattering timescale (mostly dominated by the geometry of the BLR). Then, $L_{\rm BLR} = \eta_{\rm BLR} L_D$ where $\eta_{\rm BLR} \sim 0.1$ is a disk covering factor attributable to the BLR and represents the fraction of disk radiation it re-processes \cite[e.g.][]{2008MNRAS.387.1669G,2009MNRAS.397..985G} and

\begin{equation}
r_{\rm BLR} = (10^{17}~{\rm cm})~\left(\frac{L_D}{10^{45}}\right)^{0.5} = (9.54 \times 10^{16}~{\rm
 cm})~\dot{m}^{0.5} m^{0.5}_9.
\end{equation}

The dust torus emission is expected to be thermal and dominant in the infra-red wavelengths. If we assume that $L_T = \eta_{T} L_D$ where $\eta_{T} \sim 0.5$ is a disk covering factor attributable to the torus \cite[e.g.][]{2008MNRAS.387.1669G,2016ApJ...830...94F} and a torus temperature $T_T$ below the dust sublimation temperature, the torus radius can be scaled in terms of the disk luminosity as \cite[e.g.][]{2008ApJ...685..160N,2016ApJ...830...94F}

\begin{align}
r_T &= (3.5 \times 10^{18}~{\rm cm})~\left(\frac{L_D}{10^{45}}\right)^{0.5} \left(\frac{T_T}{10^{3}
 {\rm K}}\right)^{-2.6} \\ \nonumber
&= (3.5 \times 10^{18}~{\rm cm})~\dot{m}^{0.5} m^{0.5}_9 T^{-2.6}_{T,3},
\end{align}

where $T_{T,3} = T_T/(10^3 {\rm K})$.

For the sources indicating a spectral upturn, we can approximate $L_C \geq L_X = 4 \pi D^2_L f_X$ where $f_X$ are the X-ray fluxes obtained from the spectral fits and listed in Table 3. We take $L_S$ based on the observed synchrotron peak flux densities (Fan et al. 2016 and references therein) for PKS 0235$+$164, PKS 0537$-$441, S5 0716$+$714, OJ 287, ON 231 and BL Lac and $L_S = 4 \pi D^2_L f_R$ based on a 1.4 GHz flux density of $4 \times 10^{-4}$ Jy (Tingay et al. 2003) for PKS 0426$-$380. Input parameters include $m_9 = 1$, $\dot{m} = 0.1 - 0.3$, $\delta = 6$, $\eta = 0.007$, $T_{T,3} = 0.2$ (based on a typical range of torus temperatures between 138 - 300 K; \cite{2007ApJ...660..117C}) and the resulting estimates of $B$ are presented in Table 6. In these estimates, we assume a flat cold dark matter dominated cosmology with matter energy density $\Omega_{m} = 0.308$ and Hubble constant $H_0 = 67.8~$km s$^{-1}$ Mpc$^{-1}$ (Planck Collaboration et al., 2016) to calculate the luminosity distance $D_L$ (Wright, 2006). The contribution of the dust torus to the estimated $U_{\rm ph.}$ is negligible in these cases (typically at 0.0001 - 0.001 \%) but is included in the calculation for completeness and for cases where larger $T_T$ which can result in a non-negligible contribution. The $B$ values estimated here are strictly upper limits owing to $L_C \geq L_X$ and are in the range $0.03 - 0.88$ G consistent with estimates from the core shift method \cite[e.g.][]{2015MNRAS.452.2004M,2017MNRAS.469..813A} for some of these sources (S5 0716$+$714 and BL Lac with $B = 0.22 \pm 0.36$ G and $0.02 \pm 0.06$ G respectively), eventhough the region being probed here is $\leq$ 0.05 pc compared to the pc-scale jet in the core shift method, thus indicating a similar magnetic field energy density from sub-pc--pc scale possibly implying that magnetic field structuring develops in the innermost jet and is retained upto the pc-scale jet, consistent with the canonical perspective. 
The method can thus be applied in a self consistent manner, accounting for all sources contributing to the energy density including the broad line region clouds, the torus and the microwave background, and is an independent check on estimates from core shift measurements which assume equipartition between the magnetic and particle kinetic energy densities. The above simplistic calculation is able to capture a general agreement with other methods. For a more rigourous approach, it can be extended to include the distribution function of the particles composing the jet (which can include electron-positron pairs; leptons and protons), the bulk properties of the jet \cite[e.g.][]{2010MNRAS.409L..79G}, treatment of the $\gamma$-ray regime and a careful assessment of the covering factors and various geometries of the sources of external seed photons \cite[e.g.][]{2016ApJ...830...94F}, and the relative motion between the emitting source and the observer which can introduce systematic variability \cite[e.g. in a helical jet,][]{2015ApJ...805...91M,2016MNRAS.463.1812M}.


\begin{table*}
\caption{Parameter estimates for sources indicating spectral upturn. Estimates of magnetic field strength at $r \leq 0.02 - 0.05$ pc from the central engine.}
\label{Bfield}
\setlength{\tabcolsep}{0.010in}
\begin{tabular}{lllllllllll}
\hline \hline
Source & Blazar & $D_L$& $m_9$ & $L_C$             & $L_S$             & \multicolumn{4}{l}{Radiation field energy density ($U_{ph.}$)} & $B$ \\
       &Class   & (Gpc)& ($M_\bullet/(10^9 M_\odot)$) & ($\times 10^{45}$ & ($\times 10^{45}$ & SSC & (SSC+D)   & (SSC+D+BLR) & (SSC+D+BLR+Torus) & (G) \\
       &        &      & & erg s$^{-1}$)     & erg s$^{-1}$)     & erg cm$^{-3}$ & erg cm$^{-3}$ & erg cm$^{-3}$ & erg cm$^{-3}$ & \\ \hline
PKS 0235$+$164  &LSP   & 6.29 & 1.00 & 43.38 & 35.48 & 1.13 & 3.65 & 6.33 & 6.33 & 0.31\\
PKS 0426$-$380  &LSP   & 7.72 & 0.40 & 4.65 & 0.04 & 0.001 & 2.52 & 5.19 & 5.19 & 0.03 \\
PKS 0537$-$441  &LSP   & 5.91 & 0.63 & 20.61 & 60.26 & 1.83 & 4.35 & 7.03 & 7.03 & 0.62\\
S5 0716$+$714   &ISP   & 1.66 & 0.13 & 3.16 & 10.00 & 0.15 & 2.66 & 5.34 & 5.34 & 0.56 \\
OJ 287          &ISP   & 1.63 & 0.63 & 0.94 & 7.41 & 0.11 & 2.62 & 5.30 & 5.30 & 0.88\\
ON 231          &ISP   & 0.49 & 0.10 & 0.10 & 0.50 & 0.01 & 2.52 & 5.20 & 5.20 & 0.69 \\
BL Lac          &ISP   & 0.32 & 0.50 & 0.20 & 0.60 & 0.01 & 2.52 & 5.20 & 5.20 & 0.54 \\ \hline
\end{tabular}
\end{table*}

The energy density in the magnetic field encountered by an electron composing the jet is $U_B = |B|^2/(8 \pi)$ with a median $U_B = 0.01$ erg cm$^{-3}$ and the  corresponding synchrotron cooling time is 
\begin{equation}
t_{\rm cool} = \frac{3 m_e c}{4 \sigma_T \beta^2 \gamma U_B} = (0.91~{\rm days}) \tilde{\gamma}^{-1} B^{-2}_1,
\end{equation} 
where $m_e = 9.11 \times 10^{-28}$ g is the electron mass, $\sigma_T = 6.65 \times 10^{-25}$ cm$^2$ is the Thomson cross section for electron
 scattering, $\beta \sim 1$ is the speed of the relativistic electrons, $\tilde{\gamma} = \gamma/10^3$ is the Lorentz factor $\gamma$ of the 
injected electrons scaled by $\sim 10^3$ which can typically be reached, and $B_1 = B/(1 {\rm G})$. For the inferred range of $B$, the median
 $t_{\rm cool} \sim 29$ days. The ratio between the Compton energy density from various sources contributing to EC ($U_{\rm ext.}$,
 here the disk, BLR and torus), and the total scattered photon energy density (including SSC) is $U_{\rm ext.}/U_{\rm ph.}$ = 
0.74--0.99 indicating that the EC seed photons prominently contribute to the observed luminosity and variability. 
In this scenario, the jet can undergo strong radiative cooling due to the Compton drag caused resulting in maximum limits for 
the bulk Lorentz factor and is useful in constraining the jet composition \cite[e.g.][]{2010MNRAS.409L..79G}. If the relativistic 
electrons within a spherical cloud of radius $\Delta r$ are variable over a median timescale $t_{\rm var} = \delta \tau/(1+z) = 4.6$ days 
(jet co-moving frame), $t_{\rm var} < t_{\rm cool}$ in general for these sources. The synchrotron electrons then participate in multiple 
epochs of inverse Compton-scattering before cooling down and the resultant 
variability will tend towards red noise behaviour in their power spectral density shapes (slope $\mu$ = -1 to -2). As the energy of 
scattered photons from a given epoch depends on the available energy from scattering events in previous epochs, thus exhibiting long term 
trends in the light curve within and across epochs. Here, the weighted mean $\mu = - 0.77 \pm 0.26$ indicating that in addition to red 
noise behaviour as expected from the above discussed process, random processes in the jet contribute white noise (slope $\mu = 0$) 
variability thus causing flattening of the slopes. Such jet based processes operational at $\sim$ day type timescales can include 
turbulence and shocks \cite[e.g.][]{1979ApJ...232...34B,2014ApJ...780...87M} or magnetic re-connection events 
\cite[e.g.][]{2003NewAR..47..513L} amongst others.

In the earlier studies, Giommi et al. (1999) and Tagliaferri et al. (2003) found lack of significant variability in the higher energy 
bands i.e. 3 - 5 keV and hence concluded a scenario of a variable synchrotron component and a stable IC component, on hour like time 
scales. But, Ferroro et al. (2006) found that even at higher energies i.e. above $\sim$ 3 keV, where the IC component is expected to 
dominate, the fractional variability amplitude was significantly high. They argued in a favour of a variable IC component on short 
timescales. For our sample of blazars, it might be the case, supported by the above analysis of IC emission sources. 

The hardness ratios are found to be anti-correlated with respect to the total count rate for S5 0716$+$714, ON 231 and BL Lacertae. 
It represents the redder-when-brighter trend which is different for the usual bluer-when-brighter trend for HSPs. The redder-when-brighter 
trend could be explained by the relative contribution of softer synchrotron component, getting higher during the enhanced emission 
with respect to the more stable IC component. It implies an overall steeping of the spectra, however the actual slope becomes flatter.  \\

\section*{Acknowledgements}
 We thank the referee for thoughtful and constructive comments which have improved the context of our work.
This research is based on observations obtained with XMM$-$Newton, an ESA science mission with instruments and contributions directly
 funded by ESA member states and NASA. We thank Lucasz Stawarz for critical comments and discussions which have helped improve 
the manuscript. H.G. is sponsored by the Chinese Academy of Sciences (CAS) Visiting Fellowship for Researchers from Developing 
Countries; CAS Presidents International Fellowship Initiative (PIFI) (grant No. 2014FFJB0005); supported by the National Natural 
Science Foundation of China (NSFC) Research Fund for International Young Scientists (grant No. 11450110398, 11650110434) and 
supported by a Special Financial Grant from the China Postdoctoral Science Foundation (grant No. 2016T90393). P.M. is supported by 
the CAS-PIFI (grant no. 2016PM024) post-doctoral fellowship and the NSFC Research Fund for International Young Scientists
 (grant no. 11650110438). A.W. acknowledges support by the Foundation for Polish Science (FNP).  MF.G. is supported by
the National Science Foundation of China (grants 11473054 and U1531245).

{}



\end{document}